# Heavy Quarks (Working Group 3)

## Summary Report for the HERA - LHC Workshop Proceedings

J. Baines [1], S.P. Baranov [2], O. Behnke [3], J. Bracinìk [4], M. Cacciari [5], M. Corradi [6], A. Dainese [7], S. Diglio [8], K.J. Eskola [9], T.O. Eynck [10], A. Geiser [11], G. Grindhammer [4], H. Jung [11], B.A. Kniehl [12], V.J. Kolhinen [9], S. Kretzer [13], K. Kutak [11], E. Laenen [10], Th. Lagouri [14], A.V. Lipatov [15], F. Maltoni [16], A.D. Martin [17], A. Meyer [11], A. Morsch [18], L. Motyka [12], K. Peters [11,19], F. Petrucci [3], O.I. Piskounova [2], R. Ranieri [20], H. Ruiz [18], M.G. Ryskin [21], I. Schienbein [22], J. Smith [10,23], M. Smizanska [24], H. Spiesberger [25], A. Tonazzo [8], U. Uwer [3], M. Verducci [18,30], R. Vogt [26], C. Weiser [27], N.P. Zotov [15]

[1] Rutherford Laboratory, UK
[2] P.N. Lebedev Physical Institute of the Russian Academy of Science, Moscow, Russia
[3] Universität Heidelberg, Philosophenweg 12 69120 Heidelberg, Germany
[4] Max-Planck-Institut für Physik, München, Germany
[5] LPTHE - Université P. et M. Curie (Paris 6), Paris, France
[6] INFN Bologna, via Irnerio 46, Bologna, Italy
[7] University and INFN, Padova, Italy
[8] Dipartimento di Fisica "E.Amaldi", Università Roma Tre and INFN, Sezione di Roma Tre, via della Vasca Navale 84, I 00146 Roma, Italy
[9] Department of Physics, University of Jyväskylä, Jyväskylä, Finland
[10] NIKHEF Theory Group, Kruislaan 409, 1098 SJ Amsterdam, The Netherlands
[11] DESY, Notkestrasse 85, D 22603 Hamburg, Germany
[12] II. Institut für Theoretische Physik, Universität Hamburg Luruper Chaussee 149, D-22761 Hamburg, Germany
[13] Brookhaven National Laboratory, Upton, NY 11973, USA
[14] Institute of Nuclear and Particle Physics, Charles University, Prague, Czech Republic
[15] D.V. Skobeltsyn Institute of Nuclear Physics, Moscow, Russia
[16] Institut de Physique Théorique, Université Catholique de Louvain, Chemin du Cyclotron, 2, B-1348, Louvain-la-Neuve, Belgium
[17] Institute for Particle Physics Phenomenology, University of Durham, DH1 3LE, UK
[18] CERN, Department of Physics, CH 1211 Geneva 23, Switzerland
[19] School of Physics and Astronomy, The University of Manchester, Manchester M13 9PL, UK
[20] Dipartimento di Fisica, Università di Firenze and INFN, Sezione di Firenze, I 50019 Sesto Fiorentino, Italy
[21] Petersburg Nuclear Physics Institute, Gatchina, St. Petersburg, Russia
[22] Southern Methodist University Dallas, 3215 Daniel Avenue, Dallas, TX 75275-0175, USA
[23] C.N. Yang Institute for Theoretical Physics, Stony Brook University, Stony Brook, NY 11794, USA
[24] Lancaster University, Lancaster, UK
[25] Johannes-Gutenberg-Universität Mainz, D-55099 Mainz, Germany
[26] Lawrence Berkeley National Laboratory, Berkeley, CA, USA and Physics Department, University of California, Davis, CA, USA
[29] Institut für Experimentelle Kernphysik, Universität Karlsruhe, Karlsruhe, Germany
[30] INFN CNAF, Bologna, Italy

**Contents**

## List of participants in the working group:

Baines John, Baranov Sergey, Barbagli Giuseppe, Behnke Olaf, Bertolin Alessandro, Bluemlein Johannes, Bruno Giuseppe, Buchmueller Oliver, Butterworth Jonathan, Cacciari Matteo, Carli Tancredi, Catmore James, Chiochia Vincenzo, Cholewa Axel, Cole Jo, Corcella Gennaro, Corradi Massimo, Dainese Andrea, Ellis Keith, Eskola Kari, Fabris Daniela, Flucke Gero, Geiser Achim, Grab Christoph, Grindhammer Guenter, Guernane Rachid, Gutsche Oliver, Jung Hannes, Kiesling Christian, Klanner Robert, Kolhinen Vesa, Kramer Gustav, Kretzer Stefan, Kuhr Thomas, Kurepin Alexey, Kutak Krzysztof, Laenen Eric, Lagouri Theodota, Levin Eugene, Likhoded Anatoly, Lipatov Artem, Lunardon Marcello, Maltoni Fabio, Mangano Michelangelo, Marti-Magro Lluis, Meyer Andreas, Morando Maurizio, Morsch Andreas, Panikashvili Natalia, Pavel Nikolaj, Peters Krisztian, Piskounova Olga, Ranieri Riccardo, Ruiz Hugo, Schienbein Ingo, Sharafiddinov Rasulkhozha, Shulha Siarhei, Smizanska Maria, Sridhar Krishnamoorthy, Staykova Zlatka, Thompson Paul, Thorne Robert, Tonazzo Alessandra, Turrisi Rosario, Uwre Ulrich, Villa Mauro, Vogt Ramona, Vulpescu Bogdan, Weiser Christian, Wing Matthew, Zoccoli Antonio, Zotov Nikolai, zur Nedden Martin.

# Introduction

*O. Behnke, M Cacciari, M. Corradi, A. Dainese, A. Geiser, A. Meyer, M. Smizanska, U. Uwer, and C. Weiser*

Perturbative QCD is expected to provide reliable predictions for the production of bottom and (to a lesser extent) charm quarks since their masses are large enough to assure the applicability of perturbative calculations. Anyway a direct comparison of perturbative QCD predictions to heavy flavour production data is not straightforward. Difficulties arise from the presence of scales very different from the quark masses that reduce the predictivity of fixed-order theory, from the non-perturbative ingredients needed to parametrize the fragmentation of the heavy quarks into the observed heavy hadrons and from the limited phase space accessible to present detectors. Moreover a breakdown of the standard collinear factorization approach can be expected at low-$x$. The study of heavy quark production in hadronic interactions and in $ep$ collisions at HERA has been therefore an active field, in the effort to overcome these difficulties and to get a deeper understanding of hard interactions.

Besides its intrinsic interest, a precise understanding of heavy quark production is important at LHC because charm and beauty from QCD processes are relevant backgrounds to other interesting processes from the Standard Model (e.g. Higgs to $b\bar{b}$) or beyond. Moreover, theoretical and experimental techniques developed at HERA in the heavy quark field, such as heavy-quark parton densities or $b$-tagging, are also of great value for future measurements at the LHC.

The present status of the heavy-quark production theory is critically reviewed in the first contribution. The second contribution summarises the present heavy flavour data from HERA and gives an outlook of what can be expected from HERA-II. The potential of the LHC experiments for charm and beauty physics is reviewed in the 3rd contribution. Then the relevance of saturation and low-$x$ effects to heavy quark production at HERA and at the LHC are discussed. The non-perturbative aspects of heavy-quark fragmentation and their relevance to HERA and LHC are discussed in the next contribution. Finally, a comparison of different theoretical predictions for HERA and LHC based on different approaches is presented.

# Theoretical review of various approaches in heavy quark production

*S.P. Baranov[3], M. Cacciari[1], S. Diglio[4], T.O. Eynck[2], H. Jung[5], B.A. Kniehl[6], S. Kretzer[7], E. Laenen[2], A.V. Lipatov[8], F. Maltoni[9], F. Petrucci[4], O.I. Piskounova[3], I. Schienbein[6], J. Smith[2,10], A. Tonazzo[4], M. Verducci[11], N.P. Zotov[8]*

[1]LPTHE - Université P. et M. Curie (Paris 6), France
[2]NIKHEF Theory Group, Kruislaan 409, 1098 SJ Amsterdam, The Netherlands
[3]P.N. Lebedev Physical Institute of Russian Academy of Science, Moscow, Russia
[4]Università Roma Tre, Dipartimento di Fisica "E.Amaldi" and
INFN Sezione Roma III, Via della Vasca Navale 84, 00146 Rome, Italy
[5]Deutsches Elektronen-Synchroton Hamburg, FRG
[6]II. Institut für Theoretische Physik, Universität Hamburg, Luruper Chaussee 149, 22761, Hamburg, Germany
[7] Brookhaven National Laboratory, Upton, NY 11973, USA
[8]D.V. Skobeltsyn Institute of Nuclear Physics, Moscow, Russia
[9]Institut de Physique Théorique, Université Catholique de Louvain,Chemin du Cyclotron, 2, B-1348, Louvain-la-Neuve, Belgium
[10]C.N. Yang Institute for Theoretical Physics, Stony Brook University, Stony Brook, NY 11794, USA
[11]CERN, CH-1211 Genève 23, Switzerland

**Abstract**
We review some of the main theoretical aspects of heavy quark production at HERA that will be important for understanding similar processes at the LHC.

*coordinators: M. Cacciari, and E. Laenen*

## 1  Introduction

The value for the LHC physics program of heavy quark production studies at HERA consists not only of measured quantities such as parton distributions, heavy quark masses etc. but at least as much of the theoretical ideas on heavy quark production that were developed and refined in the course of these studies. The strong experimental interest in heavy quark observables at HERA has led to a significantly increased understanding of the benefits and limitations of finite order calculations. It has stimulated theorists to deepen their insight into the issue of when a heavy quark should be treated as a parton, and it has provoked novel proposals to explain the hadronization of heavy quarks. In what follows we review and critically assess some of these ideas.

## 2  Heavy quark production

The study of heavy quarks, historically plagued by low production rates and large uncertainties, has now entered the regime of 'precision physics'. On the one hand, the larger centre-of-mass energies of the colliders running now (Tevatron, HERA) and in the near future (LHC) lead to a much more copious production yield. On the other hand, technological advances such as the introduction of microvertex detectors based on semiconductor devices allow for much better tagging of the produced heavy flavours, and hence better measurements. Needless to say, an equally substantial improvement of the theoretical calculations has been needed in order to match this progress and therefore deliver predictions with an accuracy at least as good as that of the experimental measurements. Properly testing and constraining the theoretical calculations will in turn help in refining the predictions for the LHC.

One example for which a good theoretical accuracy at the LHC is desirable is in calculating the total $Z$ boson production rate, a process which can be used as a luminosity candle and which we would like to have under control at the one per cent level. One channel contributing to this process is gluon-gluon fusion followed by bottom-antibottom annihilation, $gg \to b\bar{b} \to Z$. This channel provides about 5% of the total $Z$ yield [1]: hence, it must be under control at the 20% level in order to achieve the sought-for final 1% accuracy.

As it turns out, it is more efficient and more reliable to rewrite this in terms of a perturbatively calculated parton distribution function (PDF) for the bottom quark, i.e. as the effective process $b\bar{b} \to Z$. The theoretical tools that we use to construct such heavy quark parton distribution functions must therefore be tested by employing them in other theoretical predictions, to be compared to the available experimental data. In the following section we shall list a number of examples where this is done.

From the point of view of 'standard' perturbative QCD calculations, the situation has not changed since the beginning of the '90s: fully massive next-to-leading order (NLO) calculations were made available for hadron-hadron [2, 3, 4, 5, 6], photon-hadron [7, 8, 9] (i.e. photoproduction) and electron-hadron [10, 11, 12, 13] (i.e. Deep Inelastic Scattering, DIS) collisions. These calculations still constitute the state of the art as far as fixed order results are concerned, and they form the basis for all modern phenomenological predictions.

Over the years, and with increasing experimental accuracies, it however became evident that perturbative QCD alone did not suffice. In fact, real particles - hadrons and leptons - are observed in the detectors, not the unphysical quarks and gluons of perturbative QCD. A proper comparison between theory and experiment requires that this gap be bridged by a description of the transition. Of course, the accuracy of such a description will reflect on the overall accuracy of the comparison. When the precision requirements were not too tight, one usually employed a Monte Carlo description to 'correct' the data, deconvoluting hadronization effects and extrapolating to the full phase space. The final 'experimental' result could then easily be compared to the perturbative calculation. This procedure has the inherent drawback of including the bias of our theoretical understanding (as implemented in the Monte Carlo) into an experimental measurement. This bias is of course likely to be more important when the correction to be performed is very large. It can sometimes become almost unacceptable, for instance when exclusive measurements are extrapolated by a factor of ten or so in order to produce an experimental result for a total photoproduction cross section or a heavy quark structure function.

The alternative approach is to present (multi)differential experimental measurements, with cuts as close as possible to the real ones, which is to say with as little theoretical correction/extrapolation as possible. The theoretical prediction must then be refined in order to compare with the real data that it must describe. This has two consequences. First, one has to deal with differential distributions which, in certain regions of phase space, display a bad convergence in perturbation theory. All-order resummations must then be performed in order to produce reliable predictions. Second, differential distributions of real hadrons depend unavoidably on some non-perturbative phenomenological inputs, fragmentation functions. Such inputs must be extracted from data and matched to the perturbative theory in a proper way, pretty much like parton distribution functions of light quarks and gluons are.

In the following sections we review the state of the art of theoretical calculations of heavy quark production in a number of high energy processes, pointing out similarities and differences. In particolar, resummations aimed at improving the theoretical description of heavy quark production at large transverse momentum or large photon virtuality in DIS (Section 3), small centre-of-mass energy (Section 5) and large centre of-mass energy (Section 6) are described in some detail.

## 3 Collinear resummations and heavy quark PDFs

Perturbative calculations of heavy quark production contain badly converging logarithmic terms of quasi-collinear origin in higher orders when a second energy scale is present and it is much larger than the heavy quark mass $m$. Examples are the (square root of the) photon virtuality $Q^2$ in DIS and the transverse momentum $p_T$ in either hadroproduction or photoproduction. Naming generically $E$ the large scale, we can write schematically the cross section for the production of the heavy quark $Q$ as

$$\sigma_Q(E,m) = \sigma_0 \left(1 + \sum_{n=1}^{\infty} \alpha_s^n \sum_{k=0}^{n} c_{nk} \ln^k\left[\frac{E^2}{m^2} + \mathcal{O}\left(\frac{m}{E}\right)\right]\right), \quad (1)$$

where $\sigma_0$ stands for the Born cross section, and the coefficients $c_{nk}$ can contain constants as well as functions of $m$ and $E$, vanishing as powers of $m/E$ when $E \gg m$.

Resummation approaches bear many different names, (ZM-VFNS, ACOT, FONLL, BSMN to name but a few) but they all share the goal of resumming leading ($\alpha_s^n \ln^n(E^2/m^2)$, LL) and sometimes also next-to-leading ($\alpha_s^n \ln^{n-1}(E^2/m^2)$, NLL) logarithmic terms to all orders in the cross section above. This is achieved by discarding power suppressed $m/E$ terms, and factoring all the logarithms into a resummation factor, to be obtained via Altarelli-Parisi evolution of an initial condition set at the heavy quark mass scale,

$$\sigma_Q^{res}(E,m) = \sigma_0 C(E,\mu) f(\mu,m) = \sigma_0 C(E,\mu) E(\mu,\mu_0) f(\mu_0,m) \,, \qquad (2)$$

where $\mu$ and $\mu_0$ represent artificial factorization scales, to be taken of order $E$ and $m$ respectively. The 'products' between the various functions actually hide convolution operations with respect to momentum fractions, not explicitly shown as arguments. $C(E,\mu)$ is a perturbatively calculable coefficient function, which does not contain large logarithms thanks to the choice $\mu \simeq E$. The function $f(\mu,m)$ can represent either a parton distribution or a fragmentation function for a heavy quark, and contains the resummation of the collinear logarithms. Due to the large heavy quark mass, its initial condition $f(\mu_0,m)$ can be calculated in perturbation theory [14, 15]: this is the distinctive feature that sets heavy quark parton and fragmentation functions apart from light flavour ones, whose initial conditions are instead entirely non-perturbative and must be fitted to data.

Once a massless but resummed result, valid in the $E \gg m$ region, is obtained, one would like to interpolate it with a fixed order cross section, valid instead in the $E \simeq m$ region, so as to retain predictivity over the whole $E$ range.

The differences between the various approaches are then to be found essentially in two points:

- the perturbative order to which the initial condition $f(\mu_0,m)$ is evaluated, and the perturbative accuracy of the evolution;
- the way the matching with the fixed order calculation is performed.

We summarize below the features of the most commonly used implementations.

### 3.1 ACOT - Aivazis, Collins, Olness, Tung

This approach was the first to try to improve the prediction of the heavy quark structure functions $F_2^c(Q^2, m_c^2)$ and $F_2^b(Q^2, m_b^2)$ at large $Q^2 \gg m_c^2, m_b^2$, by moving potentially large logarithms $\ln(Q^2/m^2)$ into heavy quark parton densities [16, 17]. A general all-order analysis of factorization for the total inclusive $F_2(Q^2)$ in this context was presented in [18].

### 3.2 Simplified ACOT and ACOT($\chi$)

*author: S. Kretzer*

The original ACOT prescription [16, 17] has been simplified in [19] along lines suggested in [20, 18]. In a nutshell, diagrams with initial state heavy quark legs can be treated as if they represented massless quarks. More generally, the diagrams can be manipulated by power suppressed terms provided that higher order diagrams are regularized consistently. ACOT($\chi$) [21, 22] explores this freedom to improve on the threshold behaviour of partonic heavy quark schemes by enforcing the physical pair-production threshold on a term-by-term basis. Heuristically, it comes down to a simple re-scaling of Bjorken-$x$, i.e. in LO

$$F_2^{c\bar{c}} \propto \left. c(\chi) \right|_{\chi = x_{\rm Bj}(1 + 4m^2/Q^2)} \,. \qquad (3)$$

Physical arguments –mostly kinematic– have been given in [21, 23, 22], here we will establish the connection with the FONLL terminology of Section 1.3.3 while focusing on the inclusive DIS process. Much of the following has appeared before, in one form or another, in the literature [16, 17, 18, 24, 25, 26, 27, 28, 19].

We formulate ACOT($\chi$) as an explicit manipulation of resummed terms of the perturbative series. We follow [24] in notation and add an $\mathcal{O}\left(\alpha_s^1\right)$ fixed order (FO) calculation to an all order collinearly resummed (RS) result. In RS heavy quark mass dependence other than logarithmic is neglected. When we remove double-counting terms from FO + RS the zero mass limit (FOM0) of the FO calculation will be required as an auxiliary quantity. Just as in RS, only asymptotic mass logarithms are retained in FOM0. We write therefore, as usual,

$$\sigma^{ACOT}(Q,m) = {\rm FO} + ({\rm RS} - {\rm FOM0}) \times G \qquad (4)$$

where $G$ is an arbitrary operation which behaves like $G = 1 + \mathcal{O}\left(\frac{m^2}{Q^2}\right)$. In [24] $G$ was chosen to be an overall multiplicative factor. More generally, it can be seen as an operation which only modifies, with $\mathcal{O}(m^2/Q^2)$ power-suppressed terms, perturbative coefficients beyond those which have been explicitly calculated, and which are therefore unknown anyway. Any choice for $G$ with this behaviour is therefore legitimate.

To motivate the ACOT($\chi$) choice for $G$ we first re-write more explicitly the three terms given above in the case of inclusive DIS:

$$\text{FO} = \alpha_s\, g\, \tilde{\otimes} H(Q, m) \tag{5}$$

$$\text{FOM0} = \alpha_s \left( g \otimes P_{qg}^{(0)} \ln \frac{\mu^2}{m^2} + g \otimes C_g \right) \tag{6}$$

$$\text{RS} = c(x) + \alpha_s \left( g \otimes C_g + c \otimes C_q \right) \tag{7}$$

where $H(Q,m)$ is the massive coefficient function for the $FO$ gluon fusion process, $C_g$ and $C_q$ are the gluon and light quark coefficient functions (the $\overline{\text{MS}}$ scheme is implied), and $g$ and $c$ are the gluon and charm (i.e. heavy quark) parton distribution functions (both the coefficient functions and the PDFs depend, of course, on the factorization scale $\mu \simeq Q$). $P_{qg}^{(0)}$ is the leading order Altarelli-Parisi splitting vertex. The symbol $\tilde{\otimes} \equiv \int_\chi^1 d\xi/\xi \ldots$ denotes a threshold-respecting convolution integral. One can convince oneself that the standard convolution $\otimes$, with $x \to \chi$ in the lower limit of integration, only differs by $\tilde{\otimes}$ by power-suppressed terms, $\tilde{\otimes} = \otimes + \mathcal{O}(m^2/Q^2)$.

The combined result (4) reads now

$$\begin{aligned}\sigma^{ACOT}(Q,m) &= \text{FO} + (\text{RS} - \text{FOM0}) \times G \\ &= \alpha_s\, g\, \tilde{\otimes} H + \left[ c(x) - \alpha_s\, g \otimes P_{qg}^{(0)} \ln \frac{\mu^2}{m^2} + \alpha_s\, c \otimes C_q \right] \times G,\end{aligned} \tag{8}$$

and we recognize the Krämer-Olness-Soper simplified ACOT framework of [19][1] if we set $G = 1$. Different choices for $G$ can still be made, but natural demands are that:

- In kinematic regions where FO represents the relevant physics (i.e. $Q \sim m$), $G$ should efficiently suppress uncontrolled spurious higher order terms in the square bracket of eq.(8).
- For computational efficiency, the simple $c(x)$ term alone should provide an optimized effective $\mathcal{O}\left(\alpha_s^0\right)$ approximation.

The ACOT($\chi$) scheme implements these requests by making an implicit choice for $G$ which corresponds to writing

$$\begin{aligned}\sigma^{ACOT(\chi)}(Q,m) &= \text{FO} + (\text{RS} - \text{FOM0}) \times G \\ &= \alpha_s\, g\, \tilde{\otimes} H + \left[ c(\chi) - \alpha_s\, g\, \tilde{\otimes} P_{qg}^{(0)} \ln \frac{\mu^2}{m^2} + \alpha_s\, c\, \tilde{\otimes} C_q \right].\end{aligned} \tag{9}$$

Further details on ACOT($\chi$) can be found in [21, 22, 23]. These articles also contain a more intuitive perspective of ACOT($\chi$). Moreover, [22] describes a PDF set that is consistent with ACOT($\chi$) applications.

### 3.3 BSMN - Buza, Smith, Matiounine, van Neerven

In Refs. [29, 30, 31, 32, 33] the treatment of heavy quarks as a parton was fully explored through next-to-next-leading order (NNLO), based on a precise two-loop analysis of the heavy quark structure functions from an operator point of view. This analysis yielded a number of results. One result is important beyond the observable at hand: the authors obtained the complete set of NNLO matching conditions for parton evolution across flavor thresholds. They found that, unlike at NLO, the matching conditions are *discontinuous* at the flavor thresholds. These conditions are necessary for any NNLO calculation at the LHC, and have already been implemented in a number of evolution packages [34, 35].

---

[1] See Eqs. (7), (8) there. General choices for $G$ correspond to the discussion above these equations.

Furthermore, their two-loop calculations explicitly showed that the heavy *quark* structure functions in such a variable flavor approach are not infrared safe: one needs to either define a heavy *quark-jet* structure function, or introduce a fragmentation function to absorb the uncancelled divergence. In either case, a set of contributions to the inclusive light parton structure functions must be included at NNLO.

A dedicated analysis [36] for charm electroproduction showed that even at very large $Q^2$ one could not distinguish the fixed order NLO calculation of [10] and the NNLO VFNS calculations of [31], given the experimental data available in the year 2000. This demonstrates the possibility that the large logarithms $\ln(Q^2/m^2)$ together with small coefficients can in the end have little weight in the overall hadronic cross section.

### 3.4 FONLL - Fixed Order plus Next-to-Leading Log resummation

This approach was developed for improving the large-$p_T$ differential cross section for heavy quark production in hadron-hadron collisions [37]. It was successively extended to photoproduction [38], and in a second phase a matching to the fixed order NLO calculations was performed [24, 39]. The FONLL acronym refers specifically to the matched version.

From the point of view of perturbative logarithms, it contains a NLO-accurate initial condition and full NLL evolution. It therefore reproduces the full NLL structure of the NLO calculation, and resums to all orders the large logarithms with NLL accuracy.

The matching with the fixed order result is performed according to the following master formula (see eq.(16) of [24]):
$$\sigma_Q^{\text{FONLL}}(p_T, m) = \text{FO} + (\text{RS} - \text{FOM0})G(m, p_T) \;, \tag{10}$$
where FO stands for the NLO fixed order massive calculation, FOM0 for its $m/p_T \to 0$ limit (where however $\ln p_T/m$ terms and non-vanishing terms are kept), and RS for the massless, resummed calculation[2]. The RS−FOM0 subtraction is meant to cancel the terms which are present in both RS and FO. This difference starts therefore at order $\alpha_s^2$ with respect to the Born cross section: at large $p_T$ it resums correctly the NLL terms, at small $p_T$ it only contains spurious terms, which are suppressed by the function $G(m, p_T) = p_T^2/(p_T^2 + c^2 m^2)$, with $c = 5$, in order to ensure a physically correct behaviour. The choice of the suppression factor was motivated in [24] by the observation that the massless limit starts to approach the massive hadroproduction calculation at $\mathcal{O}(\alpha_s^3)$ only for $p_T > 5m$. Below this value the massless limit returns unreliable results, and its contribution must therefore be suppressed. It is important to realize that $G(m, p_T)$ only affects terms which are beyond the control of perturbation theory, and therefore it does not spoil the NLO+NLL accuracy. The choice to control such terms by means of an ad-hoc function might seem a somewhat unpleasant characteristic of this approach. However, it simply portrays the freedom one has in performing the matching, and does not represent a shortcoming of the approach: different matching procedures will simply make other implicit or explicit choices for $G(m, p_T)$.

For the sake of making comparisons with other approaches easier, the formula (10) can be rewritten with some more details as follows:
$$\sigma_Q^{\text{FONLL}}(p_T, m) = \sum_{ij \in \mathcal{L}} F_i F_j \, \sigma_{ij \to QX}(p_T, m)$$
$$+ \left( \sum_{ijk \in \mathcal{L}+\mathcal{H}} F_i F_j \, \hat{\sigma}_{ij \to kX}^{\overline{MS}}(p_T) D_{k \to Q} - \sum_{ij \in \mathcal{L}} F_i F_j \, \sigma_{ij \to QX}(p_T, m; m \to 0) \right) G(m, p_T) \;. \tag{11}$$

A number of ingredients needing definition have been introduced. The kernel cross sections $\sigma_{ij \to QX}(p_T, m)$ are the massive NLO calculations for heavy quark production of Refs. [2, 3, 4, 5, 6]. When convoluted with the PDFs for light flavours $F_i$ ($i \in \mathcal{L}$) they yield the FO term in eq. (10). The $\sigma_{ij \to QX}(p_T, m; m \to 0)$ terms represent the $m \to 0$ limit of the massive NLO cross sections, performed by sending to zero $m/p_T$ terms while preserving $\ln(p_T/m)$ contributions and non-vanishing constants. When convoluted with light flavour

---

[2]This term might also be referred to as a 'zero-mass variable flavour number scheme' (ZM-VFNS) contribution. However this name, while by itself completely general, has been used in the past for specific approaches with different overall perturbative accuracies. We shall therefore avoid its use. It will be understood that 'RS' in this approach has full NLL accuracy.

PDFs they give FOM0. Finally, $\hat{\sigma}^{\overline{MS}}_{ij \to kX}(p_T)$ are the massless $\overline{MS}$-subtracted NLO cross section kernels given in [40]. In addition to the light flavour PDFs, they are also convoluted with the perturbatively-calculated parton distribution functions for the heavy quarks ($i \in \mathcal{H}$) and with the fragmentation functions describing the transformation of a parton into a heavy quark, $D_{k \to Q}$ [15], to give the term RS.

The formula given above returns the differential cross section for heavy *quark* production, evaluated with NLO + NLL accuracy. In order to obtain the corresponding cross section for an observable heavy meson it must still be convoluted with the proper scale-independent non-perturbative fragmentation function, extracted from experimental data, describing the heavy quark → heavy hadron transition:

$$\sigma_H^{\text{FONLL}}(p_T, m) = \sigma_Q^{\text{FONLL}}(p_T, m) D_{Q \to H}^{NP} . \quad (12)$$

Phenomenological analyses of charm- and bottom-flavoured hadrons production within the FONLL approach have been given in [41, 42, 43, 44, 45].

### 3.5 GM-VFNS - General mass variable flavour number scheme

This approach also combines a massless resummed calculation with a massive fixed order one, for predicting $p_T$ distributions in hadron-hadron collisions. One difference with respect to FONLL is that this approach does not include the perturbative NLO parton-to-heavy-quark fragmentation functions $D_{k \to Q}$. Rather, it directly convolutes a properly $\overline{MS}$ subtracted cross section (with mass terms also included, hence the 'general mass' name) with non-perturbative fragmentation functions for heavy mesons $D_{Q \to H}^{NP, \overline{MS}}$, fitted at LEP in a pure $\overline{MS}$ scheme. The cross section can be schematically written as

$$\sigma_H^{\text{GM-VFNS}}(p_T, m) = \sum_{ij \in \mathcal{L}} F_i F_j \, \hat{\sigma}_{ij \to QX}(p_T, m) D_{Q \to H}^{NP, \overline{MS}} + \sum_{ijk \in \mathcal{L}+\mathcal{H}} F_i F_j \, \hat{\sigma}^{\overline{MS}}_{ij \to kX}(p_T) D_{k \to H}^{NP, \overline{MS}} , \quad (13)$$

where the 'massive-but-subtracted' cross section kernels $\hat{\sigma}_{ij \to QX}(p_T, m)$ are defined by

$$\hat{\sigma}_{ij \to QX}(p_T, m) \equiv \sigma_{ij \to Q}(p_T, m) - \sigma_{ij \to QX}(p_T, m; m \to 0) + \hat{\sigma}^{\overline{MS}}_{ij \to QX}(p_T) . \quad (14)$$

The new kernels $\hat{\sigma}_{ij \to QX}(p_T, m)$ defined by this operation (of the form FO-FOM0+RS) can be convoluted with an evolved $\overline{MS}$-subtracted fragmentation function, but they also retain power suppressed $m/p_T$ terms. It should also be noted that the sum in the second term of (13) only runs over contributions not already included in the first.

Recalling the way the perturbative parton-to-heavy-quark $D_{k \to Q}$ fragmentation functions are defined in [15], setting

$$D_{k \to H}^{NP, \overline{MS}} = D_{k \to Q} D_{Q \to H}^{NP}, \qquad k \in \mathcal{L} + \mathcal{H} , \quad (15)$$

and comparing eqs.(13) and (11), it can be seen that the GM-VFNS master formula is a reshuffling of the FONLL one, up to higher-orders terms.

Two comments are worth making. The first is that due to the absence of the perturbative $D_{k \to Q}$ terms, eq. (13) cannot reproduce the NLO heavy *quark* production cross section: even the normalization must be extracted from the experimental data. Eq. (11), on the other hand, can reproduce the heavy quark spectrum, and only the heavy quark → heavy meson transition is fitted to data. The second remark concerns the higher order power suppressed terms: since GM-VNFS implicitly makes a different choice for the $G(m, p_T)$ function, the results from the two approaches might differ considerably in the $p_T \sim m$ region since, while formally suppressed, such terms can be numerically important.

An example of a phenomenological application of the GM-VFNS scheme is given below.

### 3.6 Hadroproduction of heavy mesons in a massive VFNS

*authors: B.A. Kniehl and I. Schienbein*

Various approaches for next-to-leading-order (NLO) calculations in perturbative QCD have been applied to one-particle-inclusive hadroproduction of heavy mesons. The general-mass variable-flavor-number scheme (GM-VFNS) devised by us in Ref. [46, 47] is closely related to the conventional massless variable-flavor-number scheme (ZM-VFNS), but keeps all $m^2/p_T^2$ terms in the hard-scattering cross sections, where $m$ is the mass of the heavy quark and $p_T$ the transverse momentum of the observed meson, in order to achieve better accuracy in the intermediate region $p_T \geq m$. The massive hard-scattering cross sections have been constructed in such a way that the conventional hard-scattering cross sections in the $\overline{\text{MS}}$ scheme are recovered in the limit $p_T \to \infty$ (or $m \to 0$). The requirement to adjust the massive theory to the ZM-VFNS with $\overline{\text{MS}}$ subtraction is necessary, since all commonly used PDFs and FFs for heavy flavors are defined in this particular scheme. In this sense, this subtraction scheme is a consistent extension of the conventional ZM-VFNS for including charm-quark mass effects. It should be noted that our implementation of a GM-VFNS is similar to the ACOT scheme [16, 17], which has been extended to one-particle-inclusive production of $B$ mesons a few years ago [48]. As explained in the second paper of Ref. [46, 47], there are small differences concerning the collinear subtraction terms. Furthermore, in Ref. [48], the resummation of the final-state collinear logarithms has been performed only to leading logarithmic accuracy. The field-theoretical foundation of a GM-VFNS has been provided a few years ago by a factorization proof including heavy-quark masses [18]. Therefore, it is possible to extract improved universal parton distribution functions (PDFs) [22] and fragmentation functions (FFs) [49] from fits employing massive hard-scattering cross sections. From this perspective, it is important to compute massive hard-scattering cross sections in a given massive scheme for all relevant processes. Explicit calculations in the original ACOT scheme have been performed in Ref. [50, 51] for inclusive and semi-inclusive deep-inelastic scattering (DIS). Furthermore, our calculation in Ref. [46, 47] for hadronic collisions completes earlier work in the GM-VFNS on $D$-meson production in $\gamma\gamma$ and $\gamma p$ collisions [52, 53, 54], and it is planned to extend our analysis to the case of heavy-meson production in DIS.

Next, we show predictions for the cross section $d\sigma/dp_T$ of $D^\star$-meson production obtained in the GM-VFNS and the ZM-VFNS. The cross section has been scaled with $p_T^5$ in order to arrive at a flat $p_T$ distribution, which is useful for visualizing the heavy-quark mass effects. The hard-scattering cross sections are convoluted with the (anti-)proton PDFs and FFs for the transition of the final-state parton into the observed $D^\star$ meson. We use the CTEQ6M PDFs [55] and the FFs for $D^\star$ mesons from Ref. [56]. As in the experimental analysis, the theoretical results are presented for the average $(D^{\star+} + D^{\star-})/2$. We consider $d\sigma/dp_T$ at $\sqrt{S} = 1.96$ TeV as a function of $p_T$ with $y$ integrated over the range $-1.0 < y < 1.0$. We take the charm mass to be $m = 1.5$ GeV and evaluate $\alpha_s^{(n_f)}(\mu_R)$ with $n_f = 4$ and scale parameter $\Lambda_{\overline{\text{MS}}}^{(4)} = 328$ MeV, corresponding to $\alpha_s^{(5)}(m_Z) = 0.1181$. The results are presented in Fig. 1 for the GM-VFNS (black lines) and the ZM-VFNS (red lines) in comparison with CDF data [57]. The solid lines have been obtained with $\mu_R = \mu_F = \mu'_F = m_T$. The upper and lower dashed curves represent the maximum and minimum cross sections found by varying $\mu_R$, $\mu_F$, and $\mu'_F$ independently within a factor of 2 up and down relative to the central values requiring for their ratios to satisfy the inequalities $0.5 \leq \mu_F/\mu_R, \mu'_F/\mu_R, \mu_F/\mu'_F \leq 2$. As can be seen, for large values of $p_T$, the predictions of the GM-VFNS nicely converge to the corresponding results in the ZM-VFNS. Both approaches lead to reasonable descriptions of the data, but the inclusion of the positive mass effects clearly improves the agreement with the data. It should be noted that the mass effects are largest for the upper curves of the uncertainty band, which have been obtained with the smaller value of the renormalization scale implying a larger $\alpha_s(\mu_R)$. At $p_T = 5$ GeV, one observes an increase of the massless cross section by about 35%. A more detailed comparison of the GM-VFNS with CDF data [57] including $D^0$, $D^+$, and $D_s^+$ mesons can be found in Refs. [58, 59].

Residual sources of theoretical uncertainty include the variations of the charm mass and the employed PDF and FF sets. A variation of the value of the charm mass does not contribute much to the theoretical uncertainty. Also, the use of other up-to-date NLO proton PDF sets produces only minor differences. Concerning the choice of the NLO FF sets, we obtain results reduced by a factor of 1.2–1.3 when we use the NLO sets from Ref. [60], which is mainly caused by a considerably different gluon FF. A more detailed discussion can be found in Ref. [56].

## 4   A case study in collinear resummation: $b$-quark PDF from $Z + b$ production at LHC

*authors: S. Diglio, F. Maltoni, F. Petrucci, A. Tonazzo and M. Verducci*

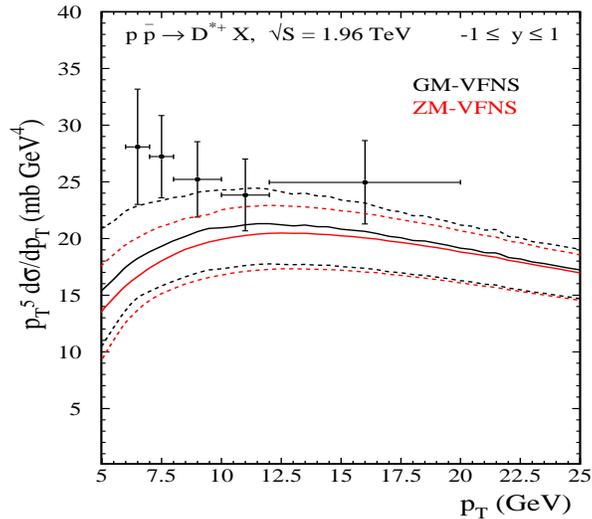

Fig. 1: QCD predictions for one-particle-inclusive production of $D^\star$ mesons at the Tevatron Run II in comparison with CDF data [57]. The results are shown for the average $D^\star = (D^{\star +} + D^{\star -})/2$. The solid lines have been obtained with $\mu_R = \mu_F = \mu'_F = m_T$. The upper and lower dashed curves represent the maximum and minimum cross sections found by varying $\mu_R$, $\mu_F$, and $\mu'_F$ independently within a factor of 2 up and down relative to the central values while keeping their ratios in the range $0.5 \leq \mu_F/\mu_R, \mu'_F/\mu_R, \mu_F/\mu'_F \leq 2$.

## 4.1 Introduction

The discovery of new physics at LHC will probably rely on the detailed understanding of standard-model background processes. Outstanding among these is the production of weak bosons ($W$, $Z$) in association with jets, one or more of which contains a heavy quark ($Q = c, b$). The prime example is the discovery of the top quark at the Fermilab Tevatron, which required a thorough understanding of the $W$+jets background, with one or more heavy-quark jets. The discovery of single-top-quark production via the weak interaction will require an even more sophisticated understanding of this background [61, 62].

For many processes involving production of heavy quarks, there are two ways (schemes) to perform the calculation in QCD: the fixed-flavor-scheme (FFS) and variable-flavor-scheme (VFS). The main practical difference between the two approaches is simple: in the VFS the heavy-quark can also be in the initial state, and in that case is assumed to be massless, while in the FFS it appears only as a final state (massive) particle. QCD factorisation tells us that if calculations could be performed at arbitrary high order, the two schemes would be equivalent. At fixed order, on the other hand, differences arise and one should choose that describing more effectively the kinematics of the process of interest. This freedom has sometimes created intense and fruitful debates among the QCD practitioners (see, *e.g.*, Ref. [63] for a detailed comparison of Higgs boson production in association with bottom quarks). Here we just recall the main two reasons for using a heavy-quark distribution function. First, it resums collinear logarithms of the form $\ln Q/m_Q$ to all orders, where $Q$ is the scale of the hard scattering and $m_Q$ is the mass of the heavy quark. Second, it simplifies the leading-order process, which often makes a higher-order calculation feasible. There are many processes in the standard model and in models beyond it, such as SUSY, that are better described using a bottom in the initial state. In Table 1, we give a non-exhaustive list of processes that will be relevant for QCD, EW and SUSY studies at the LHC, and the QCD order at which they are known.

At present the $b$ distribution function is derived perturbatively from the gluon distribution function [17, 18, 55, 34]. Recently, direct, albeit not very precise, measurements of $F_2^b$ have become available that are compatible with the perturbative determination [64, 65]. In the light of its phenomenological importance, a better direct determination of the $b$ distribution function is certainly desirable.

To this aim it has been proposed to use the associated production of a photon and a $b$-jet via $gb \to \gamma b$ at the LHC [66]. This measurement suffers from two main limitations. The first is the large contamination from charm which has a much larger cross section due to both the pdf and the electromagnetic coupling. The second is that the theoretical prediction at NLO for an isolated photon is uncertain, due to necessity of introducing a photon fragmentation function, which is at present poorly known.

| Name | LO Process | Interest | Accuracy |
|---|---|---|---|
| single-top t-channel | $qb \to qt$ | top EW couplings | NLO |
| single-top tW-associated | $gb \to tW^-$ | Higgs bckg, top EW couplings | NLO |
| Vector boson + 1 b-jet | $gb \to (\gamma, Z)b$ | b-pdf, SUSY Higgs benchmark | NLO |
| Vector boson + 1 b-jet +1 jet | $qb \to (\gamma, Z, W)bq$ | single-top and Higgs bckgs | NLO |
| Higgs inclusive | $b\bar{b} \to (h, H, A)$ | SUSY Higgs discovery at large $\tan\beta$ | NNLO |
| Higgs + 1 b-jet | $gb \to (h, H, A)b$ | SUSY Higgs discovery at large $\tan\beta$ | NLO |
| Charged Higgs | $gb \to tH^-$ | SUSY Higgs discovery | NLO |

Table 1: Process relevant for SM measurements and SUSY discoveries at the LHC which entail the use of bottom in the initial state. All of them are known at least at NLO accuracy.

In this note we follow the suggestion of Ref. [67] to use $Z$ production in association with a $b$-jet to extract information on the $b$-pdf. At leading order, it proceeds via $gb \to Zb$, as shown in Fig. 2. This process is known at NLO, including $\gamma/Z$ interference effects. The advantages of using a $\gamma/Z$ decaying into leptons with respect to a real photon are noticeable. The NLO cross section is theoretically very well known and, apart from the PDF's, free of non-perturbative inputs. In addition, the competing process $gc \to Zc$ is suppressed by the ratio of the couplings of the charm and the bottom to the $Z$, and makes the $b$-pdf determination much cleaner.

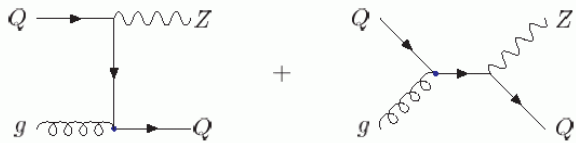

Fig. 2: Leading Order Feynman diagrams for associated production of a $Z$ boson and a single high-$p_T$ heavy quark ($Q = c, b$).

The D0 Experiment at Tevatron has recently measured the cross-section ratio $\sigma(Z+b)/\sigma(Z+jet)$ [68], and their result is consistent with the NLO calculation.

As pointed out in [67], the measurement of this process at the LHC should be even more interesting because the contribution of the leading order process, sensitive to the $b$ content of the proton, is more relevant than at the Tevatron. In addition, the total cross-section is larger by a factor 50, and the relative contribution of background processes, mainly $Z+c$, is smaller. These features are summarised in table 2, taken from Ref. [67].

Besides the possibility of extracting the $b$-pdf, $Z + b$ represents also a benchmark and in some cases a background to the search of the Higgs boson, when it is produced in association with a single high-$p_T$ $b$ quark [63]: the dominant leading-order subprocess for the production of a Higgs boson via its coupling to the $b$ is $b\bar{b} \to h$; however, if the presence of a single $b$ with high $p_T$ is demanded, the dominant process becomes $gb \to hb$, with cross-sections of the order of tens of fb. The $h$ can then decay to the same final states as the $Z$; in particular, the decay $h \to \mu^+\mu^-$ is enhanced in some models [69, 70, 71].

A preliminary analysis on the potential of the ATLAS experiment to measure the $Z+b$-jet production at the LHC is presented in the following.

| Cross sections (pb) | Tevatron | LHC |
|---|---|---|
| Process | $ZQ$ inclusive | |
| $gb \to Zb$ | $13.4 \pm 0.9 \pm 0.8 \pm 0.8$ | $1040^{+70+70\ +30}_{-60-100-50}$ |
| $gb \to Zb\bar{b}$ | 6.83 | 49.2 |
| $gc \to Zc$ | $20.3^{+1.8}_{-1.5} \pm 0.1^{+1.3}_{-1.2}$ | $1390 \pm 100^{+60+40}_{-70-80}$ |
| $gc \to Zc\bar{c}$ | 13.8 | 89.7 |
| | $Zj$ inclusive | |
| $q\bar{q} \to Zg, gq \to Zq$ | $1010^{+44+9+7}_{-40-2-12}$ | $15870^{+900+60\ +300}_{-600-300-500}$ |

Table 2: Next-to-leading-order inclusive cross sections (pb) for $Z$-boson production in association with heavy-quark jets at the Tevatron ($\sqrt{s}$ = 1.96 TeV $p\bar{p}$) and the LHC ($\sqrt{s}$ = 14 TeV $pp$). A jet lies in the range $p_T > 15$ GeV/c and $|\eta| < 2$ (Tevatron) or $|\eta| < 2.5$ (LHC). $ZQ$ indicates events containing a heavy quark, $Zj$ events which do not contain a heavy quark.

## 4.2 A study of LHC measurement potential

A sample of $Z$+jet events generated using the PYTHIA Monte Carlo [72] was processed with a fast simulation of the ATLAS detector, the ATLFAST package [73]. Only decays of the $Z$ boson to $\mu^+\mu^-$ were taken into account. The signal was defined as the sample events containing a $b$ quark with $p_T > 15$ GeV/c and $|\eta| < 2.5$. The background samples containing respectively a $c$ quark within the same cuts, or a jet originating from a light quark or a gluon in the same range, were considered separately. The NLO cross-sections computed in [67] were used for the signal and for these two classes of background, while the cross-section given by PYTHIA was taken for the other types of events.

The experimental selection of $Z$+jet events with $Z \to \mu^+\mu^-$ required the detection of two muons of opposite charge with $p_T > 20$ GeV/c and $|\eta| < 2.5$ and one hadronic jet. The presence of two high-$p_T$ muons ensures the possibility to have high trigger efficiency on this type of events. In addition, to reject the contribution from virtual photons, the invariant mass $M_{\mu\mu}$ of the muon pair was required to be close to the $Z$ mass (80 GeV/c$^2$ < $M_{\mu\mu}$ < 105 GeV/c$^2$). About 50% of signal events are retained after applying these cuts, the loss being equally due to the $\eta$ acceptance and to the $p_T$ cut.

The selection of events where the jet originates from a $b$ quark was based on two different tagging methods, as described in the following. Their complementarity is still to be studied in detail, however the comparison of two independent selections will be important to control the systematic uncertainties.

The first method to select $Z + b$ events was based solely on the presence of a third muon. Hadrons containing a $b$ quark give origin to prompt muon decays in about 12% of the cases. The efficiency of this method, therefore, cannot exceed this value, however the background is also expected to be small. The "third muon", considered to be the muon from the $b$ hadron decay, will in general be softer and closer to a jet than the muons from the $Z$ decay. The distribution of the transverse momentum of the third muon in $Z + j$ events is shown in Fig. 3. Different thresholds on the third muon $p_T$ were considered for the final selection.

The second analysis used an inclusive method for $b$-tagging, based on the presence of secondary vertices and of tracks with high impact parameter with respect to the primary vertex, originated from the decay of the long-lived $b$ hadrons. The ATLFAST package reproduces the ATLAS $b$-tagging capabilities by applying the tagging efficiency on $b$ jets and a mis-tag rate on non-$b$ jets on a statistical basis, according to the values set by the user to reproduce the actual detector performance. The efficiency of the inclusive $b$-tagging on signal events, after the selection described above, is about 30%. The mistagging probability is about 4% on $c$-quark jets and 0.5% on light jets.

The overall efficiency on signal events, the expected number of signal and background events with an integrated luminosity of 10 fb$^{-1}$ and the expected purity of the selected samples are reported in table 3. With the fast simulation, the soft muon tagging capabilities are optimistic, in that full efficiency and no mis-tag are assumed for the lepton identification; more realistic assumptions will be made when the study is carried on with the full detector simulation. The efficiency on signal events achieved with the inclusive $b$-tagging method, where the results of the fast simulation are more realistic, is higher than with the soft muon tagging, while the purity of the selected sample is still quite good. Consistent results were obtained with a full simulation of the ATLAS experiment, on a small statistics sample.

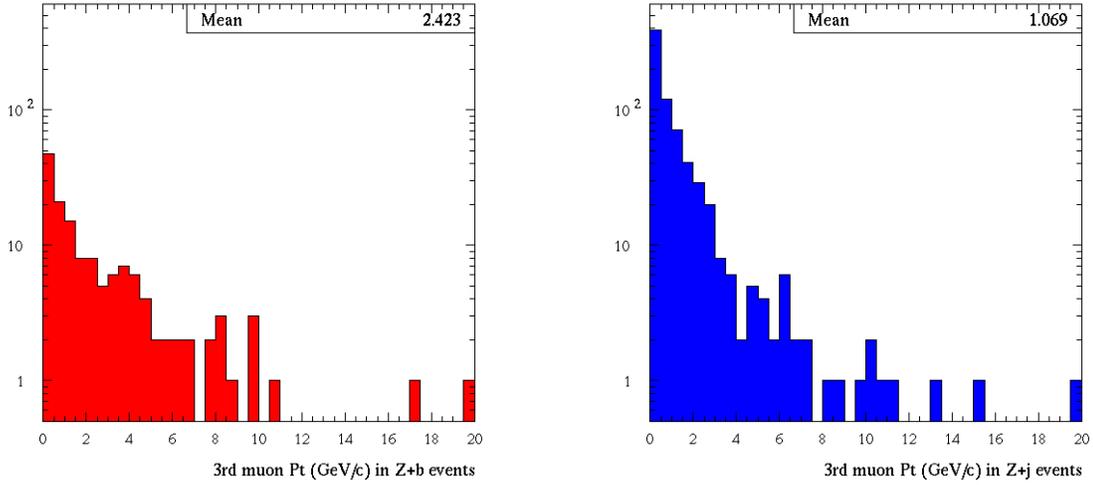

Fig. 3: Distribution of the transverse momentum of the third muon in a $Z$+jet sample, for signal events (left) and for events with no $b$ quark (right).

A better determination of the signal component in the selected sample will eventually be achieved by exploiting the information on the transverse momentum of the $b$-jet or of the third muon.

| Cut | Efficiency | $N_b^{p_T>15 \text{ GeV}, |\eta|\leq 2.5}$ | $N_c^{p_T>15 \text{ GeV}, |\eta|\leq 2.5}$ | $N_{other}$ | Purity |
|---|---|---|---|---|---|
| $p_T^\mu > 4$ GeV/c | 4% | 13990 | 6270 | 0 | 69% |
| $p_T^\mu > 5$ GeV/c | 3% | 11090 | 5210 | 0 | 69% |
| $p_T^\mu > 6$ GeV/c | 2.5% | 8430 | 4180 | 0 | 67% |
| incl. $b$-tag | 14% | 49500 | 17400 | 49600 | 43% |

Table 3: Expected efficiency, statistics and purity in a data sample corresponding to an integrated luminosity of 10 fb$^{-1}$, using the soft muon tagging with different thresholds on the muon transverse momentum and the inclusive $b$-tagging. $N_b$ denotes the number of expected signal events as defined in the text, $N_c$ the number of selected events with a $c$ jet with $p_T > 15$ GeV and $|\eta| < 2.5$, $N_{other}$ the selected events from other processes. The statistical error on efficiencies and purities, due to the limited size of the simulated sample, is at the level of 1-2%.

Given the large statistics of the available data samples, the measurement will be limited by systematic effects.

The possibility to control the systematic effects directly from data samples has been explored, in particular the evaluation of $b$-tagging performance and of the residual background.

The $b$-tagging efficiency can be checked using $b$-enriched samples. Based on previous experience at Tevatron and LEP, we can expect a relative uncertainty of about 5%.

The background in the selected sample is mainly due to mis-tagged jets from $c$ and light quarks. This can be controlled by looking at the number of $b$-tagged jets in data samples that in principle should contain no $b$-jets at first order. $W$+jet events, for example, will be available with large statistics and with jets covering the full $p_T$ range of the signal. It can therefore be expected to estimate the background from mis-tagging with a relative uncertainty at the level of few percent, as shown by the plots in figure 4.

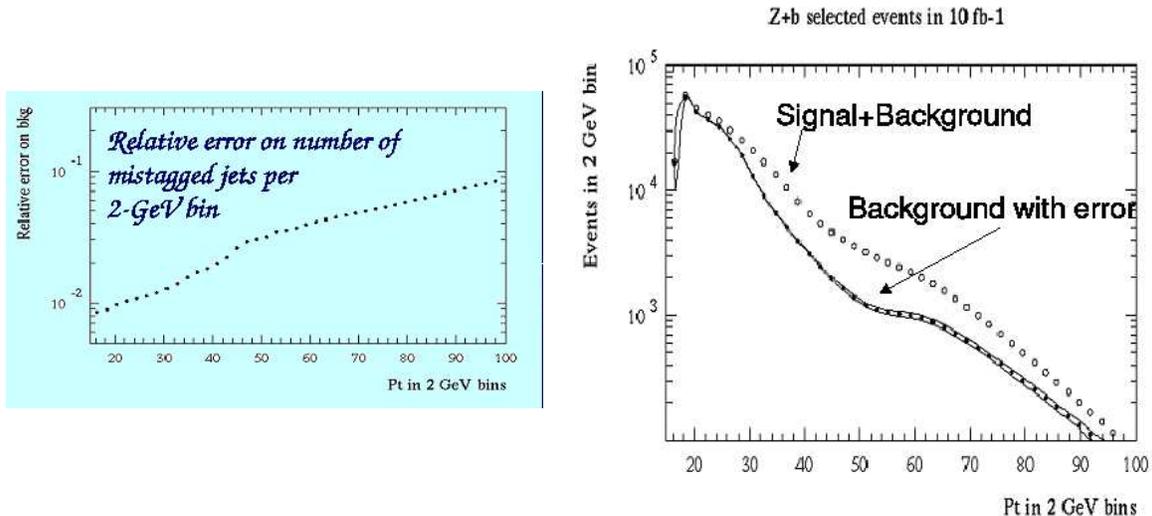

Fig. 4: Systematics due to mis-tagging of $b$-jets as evaluated from $W$+jet events. Left: relative error on background level per jet $p_T$ bin. Right: $p_T$ distribution of jets in event selected as $Z + b$; the error band on the background contribution represents the systematic uncertainty, as derived from the previous plot.

### 4.3 Conclusions and outlook

$Z$ boson production in association with a $b$-jet can provide information on the $b$-pdf.

A preliminary study of the $Z + b$ channel using a fast simulation of the ATLAS detector has shown that this type of event will be observed with very high statistics and good purity at the LHC. Given the large statistics of the samples, the precision of the $Z + b$ cross-section measurement will be limited by systematic effects. Some possibilities to evaluate systematic uncertainties directly form the data have been considered. An overall accuracy on the measurement at the level of 5% can be expected.

The availability of large samples opens interesting possibilities for the study of differential distributions: for instance, measuring the cross-section as a function of the $\eta$ and $p_T$ of the $Z$ boson would allow for the measurement of the $b$ PDF as a function of the momentum fraction carried by the quark inside the proton. These items are an additional topic for further studies.

## 5 Soft-gluon resummation

*authors: T.O. Eynck and E. Laenen*

QCD factorizes long- and short distance dynamics in inclusive cross sections with initial state hadrons into non-perturbative, but universal parton distribution functions, and perturbatively calculable hard scattering functions. Large remnants of the long-distance dynamics occur near the threshold edge of phase space in the form of logarithmic distributions that are singular at the edge. Resummation [74, 75] of these effects organizes them to all orders in perturbation theory, and thereby extends the predictive power of QCD.

Threshold resummation is now a well-established calculational scheme with systematically improvable accuracy. It allows organization of all subleading powers of the logarithmic enhancements, and can be consistently matched to finite order perturbation theory. Resummed expressions, which take the form of exponentiated integrals over functions of the running coupling, require however a prescription for their numerical evaluation to handle a Landau pole singularity of the coupling. But for this intrinsic ambiguity (which must cancel against ambiguities in power corrections), threshold resummation is just as systematically improvable as the standard coupling constant expansion.

As stated earlier, the more differential a cross section, the better suited it is for phenomenology, because one may incorporate detector-specific acceptance cuts and thereby reduce the need for extrapolation. Therefore we should like to better understand the behavior of threshold-resummed expressions for double-differential cross sections. A study for the inclusive threshold-resummed heavy quark structure function can be found in Ref. [76]. Here we examine the differential structure function for the reaction

$$\gamma^*(q) + P(p) \to Q(p_1) + X'(p_2') \tag{16}$$

which we write as

$$\frac{d^2 F_2^Q(S, T_1, U_1)}{dT_1 \, dU_1} \tag{17}$$

We define the invariants

$$\begin{aligned} S &= (p+q)^2 \equiv S' - Q^2, & T_1 &= (p-p_1)^2 - m^2, \\ U_1 &= (q-p_1)^2 - m^2, & S_4 &= S' + T_1 + U_1 \,. \end{aligned} \tag{18}$$

The invariant mass squared of the final state $X'$ is given by

$$M_{X'}^2 = m^2 + S_4 \tag{19}$$

so that the elastic (threshold) limit for the subprocess (16) is approached by $S_4 \to 0$. It may be converted to the double-differential structure function in terms of the heavy quark transverse momentum and rapidity, e.g.

$$\frac{d^2 F_k^Q}{d(p_T^Q)^2 \, dy^Q} = S' \frac{d^2 F_k^Q}{dT_1 \, dU_1}, \tag{20}$$

where e.g. [11]

$$p_T^Q = \left[ \frac{S' T_1 U_1 + Q^2 T_1^2 + Q^2 S' T_1}{S'^2} - m^2 \right]^{(1/2)}. \tag{21}$$

At the parton level one may define invariants equivalent to those in (18), which we will denote by using lower case. The order-by-order perturbation theory expansion for the partonic version of this distribution $\omega(s_4, t_1, u_1)$ and its all-order resummation have the following schematic forms

$$\begin{aligned} \omega &= 1 + \alpha_s (L^2 + L + 1) + \alpha_s^2 (L^4 + L^3 + L^2 + L + 1) + \ldots \\ &= \exp\left( \underbrace{\underbrace{L g_1(\alpha_s L)}_{LL} + g_2(\alpha_s L) + \ldots}_{NLL} \right) \underbrace{C(\alpha_s)}_{\text{constants}} \\ &\quad + \text{suppressed terms} \end{aligned} \tag{22}$$

with

$$g_1(\lambda) = \frac{C_F}{\pi b_0 \lambda} \left[ \lambda + (1-\lambda) \ln(1-\lambda) \right], \quad \lambda = b_0 \alpha_s \ln N \,. \tag{23}$$

(We have also computed $g_2(\lambda)$; by including ever more $g_i$ functions in the exponent in Eq. (22) we can increase the parametric accuracy of the resummation.) The symbol $L^i$ represents, in this case, the logarithmically singular plus-distributions

$$\left[ \frac{\ln^{i-1}(\rho)}{\rho} \right]_+ \tag{24}$$

with $\rho = s_4/m^2$, or, after a Laplace transform $\int d\rho \exp(-N\rho)$ by $\ln^i N$. The conversion to momentum space then requires a numerical inverse Laplace transform. For the case at hand one needs to compute

$$S'^2 \frac{d^2 F_2^Q(S_4, T_1, U_1)}{dT_1 \, dU_1} = \int_{c-i\infty}^{c+i\infty} \frac{dN}{2\pi i} e^{NS_4/m^2} \bar{\phi}_g\left( N \frac{S' + T_1}{m^2} \right) \omega(N, T_1, U_1), \tag{25}$$

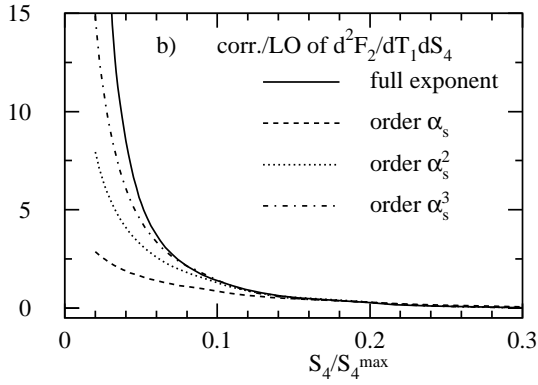

Fig. 5: Expandability of the resummed expressions for $d^2 F_2^c/dT_1 dS_4$ with NLL exponent (ratio to LO)

with $c$ the intercept of the contour with the real $N$ axis, and $\phi_g(N)$ the gluon density in moment space. We chose a toy density for the gluon PDF, and the minimal prescription [77] to perform the $N$ integral. In Fig. 5 we evaluate this expansion as a function of the recoil mass $S_4$, and compare it to its finite order expansions. We keep the variable $T_1$ fixed at the average of its minimum and maximum allowed value. Clearly, for reasonable values of $S_4$ the resummed result is already well-approximated by its 2nd and 3rd order expansions.

Another way to evaluate the resummed expression is in terms of towers of [78] $L = \ln N$.

$$\omega = h_{00}(\alpha_s) \left[ 1 + \sum_{k=1}^{\infty} \left(\frac{\alpha_s}{\pi}\right)^k \left( c_{k1} L^{2k} + c_{k2} L^{2k-1} + c_{k3} L^{2k-2} + \ldots \right) \right]. \qquad (26)$$

where the indicated coefficients $c_{kj}$ can be determined exactly. More accuracy here means including more subleading towers. This method is equivalent, but not identical to the minimal prescription method. In practice, one need only include the first 4 terms in each tower, the higher terms are vanishingly small. The ambiguities mentioned earlier are shifted to far-subleading towers in this approach. To exhibit the convergence of terms in the towers, it will be useful and illustrative to exhibit contributions of a particular order in the strong coupling and the large logarithms. We will employ the notation

$$\mathrm{N}^k\mathrm{LO} - l\mathrm{L}_N \qquad (27)$$

for finite order results, where $k$ indicates the order in the strong coupling, the subscript $N$ denotes moments, and $l$ expresses if only the leading term ($l = 1$, $\mathrm{L}^{2k}_N$), or also the next-to-leading term ($l = 2$, $\mathrm{L}^{2k-1}_N$) is included, etc. In Fig. 6 we see also in this approach a rapid convergence toward the tower-resummed result.

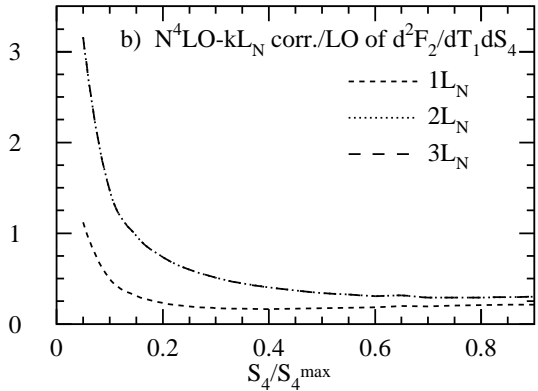

Fig. 6: Tower resummation at $\mathrm{N}^4\mathrm{LO} - k\mathrm{L}_N$, $k \in \{1, 2, 3\}$ ($\mathrm{N}^4\mathrm{LO} - 2\mathrm{L}_N$ and $\mathrm{N}^4\mathrm{LO} - 3\mathrm{L}_N$ almost coincide).

A more complete study of the relevance of threshold resummation for electroproduction of heavy quarks at HERA still awaits. We note that even if the size of the corrections does not cause much concern for the perturbative analysis of an observable, threshold resummation or its finite order approximations, often lead to a reduction of scale dependence [79], indeed also seen in Ref. [76].

# 6   $k_t$ - factorization

*authors: S.P. Baranov, H. Jung, A.V. Lipatov and N.P. Zotov*

## 6.1   Introduction

The transverse momenta of the partons initiating a hard scattering process, like heavy quark production via $\gamma g \to Q\bar{Q}$ or $gg \to Q\bar{Q}$ in lepto- (hadro-) production, respectively, is mainly generated by the QCD evolution, which can reach large values, in DGLAP up to the factorization scale, in BFKL/CCFM/LDC even larger.

The typical transverse momenta of the gluons in a process $gg \to X$ for different masses $M$ of the system $X$ are shown in Fig. 7 as a function of the momentum fraction $x$ of one of the gluons for LHC energies. The transverse momenta can become large, so that they cannot be neglected. A theoretical approach, formulated for small $x$, which takes into account the tranverse momenta is the $k_t$-factorization [80, 81] or semi-hard [82] approach.

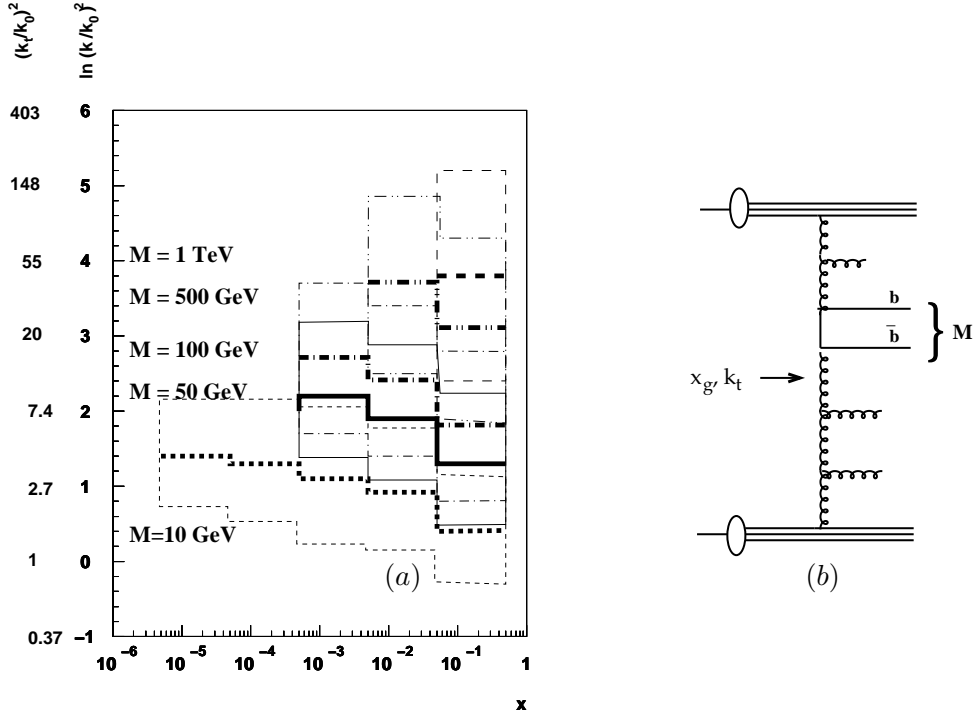

Fig. 7: Average tranverse gluon momenta $k_t$ in processes $gg \to X$ for different masses $M$ of the system $X$ as a function of the momentum fraction of one of the gluons $x$. The thin lines indicate the RMS spread of the distributions. In (b) is shown the definition of $x, k_t$ and $M$ for a gluon induced process.

In $k_t$-factorization the cross section for any process $pp \to X$ can be written as:

$$\sigma = \int dx_1 dx_2 \int dk_{t\,1} dk_{t\,2} \mathcal{A}(x_1, k_{t\,1}, q) \mathcal{A}(x_2, k_{t\,2}, q) \hat{\sigma}(x_1, x_2, k_{t\,1} k_{t\,2}, q) \qquad (28)$$

with $\mathcal{A}(x, k_t, q)$ being the un-integrated ($k_t$-dependent) parton density function uPDF, $q$ defines the factorization scale and $\hat{\sigma}$ is the partonic cross section. The off-shell matrix-elements $\hat{\sigma}$ are calculated in [80, 81].

The effects of finite transverse momenta are present independent of the evolution scheme: uPDFs can be defined also for the DGLAP evolution. A more detailed discussion on these effects can be found in [83, 84].

It is interesting to note, that the $k_t$-factorization approach (in LO $\alpha_s$) agrees very well with calculations performed in the collinear approach in NLO $\alpha_s$, which is shown in [85]. The main effect comes from a more realistic treatment of the kinematics already in LO, which otherwise has to be covered in NLO. The $k_t$ factorization approach, however, is strictly valid only at small $x$, where the virtuality of the exchanged gluons can be identified with its tranverse momentum $k^2 \sim -k_t^2$. The full expression for the virtuality is [86]:

$$k^2 = \frac{-k_t^2}{1-x} - \frac{x \cdot m^2}{1-x} \qquad (29)$$

with $m$ being the recoiling mass of the hadronic system except the hard scattering process, taking into account the history of the evolution process. For finite $x$ the mass effects can be substantial.

## 6.2 Open $b\bar{b}$ production and correlations at the LHC

Heavy quark production in the $k_t$-factorization approach at HERA and the Tevatron was considered already in many papers (see, for example, [82, 87, 88, 89, 90]). In Ref. [91] the $k_t$-factorization approach was used for a more detailed analysis of the D0 and CDF experimental data. The effect of the initial gluon tranverse momenta on the kinematics of the $b\bar{b}$ production at the LHC were investigated [92]. The renormalization and factorization scales were set equal to either the initial gluon virtualities, $\mu_R^2 = \mu_F^2 = q_{T1,2}^2$, or $\mu_F^2 = m_{bT}^2$, as is in the standard collinear QCD, and the quark mass of $m_b = 4.75$ GeV was used.

In Fig. 8a we show the transverse momentum distributions of $B$ mesons at LHC energies. The calculation was performed in the range $|\eta^B| < 1$ and the Peterson fragmentation with $\epsilon = 0.006$ using the KMS [93] parameterization for the un-integrated gluon density (see [83, 84]). The prediction for the azimuthal correlations between the muons coming from $B$ meson decays are shown in Fig. 8b with $p_t^\mu > 6$ GeV and $|\eta^\mu| < 2.5$. The azimuthal correlations indicate an important theoretical difference between the collinear and $k_t$-factorization approaches. In the collinear approximation at parton level and leading order, the $b$ quarks are be produced exactly back-to-back, which is clearly unphysical when the gluon is evolved up to a large enough scale. Only starting with NLO a significant deviation from the back-to-back scenario is found. Thus the NLO calculation has to correct for the wrong kinematics in LO together with higher order corrections, leading to large $K$ factors. In the $k_t$-factorization, the transverse momenta of the gluons are correctly treated already in LO. In the $k_t$ - factorization approach the NLO corrections are therefore expected to be much smaller, since here only the purely dynamical corrections have to be applied, whereas the kinematics are already correctly treated in LO.

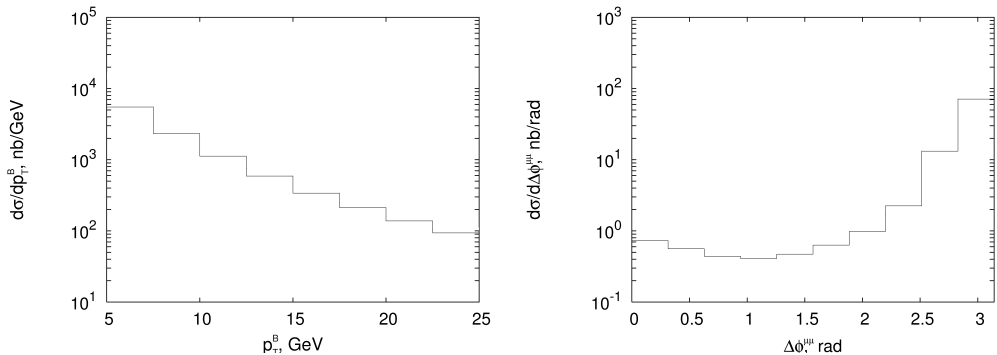

Fig. 8: Prediction for $B$-meson production at the LHC using the KMS un-integrated gluon density. In $a$ the $p_T$ distribution of $B$-mesons is shown. In $b$ the azimuthal $\mu\mu$ correlation coming from the $B$ decays is shown.

## 6.3 Quarkonium production and polarization at the LHC

Since the initial gluons have non-zero transverse momenta, they are off-shell, and they have a longitudinal component in their polarization vector. Typically, the $k_t$ values of the two colliding gluons are much different, as the parton evolution is equivalent to the random walk in the $\ln|k_t|$ plane, not in $k_t$ plane. Roughly speaking, the $k_t$ of one of the gluons can be neglected in comparison with that of the other. So, in the initial state we have one nearly on-shell (transversely polarized) gluon and one off-shell (longitudinally polarized) gluon. After the interaction, they convert into one on-shell gluon and a heavy vector meson. Simple helicity conservation arguments show that the polarization of vector meson must be longitudinal, in contrast with the ordinary parton model, where the initial gluons are both on-shell. This effect has been already studied for the HERA [94] and Tevatron [95] conditions. Fig.9a shows the predictions for the LHC energy obtained with KMS [93] parameterization for un-integrated gluon densities. The calculations are restricted to the pseudorapidity interval $-2.5 < \eta_\Upsilon < 2.5$ and assume ATLAS "$\mu 6\mu 3$" trigger cut, which means one muon with $p_t > 6$ GeV and another muon with $p_t > 3$ GeV.

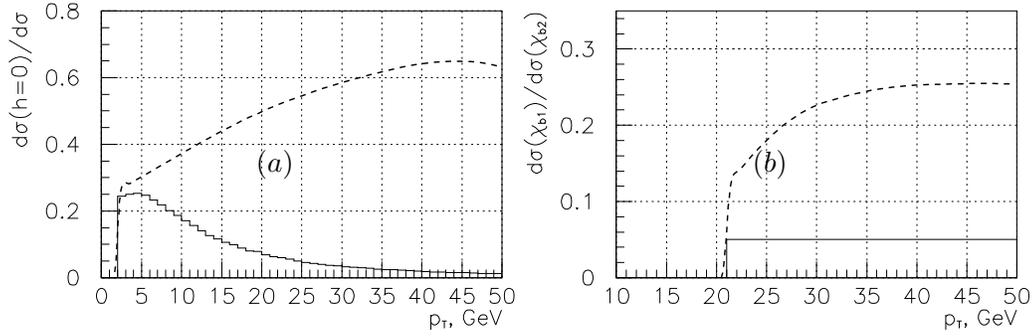

Fig. 9: Predictions of different theoretical approaches for quarkonium production. In (a) the fraction of longitudinally polarized $\Upsilon$ mesons is shown: solid histogram - collinear parton model, singlet + octet; dashed - $k_t$-factorization with KMS u.g.d.. In (b) the ratio of the production rates $\chi_{b1}/\chi_{b2}$ is shown: solid histogram - collinear parton model, singlet + octet; dashed - $k_t$ factorization with KMS u.g.d.

Important effects are also seen in the production of $P$-wave bottomium states with different spins $\chi_{b1}$ and $\chi_{b2}$. At the Tevatron energies, this process has been considered in Ref. [96], and the predictions for the LHC are presented in Fig.9b. The $P$-wave states are assumed to be detected via the decay $\chi_b \to \Upsilon + \gamma$, with an additional requirement that the energy of the decay photon be greater than 2 GeV. The ratio of the production rates $\sigma(\chi_1)/\sigma(\chi_2)$ is qualitatively different in the $k_t$-factorization and the collinear parton model. The underlying physics is clearly connected with gluon off-shellness. In the collinear parton model, the relative suppression of $\chi_1$ states becomes stronger with increasing $p_T$ because of the increasing role of the color-octet contribution: in this approach, the leading-order fragmentation of an on-shell transversely polarized gluon into a vector meson is forbidden. In contrast with that, in the $k_t$-factorization approach, the increase in the final state $p_T$ is only due to the increasing transverse momenta (and virtualities) of the initial gluons, and, consequently, the suppression motivated by the Landau-Yang theorem becomes weaker at large $p_T$.

### 6.4 Associated Higgs + jets production at the LHC

The dominant mechanism for Higgs production at the LHC is gluon-gluon fusion, and the calculations can be significantly simplified in the large top mass limit ($M_H \leq 2M_{top}$) [97]. The differential cross section of the inclusive Higgs production $p\bar{p} \to H + X$ in the $k_t$-factorization approach has been calculated in [98, 99] and can be written as:

$$\frac{d\sigma(p\bar{p} \to H + X)}{dy_H} = \int \frac{\alpha_s^2(\mu^2)}{288\pi} \frac{G_F\sqrt{2}}{x_1 x_2 m_H^2 s} \left[m_H^2 + \mathbf{p}_T^2\right]^2 \cos^2(\Delta\varphi) \times$$
$$\times \mathcal{A}(x_1, \mathbf{k}_{1T}^2, \mu^2) \mathcal{A}(x_2, \mathbf{k}_{2T}^2, \mu^2) d\mathbf{k}_{1T}^2 d\mathbf{k}_{2T}^2 \frac{d(\Delta\varphi)}{2\pi}, \quad (30)$$

where $G_F$ is the Fermi coupling constant, $\mathcal{A}(x, \mathbf{k}_T^2, \mu^2)$ is the un-integrated gluon distribution, $\Delta\varphi$ the azimuthal angle between the momenta $\mathbf{k}_{1T}$ and $\mathbf{k}_{2T}$, and the transverse momentum of the produced Higgs boson is $\mathbf{p}_T = \mathbf{k}_{1T} + \mathbf{k}_{2T}$. It should be noted, that this process is particularly interesting in $k_t$-factorization, as the transverse momenta of the gluons are in the same order as their longitudinal momenta ($\sim \mathcal{O}(10 \text{ GeV})$) [100].

The total inclusive Higgs production cross section at the LHC energies ($\sqrt{s} = 14$ TeV) is plotted in Fig. 10(a) as a function of the Higgs mass in the mass range $m_H = 100 - 200$ GeV. The solid line is obtained by fixing both the factorization and renormalization scales at the default value $\mu = m_H$ with the J2003 (set 1) CCFM un-integrated gluon distribution [101]. In order to estimate the theoretical uncertainties, we take $\mu = \xi m_H$ and vary the scale parameter $\xi$ between $1/2$ and 2 about the default value $\xi = 1$. The uncertainty band is presented by the upper and lower dashed lines. We find that our central values agree very well with recent NNLO results [102].

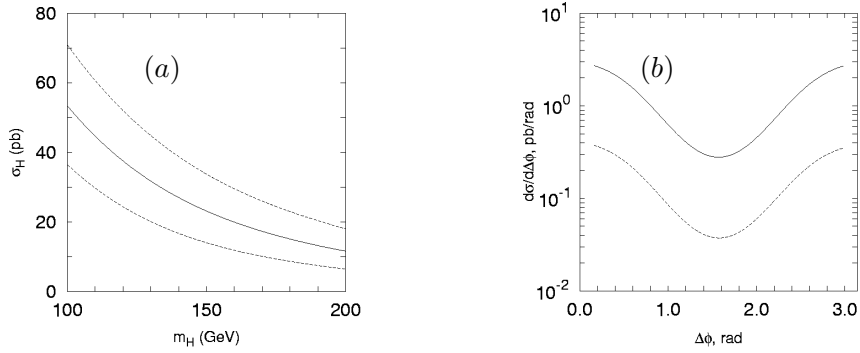

Fig. 10: Higg production at the LHC. In $(a)$ the total cross section for Higgs boson production as a function of Higgs mass is shown: the solid curve corresponds to the default scale $\mu = m_H$, upper and lower dashed curves - $\mu = m_H/2$ and $\mu = 2m_H/2$ respectively. In $(b)$ the jet-jet azimuthal angle distribution in the Higgs+jet+jet production at $\sqrt{s} = 14$ TeV. The kinematical cut $|\mathbf{p}_{\text{jet }T}| > 20$ GeV was applied for both jets. Solid and dashed lines correspond to the J2003 (set 1) and J2003 (set 2) [101] u.g.d. respectively.

To demonstrate the capabilities of the $k_t$-factorization approach, we calculate the azimuthal angle $\Delta\phi$ distribution between the two final jets transverse momenta in the Higgs+jet+jet production process. Our results are shown in Fig. 10($b$). The dip at $\Delta\phi = \pi/2$ comes from the $\cos(\Delta\varphi)$ in eq.(30). In the approach presented here, the $k_t$ of the initial gluons is approximately compensated by the transverse momenta of the jets [103]: $\mathbf{k}_T \simeq -\mathbf{p}_{T,jet}$, and, consequently, $\Delta\phi \simeq \Delta\varphi$ applying a cut-off $|\mathbf{p}_{\text{jet }T}| > 20$ GeV. This dip is characteristic for the loop-induced Higgs coupling to gluons in the framework of fixed-order perturbative QCD calculations [102]. Thus, we illustrate that the features usually interpreted as NNLO effects are reproduced in the $k_t$-factorization with LO matrix elements.

However, we see a very large difference (about one order of magnitude) between the predictions based on the J2003 gluon densities set 1 and set 2 [101], showing the sensitivity to the shape of the un-integrated gluon density.

### 6.5 Conclusions

The finite $k_t$ of the initial state gluons significantly modifies the kinematics of the gluon-gluon fusion process and leads to nontrivial angular correlations between the final state heavy quarks. The longitudinal polarization of the initial off-shell gluons manifests in the longitudinal polarization of $J/\psi$ and $\Upsilon$ mesons at moderate $p_T$ and, also, affects the production rates of $P$-wave quarkonia.

The predictions in $k_t$-factorization are very close to NNLO pQCD results for the inclusive Higgs production at the LHC, since the main part of high-order collinear pQCD corrections is already included in the $k_t$-factorization. In the $k_t$-factorization approach the calculation of associated Higgs+jets production is much simpler than in the collinear factorization approach. However, the large scale dependence of our calculations (of the order of $20 - 50\%$) probably indicates the sensitivity to the unintegrated gluon distributions.

## 7 Baryon charge transfer and production asymmetry of $\Lambda^0/\bar\Lambda^0$ in hadron interactions

*author: O.I. Piskounova.*

### 7.1 Introduction to the QGSM

The phenomenon of nonzero asymmetry of baryon production with nonbaryonic beams ($\pi$,$\mu$,e) was mentioned and explained in a few theoretical papers. Baryon charge can be transferred from proton or nucleus targets through the large rapidity gap with the string junction. In baryonic beam reactions ($p$,$A$, etc.) this effect is displayed in the valuable baryon/antibaryon spectrum asymmetry at $y = 0$. Every theoretical discourse on baryon charge transfer appeals to the value of the intercept, $\alpha_{SJ}(0)$, that is an intercept of the Regge-trajectory of imaginary particles which consists only of string junctions from baryon and antibaryon. Practically, the models that can account for this effect are only non perturbative QCD phenomenological models: the Dual Parton Model (DPM) [104] and the Quark Gluon String Model (QGSM) [105] as well as the DPMJET Monte Carlo expansion of these two models. Both analytical models are similar and they were based on the common Regge asymptotic presentation of constituent quark structure functions and string (quark) fragmentation functions. Here we are considering QGSM. In the comparison to the other models, QGSM accounts for many Pomeron exchanges. This approach works very well to give us the correct description of particle production cross sections at very high energies. The QGSM procedures of constructing of quark/diquark structure functions and fragmentation functions were presented in many previous publications. We take into consideration the $\pi$-$p$ reaction that gives similar asymmetries as the $\gamma$-$p$ reaction. The spectra in this reaction are more sensitive to the baryon excess in the region of positive $x_F$ than the spectra of baryons in $p$-$p$ collisions.

### 7.2 Comparison with Experimental Data

The asymmetry $A(y)$ between the spectra of baryons and antibaryons is defined as:

$$A(x) = \frac{dN^{\Lambda^0}/dx - dN^{\bar{\Lambda}^0}/dx}{dN^{\Lambda^0}/dx + dN^{\bar{\Lambda}^0}/dx}, \tag{31}$$

The EHS and the NA49 experiments have presented evidence for a nonzero baryon production asymmetry in proton-proton fixed target interactions, measuring at $y = 0$ values of the order of 0.5 - 0.3. In pion-proton interactions (E769) we can see the $y$ dependence of the asymmetry and the measured asymmetry, which was multiplied by a factor of 2 in order to be compared with the $pp$ asymmetry.

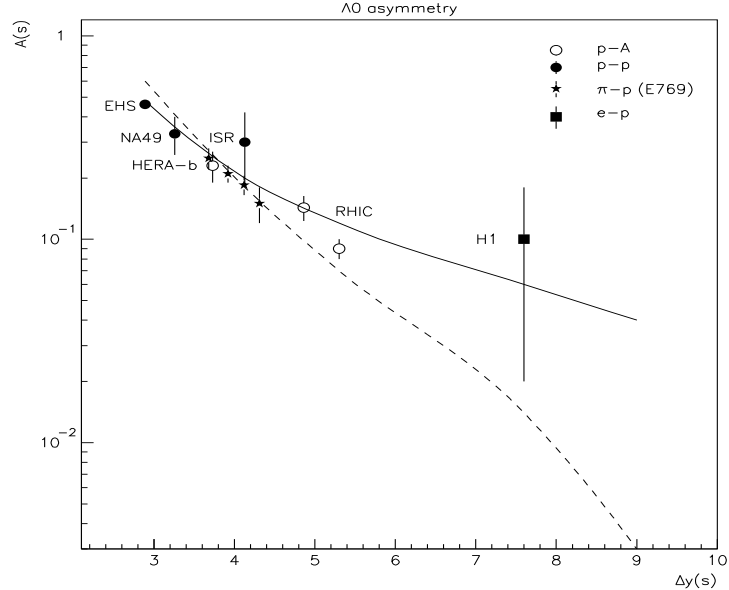

Fig. 11: Asymmetry in $\Lambda^0$ and $\bar{\Lambda}^0$ production and QGSM curves: $\alpha_{SJ}(0)$=0.5(dashed line) and $\alpha_{SJ}(0)$=0.9 (solid line).

The data from these experiments can be presented in one plot for $A(\Delta y)$, where $\Delta y$ is the rapidity distance from interacting target-proton (see Fig. 11). It is seen that the points are situated on the same line. If we add the data from proton-nucleus experiments (HERA-B and RHIC) they are still approximately on this line. Only the STAR asymmetry point at $\sqrt{s} = 130$ GeV can be interpreted as a sign that the curve is bent. And the result of the H1 experiment at HERA [106] calls certainly for a steeper curve. What means do we have in QGSM to describe this dependence? In Ref. [107] it was shown that the data of the E769 and H1 experiments can not be described with the same value of $\alpha_{SJ}(0)$: the points at lower energy correspond to $\alpha_{SJ}(0)$=0.5, while the H1 point requires $\alpha_{SJ}(0)$=0.9.

### 7.3 Summary

The purpose of this contribution is to show the band of asymmetries that can be predicted for the LHC experiments between the two possibilities given above for $\alpha_{SJ}(0)$. The results are shown in Fig. 11. The solid line represents the case of $\alpha_{SJ}(0)$=0.9. This curve fits the data at low energies (small $\Delta y$) due to varying the energy splitting between string junction and diquark configuration: 0.1*SJ+0.9*DQ. What we had to tune also was the fragmentation parameter af=0.15 instead of 0.2 accepted in previous papers. Also the curve for $\alpha_{SJ}(0)$=0.5 is shown in Fig. 11 with a dashed line. This line certainly doesn't fit the H1 point and gives a negligible asymmetry at the energy of the LHC experiments. Finally, we have the prediction for strange baryon asymmetries at the LHC within the range: $0.003 < A < 0.04$. The same procedure has to be applied to charmed baryon asymmetry to get the predictions at LHC energy.

# Experimental overview of heavy quark measurements at HERA

O. Behnke[1], A. Geiser[2], A. Meyer[2]

[1] Universität Heidelberg, Germany
[2] DESY Hamburg, Germany

**Abstract**
Experimental results on heavy flavour production at HERA are reviewed in the context of their relevance for future measurements at the LHC.

## 1 Introduction

Measurements of heavy flavour production at HERA can have significant impacts on the preparation and understanding of heavy flavour measurements at the LHC, and on the understanding of background processes to new physics discoveries [1]. The purpose of this contribution is to summarize the current status of heavy flavour measurements at HERA, and provide an outlook on how they might improve in the near future. The relation of these measurements to measurements at the LHC will be covered in more detail in subsequent contributions [2, 3, 4]. Since the top quark is too heavy to be produced at HERA with a significant rate, the term "heavy flavour" refers to $b$ and $c$ quarks only. The dominant diagram for heavy flavour production at HERA is shown in Fig. 1. The theory of heavy quark production at HERA is covered in the theoretical review section [5].

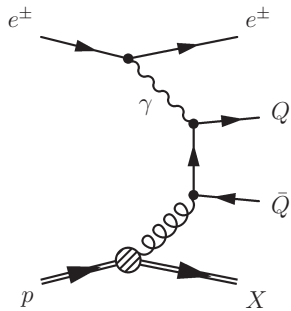

Fig. 1: Feynman graph for the production of a heavy quark pair via the leading order boson-gluon-fusion (BGF) process.

The interest in heavy flavour production arises from several aspects.

- Tagging a heavy flavour particle, e.g. inside a jet, establishes that this jet arises from a quark rather than a gluon. The number of possible QCD diagrams is thus reduced, and specific QCD final states can be studied more precisely than in inclusive measurements. This is even more true when *both* quarks of a heavy flavour quark pair can be tagged.
- The charm and beauty masses ($m_b, m_c \gg \Lambda_{QCD}$) provide energy scales which are large enough to allow perturbative calculations using a "massive" scheme [6, 7]. *All* QCD-processes involving heavy quarks should thus be reliably calculable. However, these mass scales often compete with other scales occurring in the same process, such as the transverse momentum ($p_T$) of the heavy quarks, or the virtuality of the exchanged photon, $Q^2$. Since the perturbative expansion can not be optimized for all scales at once, additional theoretical uncertainties enter which reduce the reliability of the predictions. If one of the competing scales ($p_T, Q^2$) is much larger than the quark mass, approximations in which the heavy quarks are treated as massless [8, 9, 10, 11, 12, 13, 14] can improve the reliability. Mixed schemes [15, 16, 17] are also possible. Understanding and resolving these difficulties should contribute to the understanding of multi-scale problems in general.
- Tagging the final state also constrains the initial state. Therefore, heavy flavour measurements can be used to measure or constrain parton density functions. In particular, Fig. 1 illustrates the direct sensitivity to the gluon density in the proton. Alternatively, in appropriate kinematic ranges, the initial state splitting of the gluon or photon into a heavy quark pair can be absorbed into the parton density definition. If the mass can be neglected, the same diagram (or higher order variants of it) can be reinterpreted as a way to measure the heavy flavour content of the proton or of the photon. These can in turn be used to calculate cross sections for other processes, such as Higgs production at the LHC.

- The production of "hidden" heavy flavour states (quarkonia) yields further insights into the interplay of (perturbative) heavy quark production and (non-perturbative) binding effects.

At HERA, the fraction of charm production vs. inclusive QCD processes is of order 10% in the perturbative QCD regime. Reasonably large samples can therefore be obtained despite the partially rather low tagging efficiency. Beauty production is suppressed with respect to charm production by the larger $b$ mass, and by the smaller coupling to the photon. The resulting total cross section is two orders of magnitude smaller than the one for charm. Beauty studies at HERA are thus often limited by statistics. Kinematically, beauty production at HERA is similar to top production at LHC ($m_b/\sqrt{s_{HERA}} \sim m_t/\sqrt{s_{LHC}}$). On the other hand, in the "interesting" physics region beauty is produced as copiously at the LHC as charm is at HERA.

## 2 Open charm production

Charmed mesons are tagged at HERA in different ways. A typical mass distribution for the "golden" channel $D^{*+} \to D^0 \pi^+, D^0 \to K^- \pi^+$ (+ c.c.) is shown in Fig. 2 [20]. Despite the low branching ratio, this channel yields large statistics charm samples of high purity. Fig. 3 [21] shows a corresponding D* production cross section in photoproduction for different kinematic variables. In general, D* production is well described by next-to-leading order QCD predictions, although the data often prefer the upper edge of the theoretical error band. Some deviations are observed in particular regions of phase space. For instance, there are indications that forward (i.e. in the direction of the proton) charm production might be slightly larger than theoretical expectations (Fig. 3b). Also, there are regions of phase space which effectively require four-body final states which are not covered by NLO calculations (see Fig. 5 in [1]). In order to describe such phase space regions, either NNLO calculations, or parton shower extensions to NLO calculations such as MC@NLO [18, 19] will be needed.

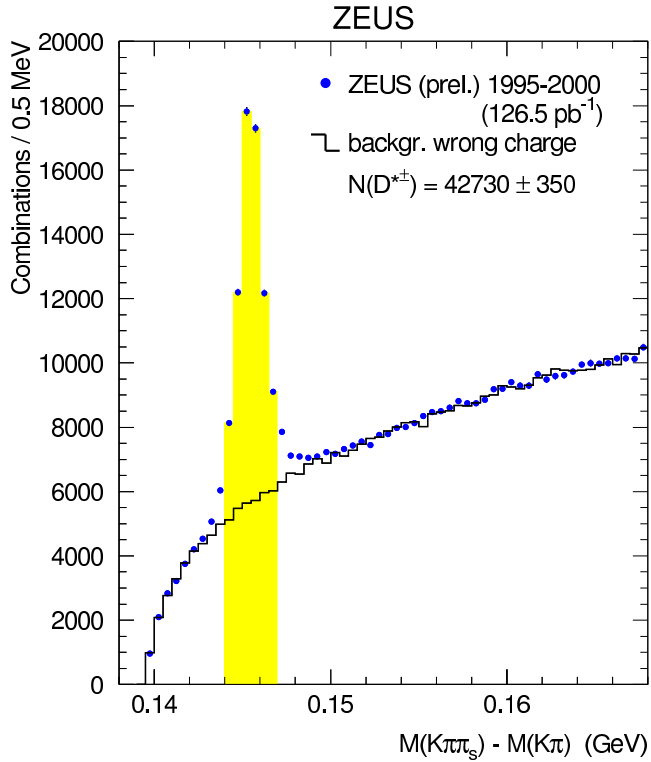

Fig. 2: Total inclusive $D^{*\pm}$ sample obtained by ZEUS for the HERA I data period in the golden decay channel $D^{*+} \to D^0 \pi_s^+ \to (K^- \pi^+) \pi_s^+$.

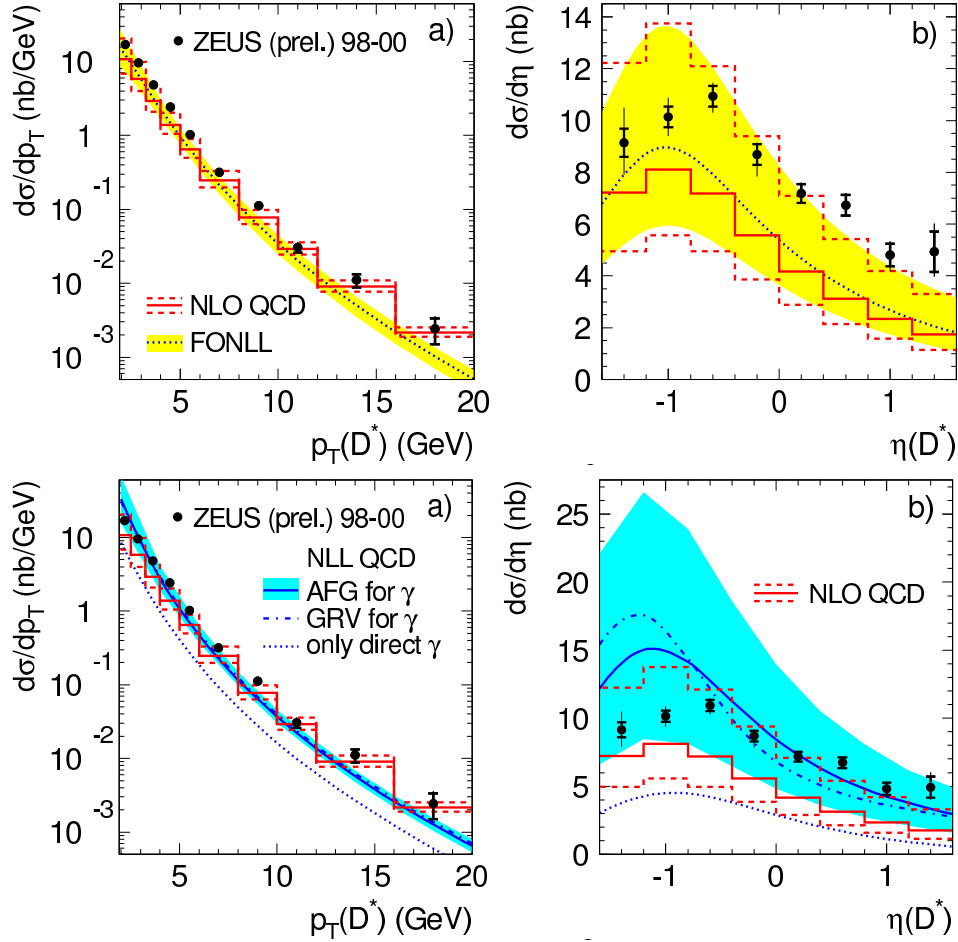

Fig. 3: $D^{*\pm}$ single differential cross sections in photoproduction as function of the $D^{*\pm}$ transverse momentum and pseudorapidity. The measurements are compared to NLO calculations in the massive (NLO), massless (NLL) and mixed (FONLL) scheme.

Other ways to tag charm include the reconstruction of a secondary vertex in a high resolution Micro-Vertex-Detector (MVD) in addition to the reconstruction of a charmed meson mass (Fig. 4) [22], or the reconstruction of inclusive multiple impact parameters resulting from the finite charm meson lifetime. A resulting cross section for $D^+$ production is shown in Fig. 5.

Since the charm mass of approximately 1.5 GeV is not very much above the threshold at which perturbative calculations are believed to produce reliable results, the generally good agreement of perturbative QCD predictions with the data is highly nontrivial, and encouraging concerning the validity of corresponding predictions for the LHC.

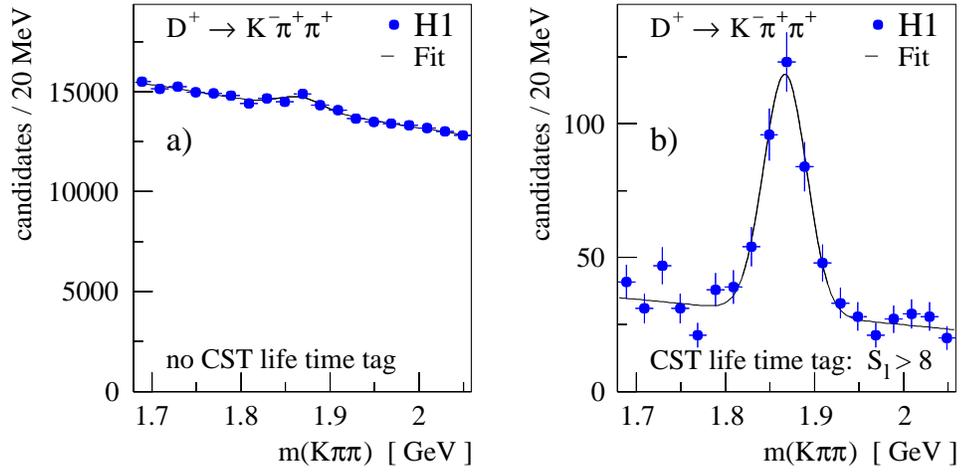

Fig. 4: $D^+$ mass peak in H1 before (left) and after (right) a cut on the decay length significance.

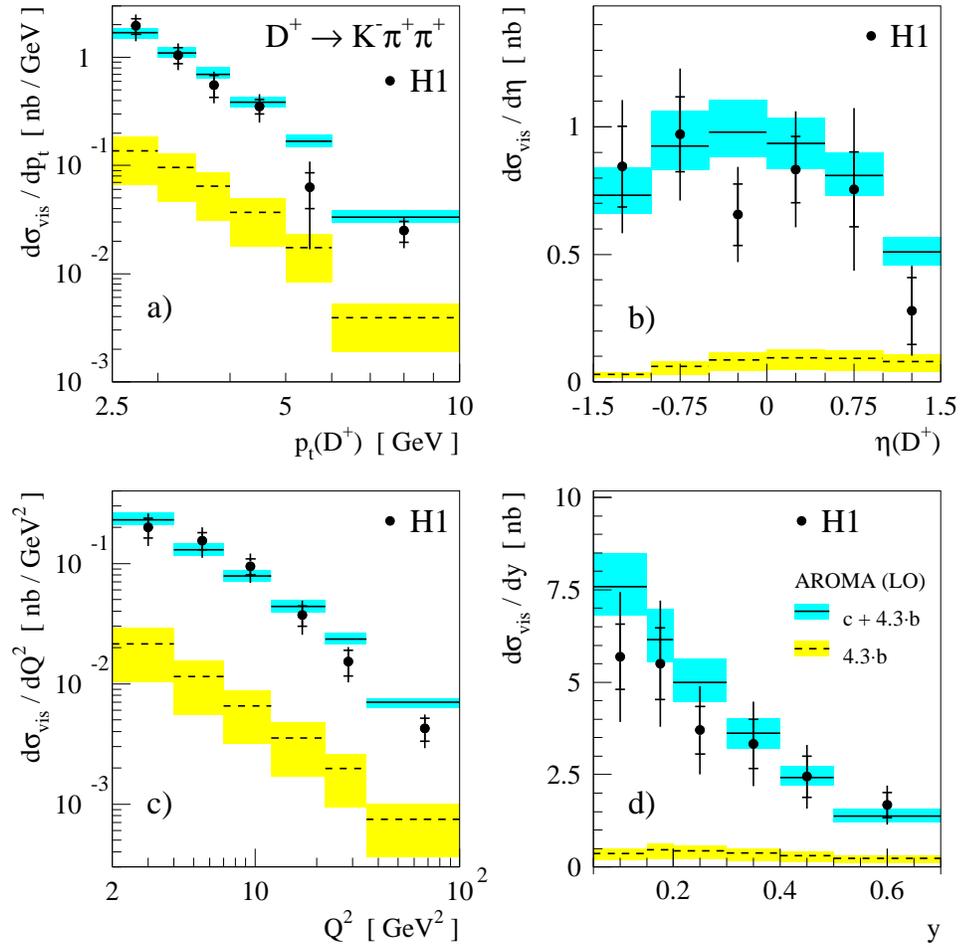

Fig. 5: Cross sections for $D^+$ production in DIS. A leading order + parton shower QCD prediction (AROMA) is shown for comparison.

## 3 Open beauty production

Open beauty production is detected at HERA using essentially three different methods, related to the large $b$ mass or long $b$ lifetime.

- If a jet is built out of the fragmentation and decay products of a $b$ meson/quark, the transverse momentum of the decay products with respect to the jet axis will be of order half the $b$ mass. This is significantly larger than for decay products of charm particles, or the transverse momenta induced by non-perturbative fragmentation effects, which are both of order 1 GeV or less. This distribution of this transverse momentum, called $p_T^{rel}$, can thus be used to measure the beauty contribution to a given jet sample.
- Due to the CKM-suppressed weak deacy of the $b$ quark, the lifetime of $b$ hadrons is longer than that of charmed particles. Furthermore, the larger decay angle due to the larger mass results in a higher significance of the decay signature. Detectors with a resolution in the $100\ \mu$ region or better can thus separate the beauty contribution from charm and light flavour contributions.
- Again due to their mass, $b$ hadrons take a larger fraction of the available energy in the fragmentation process. Furthermore, they produce decay products with sizeable transverse momentum even when produced close to the kinematic threshold. Simple lower cuts on the transverse momentum of such decay products therefore enrich the beauty content of a sample. Applying such cuts on two different decay products (double tagging) often sufficiently enriches the beauty content such that the remaining background can be eliminated or measured by studying the correlation between these decay products.

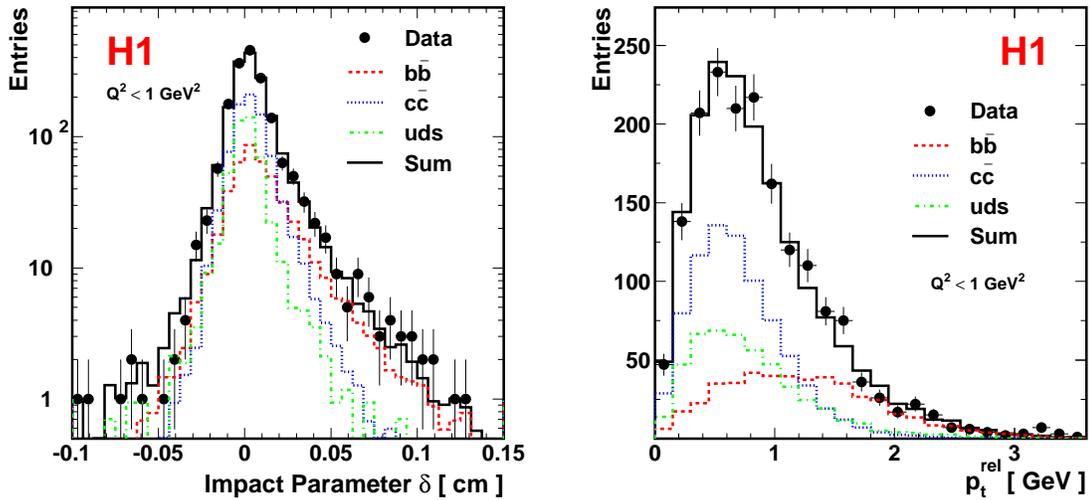

Fig. 6: Distributions of the impact parameter $\delta$ of the muon track (left) and the transverse muon momentum $p_t^{rel}$ relative to the axis of the associated jet (right) in H1. Also shown are the estimated contributions of events arising from $b$ quarks (dashed line), $c$ quarks (dotted line) and the light quarks (dash-dotted line).

An example for an analysis using the first two methods with muons from semileptonic $b$ decays is shown in Fig. 6 [23]. Some cross sections resulting from this type of analysis are shown in Fig. 3 of [1]. In general, reasonable agreement is observed between the data and corresponding NLO QCD predictions, although, as in the charm case, the data tend to prefer the upper edge of the theoretical error band. In some regions of phase space, e.g. at low $p_T^\mu$ or high $\eta^\mu$ differences of up to two standard deviations are observed. More precise measurements (section 8) will be needed to decide whether these deviations are really significant.

An example for an analysis using the 2nd method only is shown in Fig. 7 [24], while an example for an analysis using the third method is shown in Fig. 8 [25].

Fig. 9 shows a summary of the data/theory comparison for all HERA beauty results as a function of $Q^2$. For the measurements sensitive to $b$ quarks with $p_T^b \sim m_b$ or lower (black points) there is a trend that the "massive" NLO QCD predictions [7] tend to underestimate the $b$ production rate at very low $Q^2$. Depending on the chosen set of structure functions and parameters, a "mixed" prediction (VFNS) [16, 17] might describe the data better. For the higher $p_T$ measurements (red/grey points), no clear trend is observed. Note that theoretical errors, which are typically of order 30%, are not shown. Fig. 10 shows a similar compilation for all HERA measurements in photoproduction ($Q^2 < 1$ GeV), now as a function of the $b$ quark $p_T$. A similar trend is observed towards low $p_T$ (but note that several measurements appear in both figures). Again, more precise measurements are needed to determine whether these trends can be confirmed.

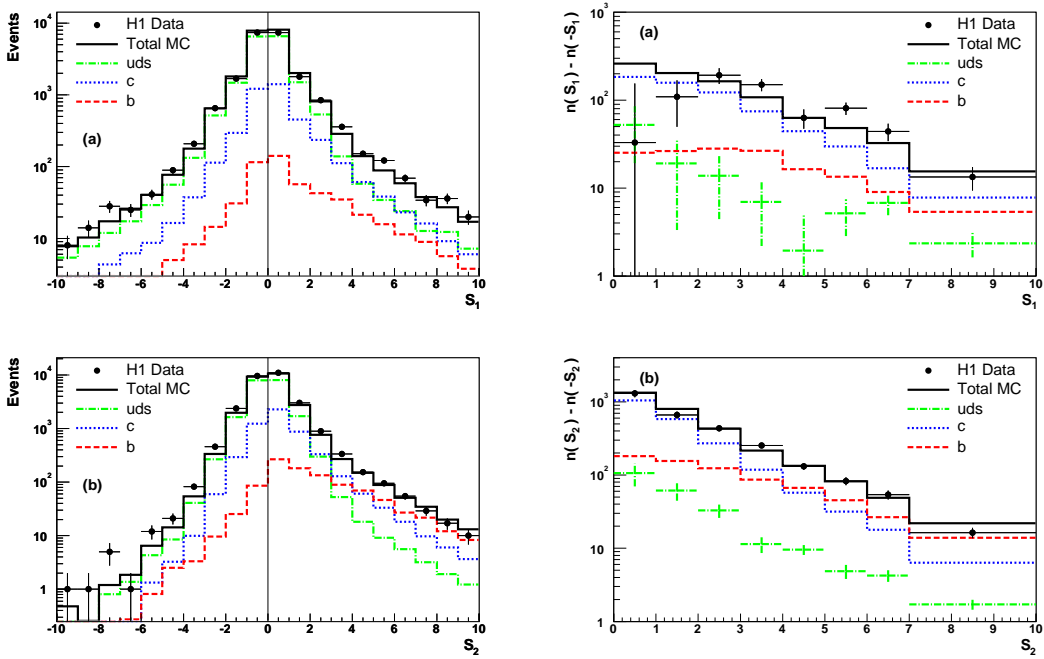

Fig. 7: Upper left: significance $S_1 = \delta/\sigma(\delta)$ distribution per event for events that contain one selected track associated to the jet axis. Lower left: significance $S_2 = \delta/\sigma(\delta)$ distribution per event of the track with the second highest absolute significance for events with $\geq 2$ selected tracks associated to the jet. Right: $S_1$ and $S_2$ distributions after subtracting the negatvie bins in the $S_1$ and $S_2$ distributions from the positive.

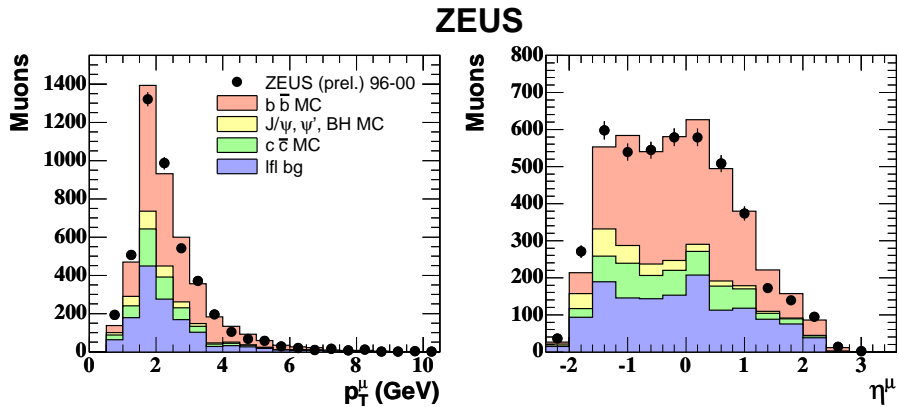

Fig. 8: Muon transverse momentum (a) and pseudorapidity (b) distributions for nonisolated low transverse momentum muon pairs in ZEUS (two entries per event). The beauty contribution is dominant.

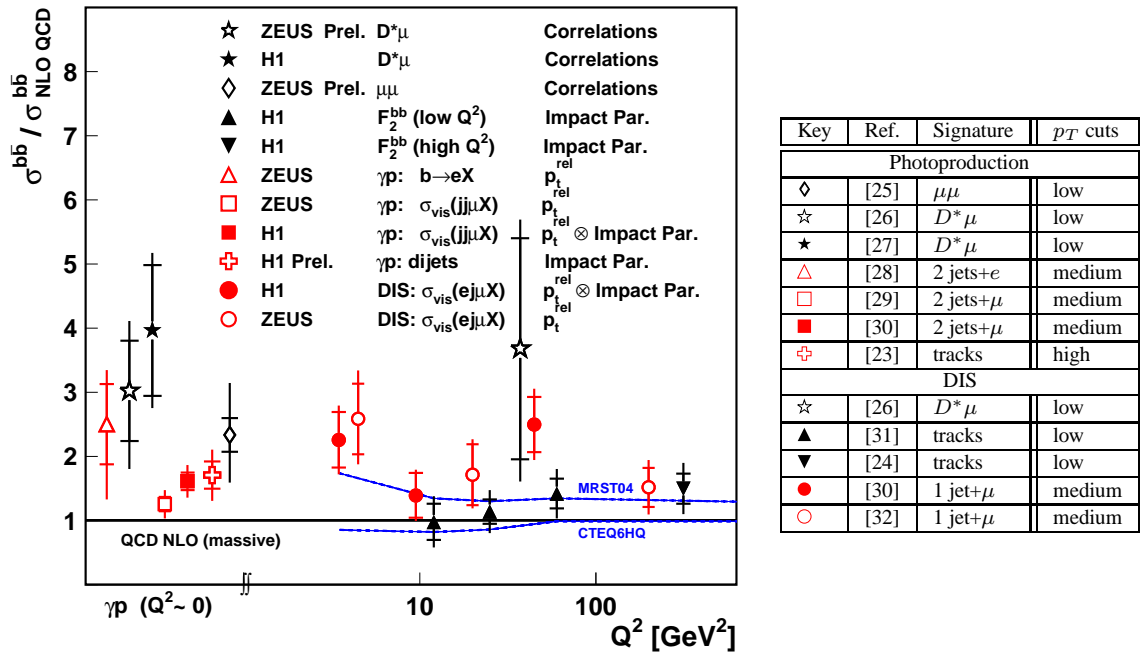

Fig. 9: Ratio of beauty production cross section measurements at HERA to NLO QCD predictions in the massive scheme as function of the photon virtuality $Q^2$. Measurements with low $p_T$ cuts are shown in black, while measurements with medium or high $p_T$ cuts are shown in red/grey. For more details see Table. The predictions from the VFNS NLO calculations by MRST and CTEQ for the DIS kinematic regime $Q^2 > 2$ GeV$^2$ are also shown (valid for comparison with the black low threshold points). Since theoretical errors are different for each point, they are not included in this plot.

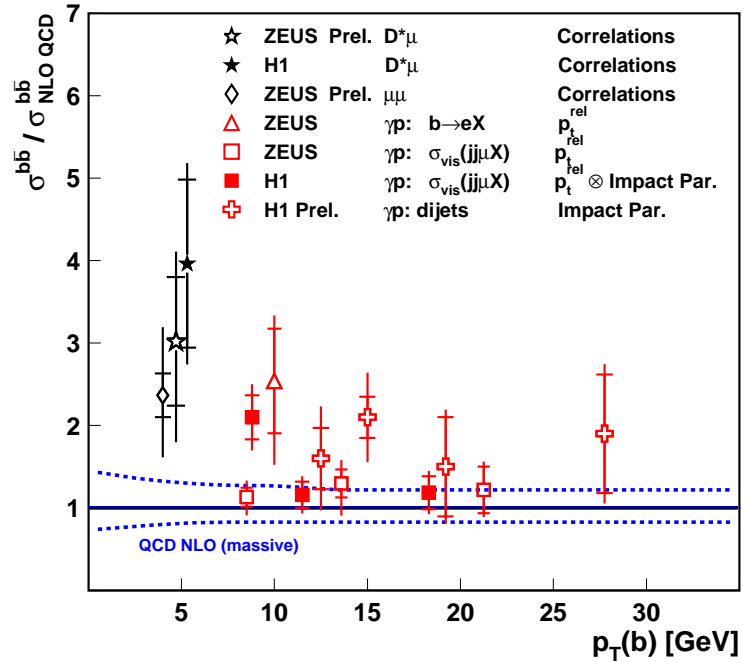

Fig. 10: Ratio of beauty production cross section measurements in photoproduction at HERA to NLO QCD predictions in the massive scheme as function of the transverse momentum of the $b$ quark $p_{Tb}$. The dashed line gives an indication of the size of the theoretical uncertianties.

## 4 Quarkonium production

Inelastic heavy quarkonia, like open charm and beauty production, are produced at HERA via the process of photon-gluon fusion. The two charm or beauty quarks hadronize to form a charmonium or bottomonium state.

A number of models have been suggested to describe inelastic quarkonium production in the framework of perturbative QCD, such as the color-singlet model (CSM) [33, 34], the color-evaporation model [35, 36] and soft color interactions [37]. Comprehensive reports on the physics of charmonium production are available [38, 39].

Recently the ansatz of non-relativistic quantum chromodynamics (NRQCD) factorization was introduced. In the NRQCD approach non-perturbative effects associated with the binding of a $q\bar{q}$ pair into a quarkonium are factored into NRQCD matrix elements that scale in a definite manner with the typical relative velocity $v$ of the heavy quark in the quarkonium. This way, colour octet quark anti-quark states, carrying different angular momenta and color charges than the quarkonium, can contribute to the charmonium production cross section. Theoretical calculations based on the NRQCD factorization approach [40, 41, 42] are available in leading order [43, 44, 45, 46, 47, 48]. In the NRQCD factorization approach the size of the color octet contributions, which are described by long distance matrix elements (LDME), are additional free parameters and have been determined in fits to the Tevatron data [49]. The NRQCD factorization approach incorporates the color singlet model i.e. the state $q\bar{q}[1,^3S_1]$ which is recovered in the limit in which the long distance matrix elements for other $q\bar{q}$ states tend to zero.

At HERA, cross sections measurements for photoproduction of $J/\psi$ and $\psi(2S)$ and for electroproduction of $J/\psi$ mesons have been performed [52, 53, 54, 55]. Bottomonium data are not available due to statistical limitations of the data.

For $J/\psi$ and $\psi(2S)$ photoproduction, calculations of the color-singlet contribution are available to next-to-leading order perturbation theory [50, 51]. Calculations which include the color octet contributions as predicted by NRQCD are available in leading order.

Figure 11 shows the measurements of the $J/\psi$ photoproduction cross section by the H1 collaboration [52] and the ZEUS collaboration [53] which are in good agreement with each other. The variable $z$ (left figure) denotes the fraction of the photon energy in the proton rest frame that is transferred to the $J/\psi$. Reasonable agreement is found between the HERA data and the NRQCD factorization ansatz in leading order (LO, CS+CO). The uncertainty indicated by the open band is due to the uncertainty in the color-octet NRQCD matrix elements. In contrast, the shaded band shows the calculation of the color-singlet contribution (NLO, CS) which is performed to next-to-leading order in $\alpha_s$ [50, 51]. This NLO, CS contribution alone describes the data quite well without inclusion of color-octet contributions. Comparison between the NLO,CS prediction (shaded band) and the LO,CS prediction (dotted line) shows that the NLO corrections are crucial for the description of the HERA photoproduction data.

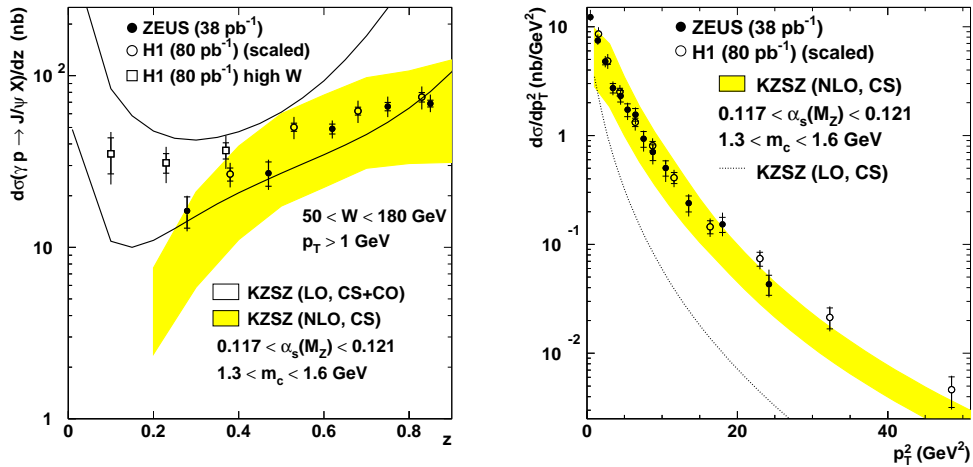

Fig. 11: Differential charmonium photoproduction cross sections as measured by H1 and ZEUS in comparison to calculations from LO NRQCD factorization (open band), NLO Color singlet contribution (shaded band) and LO color-singlet contribution (dotted line).

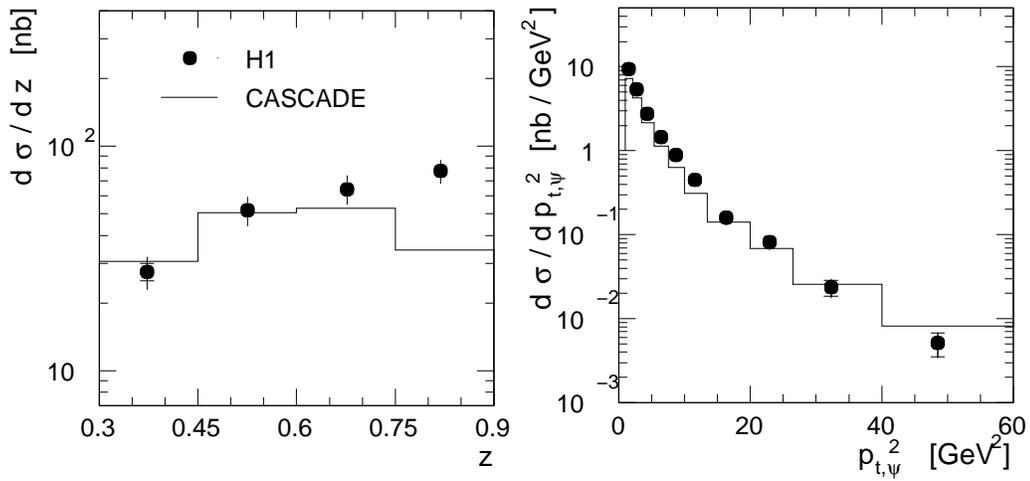

Fig. 12: Differential Charmonium photoproduction in comparison with a prediction using the color singlet-model and $k_t$ factorization as implemented in the Monte Carlo generator CASCADE.

Charmonium production cross sections have also been calculated in the $k_t$ factorization approach [56, 57, 58]. In these calculations the color-singlet model is used to describe the formation of the charmonium state. Figure 12 shows a comparison of the H1 data with the predictions from the $k_t$ factorization approach as implemented in the Monte Carlo generator CASCADE [59]. Good agreement is observed between data and predictions for $z < 0.8$. At high $z$ values, the CASCADE calculation underestimates the cross section. The CASCADE predictions for the the $p_{t,\psi}^2$ dependence of the cross section fit the data considerably better than the LO,CS calculation in the collinear factorization approach (dotted curve in fig. 11).

In fig 13 the differential cross sections for electroproduction of $J/\psi$ mesons as measured by H1 [54] and ZEUS [55] are shown as a function of $z$ and compared with predictions from the color singlet model (shaded band), with the NRQCD calculation [60] (CS+CO, open band), and also with calculations in the $k_t$ factorization approach (dotted line) as provided by [58] and implemented in the Monte Carlo generator CASCADE (dash-dotted line).

In the left figure the data are seen to agree well with the predictions using the color singlet model (shaded band and lines) while the full NRQCD calculation (open band), including color-octet contributions is wrong in shape and normalization. The agreement deteriorates when the cut $p_{t,\psi}^{*2} > 1$ GeV is applied (right fig. 13). This cut is justified, however, as towards small $p_{t,\psi}^{*2}$ perturbation theory becomes increasingly unreliable due to collinear singularities for the contributions $e + g \rightarrow e + c\bar{c}[n] + g$ with $n={}^1S_0^{(8)}$ and ${}^3P_J^{(8)}$ [60].

In conclusion, NRQCD, as presently available in leading order, does not give a satisfactory description of the HERA data. In contrast, the color singlet model shows a reasonable description of the HERA data, when implemented in calculations to next-to-leading order perturbation theory or in calculations in which the $k_t$-factorization approach is used.

## 5 Charm and Beauty contributions to structure functions

To a good approximation, except at very high $Q^2$, the cross section for inclusive deep inelastic electron scattering off the proton at HERA can be described in terms of a single proton structure function $F_2$ (for formula see [1]). This structure function only depends on the photon virtuality, $Q^2$, and on the Bjorken scaling variable $x$. Assuming that the electron scatters off a single quark in the proton (0th order QCD, quark-parton model) $x$ can be reinterpreted as the fraction of the proton momentum carried by the struck quark. This is a reasonable approximation for the light quark content of the proton.

For heavy quarks, the situation is a bit more complicated. Due to the heavy quark mass, on-shell heavy quarks can not exist within the proton. Rather, the dominant process for heavy quark production is the 1st order QCD BGF process depicted in Fig. 1. However, this process (and other higher order processes) still contributes to electron scattering, and hence to $F_2$. This can be interpreted in two ways.

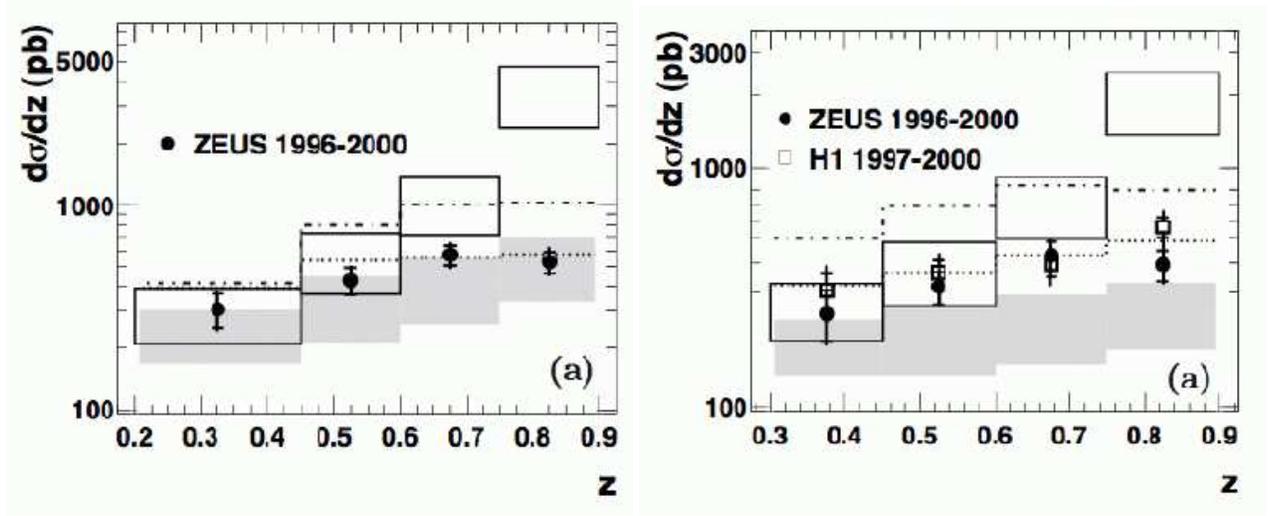

Fig. 13: Differential cross sections $d\sigma/dz$ without (left) and with (right) a cut on $p_{t,\psi}^{*2} > 1$ GeV. The data are compared to the NRQCD calculation (CS+CO, open band), the color-singlet contribution (CS, shaded band), with a prediction in the $k_t$ factorization approach assuming the CSM (dotted line) and with the Monte Carlo generator CASCADE (dash-dotted line).

In the massive approximation, heavy quarks are treated as being produced dynamically in the scattering process. The heavy quark contribution to $F_2$, frequently denoted as $F_2^{c\bar{c}}$ and $F_2^{b\bar{b}}$, therefore indirectly measures the *gluon* content of the proton. If $Q^2$ is large enough such that the quark mass can be neglected ($Q^2 \gg (2m_Q)^2$), the splitting of the gluon into a heavy quark pair can be reinterpreted to occur *within* the proton. $F_2^{c\bar{c}}$ and $F_2^{b\bar{b}}$ then measure the occurrence of *virtual* heavy quark pairs in the proton, or the "heavy quark structure" of the proton.

For charm production, the condition $Q^2 \gg (2m_Q)^2$ is valid for a large part of the HERA phase space. For beauty, it is only satisfied at very large $Q^2$. This is also the region most interesting for physics at the LHC.

Similar arguments hold for the heavy quark structure of the photon.

As an example, Fig. 14 [61, 62, 63, 24, 31] shows $F_2^{c\bar{c}}$ as measured by the ZEUS and H1 collaborations. A different representation of these results is shown in Fig. 6 of [1]. There, also $F_2^{b\bar{b}}$ is shown. Good agreement is observed with QCD predictions. Parametrizations of heavy quark densities of the proton at LHC energies should therefore be valid within their respective errors.

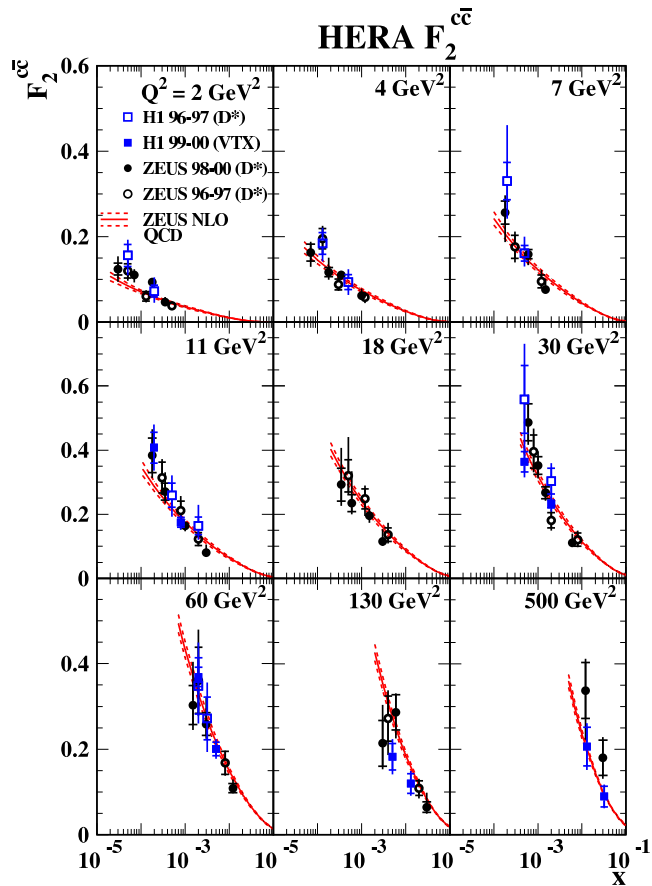

Fig. 14: $F_2^{c\bar{c}}$ results as a function of $x$ in bins of $Q^2$, from the H1 and ZEUS $D^{*\pm}$ analyses and from the H1 inclusive lifetime tagging measurements The data are compared to a NLO prediction using the ZEUS NLO fit results for the proton parton densities.

## 6 Charm fragmentation

The large cross section for charm production at HERA allows measurements of charm fragmentation which are very competitive with $e^+e^-$ measurements. As this topic is covered very nicely in [1] and [2] it is not treated further here.

## 7 Quark-antiquark correlations

Heavy quarks are always produced in pairs. An interesting way to check QCD is thus to verify whether the kinematic correlations between the quark pair are correctly described by QCD.

Fig. 15 shows different interpretations of the same higher order beauty production process. These different interpretations partially manifest themselves in different kinematic regions of beauty production phase space. If the highest virtuality part of the process occurs in the leading order BGF-like subprocess (left), the two $b$ quarks will be almost back-to-back in the detector transverse plane. The two extra gluons can either be reabsorbed into the proton structure, recovering the original BGF graph, or manifest themselves as visible "parton shower" activity in the direction of the proton. Alternatively, if the dominant leading order subprocess is gluon exchange with one of the $b$ quarks (right, "flavour excitation in the photon"), this $b$ quark will recoil against a gluon jet. At sufficiently large momentum transfer (rare at HERA), the second $b$ quark can be treated as a "spectator", and will approximately follow the initial photon direction. At next-to-leading order, contributions to both processes are described by the same Feynman graph, but the two extreme kinematic cases (and all variants in between) are still included. If both heavy quarks are tagged, these different kinematic regions can be distinguished by measuring the momentum and angular correlations between the two quarks.

Fig. 16 [25] shows the angular correlations between the two muons originating from different $b$ quarks of a $b\bar{b}$ pair. Reasonable agreement is observed with QCD predictions. The predominantly back-to-back topology

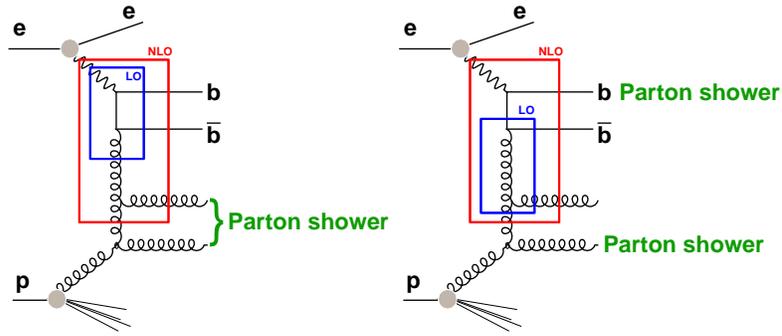

Fig. 15: Example for higher order Feynman graph for beauty production. Different interpretations of the graph in terms of NLO or LO matrix elements plus parton showers are highlighted. Depending on the kinematics and the scheme chosen, part of the gluons could also be reabsorbed into the proton structure function definition, and/or the $\gamma b \bar{b}$ vertex could be interpreted as part of the photon structure.

comfirms the dominance of the BGF-like contribution.

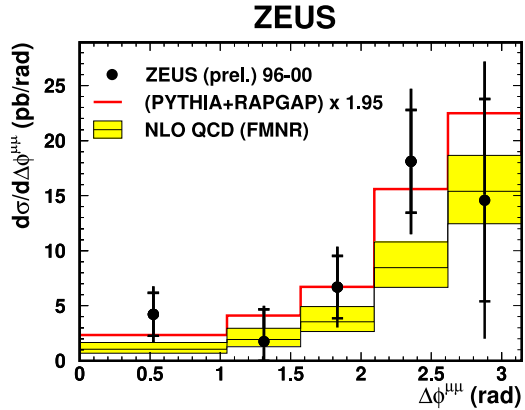

Fig. 16: Differential cross section $d\sigma/d\Delta\phi^{\mu\mu}$ for dimuon events from $b\bar{b}$ decays in which each muon originates from a different $b(\bar{b})$ quark. The data (solid dots) are compared to the leading order + parton shower generators PYTHIA and RAPGAP (histogram) and to massive NLO QCD predictions (shaded band).

## 8  HERA II prospects

Both the HERA collider and its detectors have been upgraded in 2001/2 to provide more luminosity with polarized electron beams, and improve heavy flavour detection. This program is called HERA II. The luminosity accumulated so far already exceeds the HERA I luminosity. An integrated luminosity up to 700 $pb^{-1}$ is expected at the end of the HERA program in 2007. This enhances the statistics for many studies by almost an order of magnitude with respect to HERA I. The improved detectors offer further handles for improved heavy flavour measurements. H1 has improved the forward coverage of its Micro-Vertex-Detector [68], and added a Fast Track Trigger [67]. ZEUS has implemented a Micro-Vertex-Detector (MVD) [65] for the first time for HERA II, and has added an upgraded forward tracking detector [66]. These improvements allow the application of measurement techniques which could not be used at HERA I, and can be used to improve the data quality, add additional statistics, and/or cover new phase space regions.

New detectors require time to fully understand their systematics, but first preliminary results have already been obtained. Fig. 17 [64] shows the cross section for beauty production obtained using the new ZEUS MVD with the first 33 $pb^{-1}$ of HERA II data, compared to the HERA I result. Good agreement is observed.

The measurements which will profit most from the improved HERA II data sets include double differential measurements such as the beauty and charm contributions to the proton structure function $F_2$, and multi-tag measurements to explicitly study quark-quark correlations. Statistical improvements of at least one order of magnitude can be expected when the increased luminosity and improved measurement techniques are combined.

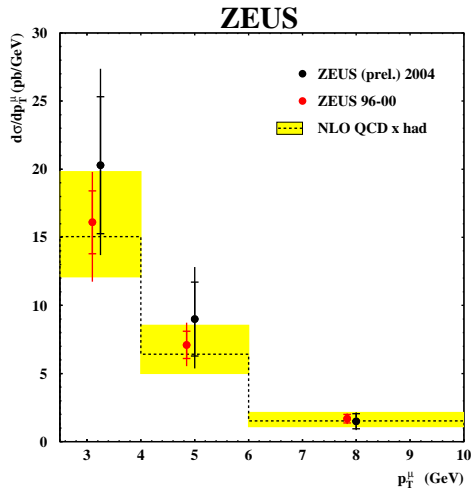

Fig. 17: Differential cross section as function of muon $p_T$ for dimuon + jet events in photoproduction. Preliminary results from the first 33 $pb^{-1}$ of HERA II data are compared to HERA I results and QCD predictions.

## 9 Conclusions

Heavy flavour production at HERA is a very active field of research yielding multiple insights into the applicability of perturbative QCD. The problem of multiple scales complicates the perturbative expansions and limits the achievable theoretical precision. In general, QCD predictions agree well with the data, although indications for deviations persist in specific regions of phase space. Some of these might be attributable to missing NNLO or even higher order contributions.

The overall reasonable agreement, as well as the self-consistency of the structure functions tested by or derived from heavy flavour production at HERA, enhances confidence in corresponding cross-section predictions at LHC, within their respective theoretical uncertainties.

# Experimental aspects of heavy flavour production at the LHC

*J. Baines[a], A. Dainese[b], Th. Lagouri[c], A. Morsch[d], R. Ranieri[e], H. Ruiz[d], M. Smizanska[f], and C. Weiser[g]*

[a] Rutherford Laboratory, UK
[b] University and INFN, Padova, Italy
[c] Institute of Nuclear and Particle Physics, Charles University, Prague, Czech Republic
[d] CERN, Geneva, Switzerland
[e] University and INFN, Firenze, Italy
[f] Lancaster University, Lancaster, UK
[g] Institut für Experimentelle Kernphysik, Universität Karlsruhe, Karlsruhe, Germany

**Abstract**
We review selected aspects of the experimental techniques being prepared to study heavy flavour production in the four LHC experiments (ALICE, ATLAS, CMS and LHCb) and we present the expected performance for some of the most significative measurements.

*coordinators: A. Dainese, M. Smizanska, and C. Weiser*

## 1 Introduction

*author: M. Smizanska*

Unprecedently large cross sections are expected for heavy-flavour production in proton–proton collisions at the LHC energy of $\sqrt{s} = 14$ TeV. Next-to-leading order perturbative QCD calculations predict values of about 10 mb for charm and 0.5 mb for beauty, with a theoretical uncertainty of a factor 2–3. Despite these large cross sections, the LHC experiments, ALICE [1, 2], ATLAS [3], CMS [4], and LHCb [5], will have to deal with rejection of background coming from non-heavy flavour inelastic interactions for which the predicted cross is about 70 mb. The four experiments will work at different luminosity conditions. ATLAS and CMS are designed to work in a wide range of luminosities up to nominal $10^{34}$ cm$^{-2}$s$^{-1}$, while the LHCb optimal luminosity will vary in the range $(2–5) \times 10^{32}$ cm$^{-2}$s$^{-1}$ and ALICE is designed to work at $3 \times 10^{30}$ cm$^{-2}$s$^{-1}$ in proton–proton collisions. Luminosity conditions in ATLAS, CMS and LHCb allow multiple interactions per bunch crossing, thus leading to requirements of even stronger identification and selection of heavy-flavour events already at trigger level. The first task will be the measurement of integrated and differential charm and beauty production cross sections in the new energy domain covered at the LHC. ALICE will play an important role, having acceptance down to very low transverse momentum, as we discuss in Section 4. These measurements can be performed within a relatively short period of running. Afterwards, the heavy-flavour studies will focus on less inclusive measurements addressing specific production mechanisms that allow to test higher order perturbative QCD predictions, as well as on rare decays of heavy-flavour hadrons, that may carry information on New Physics beyond the Standard Model. In order to meet these requirements dedicated and sophisticated trigger strategies have been prepared by the LHC experiments.

## 2 Heavy flavour detection in the LHC experiments

*author: A. Dainese*

The four detectors that will take data at the LHC have different features and design requirements, but all of them are expected to have excellent capabilities for heavy-flavour measurements. Their complementarity will provide a very broad coverage in terms of phase-space, decay channels and observables.

Experimentally, the key elements for a rich heavy-flavour program are track and vertex reconstruction and particle identification (PID).

Open charm and beauty mesons have typical life-times in the order of a $ps$ ($c\tau$ values are about 125–300 $\mu$m for D mesons and 500 $\mu$m for B mesons) and the most direct detection strategy is the identification of single tracks or vertices that are displaced from the interaction vertex. The detector capability to perform this task can be characterized by the transverse impact parameter[1] ($d_0$) resolution. All experiments will be equipped with high position-resolution silicon-detector layers, including pixel detector for the innermost layers, for precise tracking and impact parameter measurement. Tracking is done in the central (pseudo)rapidity region for ALICE ($|\eta| < 0.9$), ATLAS and CMS ($|\eta| \lesssim 2.5$), and in the forward region for LHCb ($2 \lesssim \eta \lesssim 5$). In Fig. 1 we show the $d_0$ resolution, which is similar for the different experiments, and better than 50 $\mu$m for $p_T \gtrsim 1.5$–3 GeV. The inner detector systems of ATLAS, CMS and ALICE will operate in different magnetic fields: The ALICE magnetic field will vary within low values (0.2–0.5 T) leading to a very low $p_T$ cutoff of 0.1–0.2 GeV, while ATLAS (2 T) and CMS (4 T) have higher cutoffs of 0.5 and 1 GeV, respectively, but better $p_T$ resolution at high $p_T$ (e.g., at $p_T = 100$ GeV, $\delta p_T/p_T \approx 1$–2% for ATLAS/CMS at central rapidity and $\approx 9\%$ for ALICE).

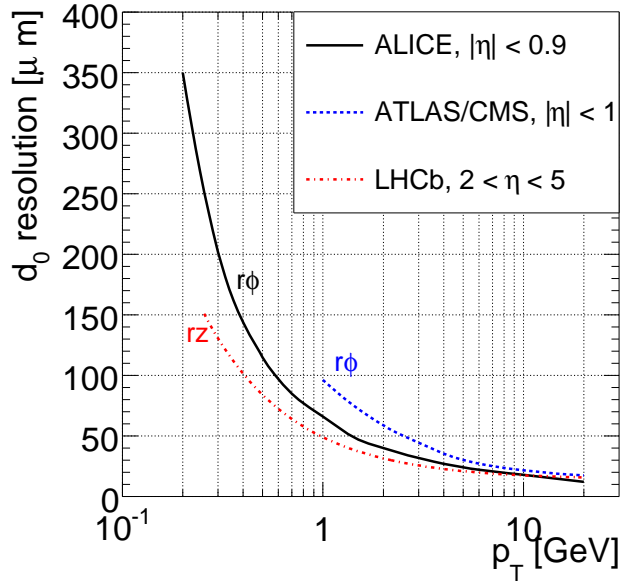

Fig. 1: Track impact parameter resolutions for the four LHC experiments. Note that for ALICE, ATLAS and CMS the impact parameter is defined in the $r\phi$ plane, while for LHCb it is defined in the $rz$ plane.

Both lepton and hadron identification are important for heavy-flavour detection. D and B mesons have relatively large branching ratios (BR) in the semi-leptonic channels, $\simeq 10\%$ to electrons and $\simeq 10\%$ to muons, and inclusive cross-section measurements can be performed via single leptons or di-leptons. Alternatively, high-$p_T$ leptons can be used as trigger-level tags to select B $\to$ J/$\psi$ + X candidate events, that provide more accurate cross section measurements. ALICE can identify electrons with $p_T > 1$ GeV and $|\eta| < 0.9$, via transition radiation and d$E$/d$x$ measurements, and muons in the forward region, $2.5 < \eta < 4$, which allows a very low $p_T$ cutoff of 1 GeV. CMS and ATLAS have a broad pseudorapidity coverage for muons, $|\eta| < 2.4$ and $|\eta| < 2.7$, respectively, but they have a higher $p_T$ cutoff varying between 4 and 6 GeV, depending on $\eta$. Both CMS and ATLAS have high-resolution electro-magnetic calorimeters that will be used to identify electrons. Semi-leptonic inclusive measurements do not provide direct information on the D(B)-meson $p_T$ distribution, especially at low $p_T$, because of the weak correlation between the lepton and the meson momenta. Therefore, for charm in particular, the reconstruction of exclusive (hadronic) decays is preferable. In this case, hadron identification allows a more effective rejection of the combinatorial background in the low-$p_T$ region. ALICE disposes of $\pi$/K/p separation via d$E$/d$x$ and time-of-flight measurements for $p < 3$–4 GeV and $|\eta| < 0.9$.

Figure 2 shows schematically the $p_T$ vs. $\eta$ acceptances for charm (c) and beauty (b) hadrons in the four experiments, as expected for one year of running at nominal luminosity (note that the value of the luminosity is different for each experiment, as previously discussed). ATLAS and CMS have similar acceptances for beauty measurements; the minimum accessible $p_T$ is relatively large because of the strong magnetic fields, which in turn, together with the high luminosity, allow to cover transverse momenta up to 200–300 GeV. The acceptance of LHCb, although centred at forward rapidity, has a significant overlap, with those of ATLAS and CMS. The acceptance of ALICE for beauty overlaps with ATLAS and CMS at central rapidity and with LHCb at forward rapidity. The moderate magnetic field allows measurements down to transverse momenta of about 2 GeV for B mesons in the forward muon arm and in the barrel, and down to about 1 GeV for D mesons in the barrel.

---

[1]We define as impact parameter the distance of closest approach to the interaction vertex of the track projection in the plane transverse to the beam direction.

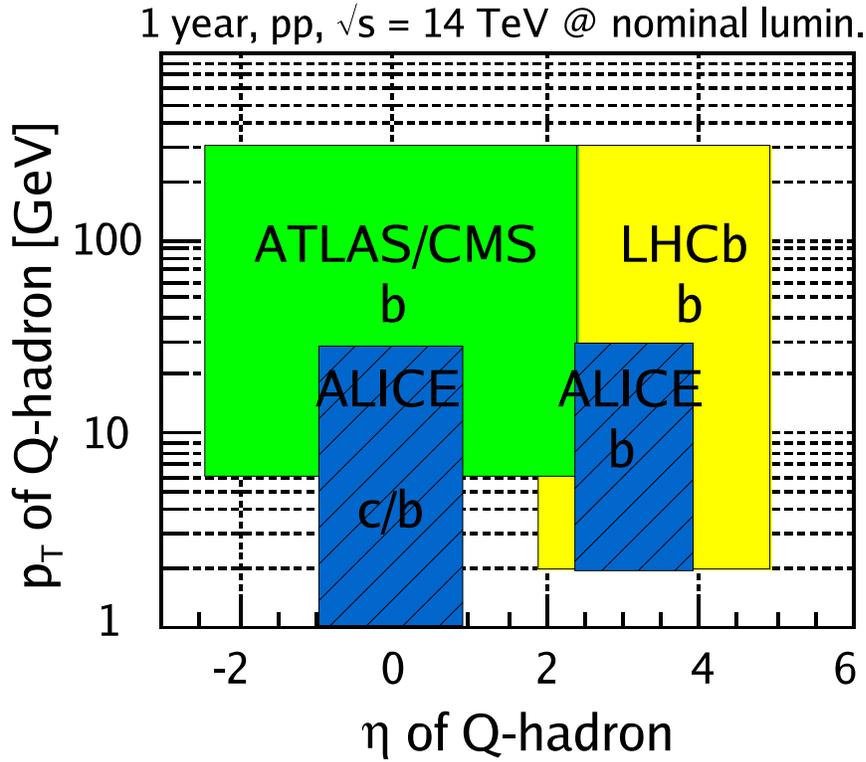

Fig. 2: Schematic acceptances in transverse momentum and pseudorapidity for open heavy flavour hadrons (indicated as 'Q-hadrons') in the four LHC experiments. The high-$p_T$ coverages correspond to one year (i.e. 7 months) of running at nominal luminosity (see beginning of this section).

## 3  Beauty triggers at the LHC

### 3.1  ATLAS beauty trigger

*author: J. Baines*

The ATLAS trigger consists of three levels [6]. Level-1 is implemented in hardware, whilst the higher level triggers (level-2 and the Event Filter, EF) are based on general-purpose processors. The level-1 triggers are based on information from the calorimeter and muon trigger chambers. At higher trigger levels, information from the Inner Detector (ID) and precision muon detector is included. The size of the level-2 and EF processor farms is limited, which in turn limits the amount of data processing that can be performed in the trigger. The B-trigger must, therefore, have the flexibility to adapt selections both as the luminosity falls during a beam-coast and, over a longer time-scale, as the peak luminosity of the LHC increases. This is achieved by using a di-muon trigger at the start of higher luminosity fills and introducing additional triggers for lower luminosity fills or as the luminosity falls during a beam coast [7].

A di-muon trigger provides a very effective selection for a range of important channels, e.g. $B_d^0 \to J/\psi(\mu^+\mu^-)K_s^0$, $B \to K^{0\star}\mu\mu$ and $B \to \rho^0\mu\mu$. The Level-1 muon trigger is efficient down to a $p_T$ of about 5 GeV in the barrel region and about 3 GeV in the end-caps. However the actual thresholds used for the di-muon trigger will be determined by rate limitations. For example, a $p_T$ threshold of 6 GeV would give a di-muon trigger rate of about 600 Hz after level-1 at a luminosity of $2 \times 10^{33}$ cm$^{-2}$s$^{-1}$. These triggers are mostly due to muons from heavy flavour decays plus some single muons which are doubly counted due to overlaps in the end-cap trigger chambers. The later are removed when the muons are confirmed at level-2 using muon precision chambers and ID information from inside the level-1 Region of Interest (RoI). At the EF tracks are refit, inside regions identified by level-2, and specific selections made on the basis of mass and decay length cuts. These consist of semi-inclusive selections, for example to select $J/\psi(\mu^+\mu^-)$ decays with a displaced vertex, and in some cases exclusive selections such as for $B \to \mu^+\mu^-$. The final trigger rate, after the EF, is about 20 Hz at a luminosity of $2 \times 10^{33}$ cm$^{-2}$s$^{-1}$.

At lower luminosities, additional triggers are introduced based on a single muon trigger ($p_T \gtrsim 8$ GeV) together with a calorimeter trigger. The calorimeter trigger identifies clusters of energy deposition in the electromagnetic and hadronic calorimeter consistent with an electron or photon (EM RoI) or a jet (Jet RoI). For hadronic final states, such as $B_s^0 \to D_s^- \pi^+$ and $B_s^0 \to D_s^- a_1^+$ track are reconstructed in the Inner Detector in RoI of about $\Delta\eta \times \Delta\phi = 1.0 \times 1.5$. By limiting track reconstruction to the part of the ID lying within the RoI, about 10% on average, there is potential for up to a factor of ten saving in execution time compared to reconstruction in the full Inner Detector. Preliminary studies of efficiency and jet-cluster multiplicity have been made using a fast simulation which includes a detailed parameterization of the calorimeter. These studies indicate that a threshold on the jet cluster energy of $E_T > 5$ GeV gives a reasonable multiplicity, i.e. a mean of about two RoI per event for events containing a muon trigger. This threshold would give a trigger that is efficient for $B_s^0 \to D_s^- \pi^+$ events with a $B$-hadron $p_T$ above about 15 GeV.

Track reconstruction inside e/gamma RoI can be used to select channels such as $B_d^0 \to J/\psi(e^+e^-)K_s^0$, $B_d \to K^{0\star}\gamma$ and $B_s \to \phi\gamma$. Preliminary studies show that a reasonable compromise between RoI multiplicity and electron efficiency might be obtained with a cluster energy threshold of $E_T > 2$ GeV. This gives a mean RoI multiplicity of about one for events containing a muon trigger and is efficient for channels containing an electron with $p_T > 5$ GeV. Following the ID track reconstruction further selections are made for specific channels of interest. These are kept as inclusive as possible at level-2 with some more exclusive selections at the EF.

In LHC running, there will be competing demands for resources in the level-2 and EF trigger farms and for trigger band-width. By adopting a flexible strategy and making the maximum use of RoI information to guide reconstruction at level-2 and the EF, the B-physics coverage of ATLAS can be maximized.

## 3.2 CMS beauty trigger

*author: R. Ranieri*

The Large Hadron Collider (LHC) will provide 40 MHz proton-proton collisions at the centre of mass energy of 14 TeV. At the beginning a luminosity of $2 \times 10^{33}$ cm$^{-2}$s$^{-1}$ is expected, corresponding to 20 fb$^{-1}$ collected per year. Assuming the $b\bar{b}$ production cross section to be 0.5 mb, $10^{13}$ b-physics events per year are foreseen: all kind of b-particles will be produced and studies will be performed not only in $B_d^0$, but also in $B_s^0$ meson system. A wide b-physics programme, including CP violation, $B_s^0 - \overline{B}_s^0$ mixing and rare decays can therefore be covered by the CMS experiment. The apparatus will be equipped with a very precise tracking system made with silicon microstrip and pixel detectors [8, 9].

The rate at which events can be archived for offline analyses is 100 Hz [10, 11]. The trigger thresholds are optimized for a wide physics discovery program with selection of high transverse momentum ($p_T$) processes. Low-$p_T$ events, as required for b-physics, are selected mainly by the first level muon trigger, then an exclusive reconstruction of few relevant *benchmark* channels can separate interesting events from the background. The b-physics programme could evolve with time following both the theoretical developments and the results which will be obtained in the next years by b-factories and Tevatron experiments.

The lowest trigger level (Level-1) is based on the fast response of calorimeters and muon stations with coarse granularity. No information on secondary vertices is available, hence the Level-1 selection of b-physics events exploits the leptonic signatures from beauty hadron decays, therefore a single muon or a di-muon pair is required. The Level-1 output at start-up will be 50 kHz. Several studies have been done to optimize the trigger thresholds in order to have the possibility of selecting most of the interesting physics signatures. A total of 3.6 kHz rate is dedicated to the Level-1 muon selection. It is obtained by requiring a single muon with $p_T > 14$ GeV or at least two muons with $p_T > 3$ GeV.

A further selection is made during the High-Level trigger (HLT) by using also the information from the tracking system. The CMS High-Level trigger is entirely based on a CPU farm with some thousand CPUs. Each processor analyses a single event; in principle offline event reconstruction can be performed, but in order to reduce the processing time fast track reconstruction has to be done. Some algorithms will be dedicated to the fast reconstruction and identification of physics processes, thus allowing to start the offline analysis directly from the online selection. They have to fulfill the HLT time constraint, hence they have to be able to analyze and accept (or reject) data within the time limits imposed by the HLT latency. To lower the execution time, which is due mainly to the processing of tracking system signals, track reconstruction is preferably performed only in limited regions of the space (*regional track reconstruction*) and stopped when a certain precision is reached in the measurement of some track parameters, such as transverse momentum and impact parameter (*conditional track finding*). Invariant mass of b-hadrons can thus be measured online with good resolution, allowing to select the searched event topologies.

An additional trigger strategy, which relies on the possibility of lowering the trigger thresholds during the LHC beam coast or lower luminosity fills to collect more b-physics events is under study.

The rare decay $B^0_{s,d} \to \mu^+\mu^-$ is triggered at Level-1 with 15.2% efficiency. At HLT, the two muons are required to be opposite charged and isolated, to come from a displaced common vertex and have an invariant mass within 150 MeV from the $B^0_s$ mass. The estimated background rate is below 2 Hz and nearly 50 signal events are expected with 10 fb$^{-1}$.

The determination of $\Delta m_s$ and $\Delta\Gamma_s$ will be a valuable input for flavour dynamics in the Standard Model and its possible extensions. The measurement of $\Delta m_s$ is allowed by the $B^0_s \to D_s^-\pi^+$ decay followed by $D_s^- \to \phi\pi^-$ and $\phi \to K^+K^-$. The $B^0_s$ CP state at decay time is tagged by the charge of the pion associated to the $D_s$ (in this case the $\pi^+$). The only way to trigger on these hadronic events is to search for the muon coming from the decay of the other b quark in the event. In addition to the single muon Level-1 trigger, it was studied the possibility of a combined trigger with a low-$p_T$ muon and a soft jet. The CMS High-Level trigger algorithm reconstructs the charged particle tracks with only three points by using the precise pixel detector. Topological and kinematical cuts are applied to reconstruct the three resonances $\phi$, $D_s$ and $B^0_s$. A 20 Hz output rate is achieved with about 1000 signal events in 20 fb$^{-1}$. Since the overall possible rate on tape is 100 Hz, the bandwidth allocated to this channel probably could not exceed 5 Hz. If the fraction of events written to tape is scaled accordingly, more than 300 signal events are expected for 20 fb$^{-1}$. In order to fully cover the range allowed by the Standard Model, about 1000 events are needed.

The decay channel $B^0_s \to J/\psi\phi$ is very important because it can not be studied with large accuracy before LHC and can reveal hints for physics beyond the Standard Model. Events with a couple of muons are passed to the HLT. The inclusive selection of $J/\psi \to \mu^+\mu^-$ decays, obtained with mass requirements on the di-muon system, leads to a total of 15 Hz rate, 90% of which is made of $J/\psi$ from b quarks. With an additional amount of CPU time, perhaps sustainable by the HLT computing power, about 170 000 events are expected in 20 fb$^{-1}$ with less than 2 Hz rate.

### 3.3 LHCb beauty trigger

*author: H. Ruiz*

The LHCb detector [5] is optimized for exploiting the B-physics potential of LHC. Together with excellent vertexing and particle identification, an efficient trigger on a wide variety of B decays is one of the main design requirements of the experiment.

The LHCb trigger system [12] is organized in three levels. The first one (L0) runs on custom electronics and operates synchronously at 40 MHz, with a 4 $\mu$s latency. The remaining two trigger levels (L1 and HLT) run on a shared farm of 1400 commercial CPUs. A brief description of the three trigger levels and their performance follows.

L0 exploits the relatively high $p_T$ of B decay products. High $p_T$ candidates are identified both in the calorimeter and in the muon system, with $p_T$ thresholds of about 3 and 1 GeV respectively. Complicated events that would consume unreasonable time at higher levels are promptly vetoed in two different ways. First, multiple primary vertex topologies are rejected by using two dedicated silicon layers of the vertex detector. Secondly, events with large multiplicity, measured at a scintillating pad layer, are vetoed. The input rate of events visible in the detector is about 10 MHz, with a $b\bar{b}$ content of 1.5%. L0 reduces this rate by a factor of 10 while increasing the $b\bar{b}$ content to 3%. The typical efficiency of L0 is 90% for channels with dimuons, 70% for radiative decays and 50% for purely hadronic decays.

At the 1 MHz input rate of L1 it becomes feasible to use tracking information, allowing for the search for B vertex displacement signatures. Tracks are first searched at the vertex detector, and then confirmed in two dedicated tracking layers (trigger tracker or TT) which provide a rough estimation of the momentum of the tracks ($\delta p_T/p_T \sim 25\%$). The generic L1 decision is based on the presence of two tracks with an impact parameter higher than 0.15 mm with a sufficiently high value of $log(p_{T1} + p_{T2})$. Alternative selection criteria are applied, based on the presence of tracks matched to L0 neutral calorimeter objects and muon candidates. The output rate of L1 is 40 kHz with a $b\bar{b}$ content of 15%. The efficiencies are at the level of 90, 80 and 70% for channels with di-muons, only hadrons and radiative decays respectively.

The HLT [13] consists of two sequential *layers*. The first one refines the L1 decision with the all the tracking information from the detector, improving the $p_T$ measurement to the level of $\delta p_T/p_T \sim 1\%$. The rate is reduced to 13 kHz and the $b\bar{b}$ content is enriched to 30%. The second layer consists on a series of alternative selections. A first group of them aims for maximal efficiency on the base-line physics channels and the corresponding control samples, by making use of the complete reconstruction of the decay vertex and its kinematical properties. These selections fill 200 Hz of bandwidth, while providing efficiencies typically higher than 90%. The rest of selections aim for more generic signatures that will provide robustness and flexibility to the trigger system. In addition, the samples selected will be useful for calibration and systematic studies. The selections aim for generic J/$\psi$ and D$^\star$ (600 Hz and 300 Hz respectively), and generic B decays (900 Hz). The latter is based on the detection of single muons with high $p_T$ and impact parameter.

In total, 2 kHz of events will be written on tape, with an expected overall efficiency ranging between 75% for channels with di-muons to 35% for purely hadronic final states.

## 4  Measurements in preparation at the LHC and expected performance

In the following we present, as examples, the expected performance for the detection of D and B mesons in ALICE[2], and for the study of $b\bar{b}$ azimuthal correlation in ATLAS. We also include a summary of the capability of ALICE of quarkonia measurements ($\psi$ family and $\Upsilon$ family).

### 4.1  Charm reconstruction in ALICE

*author: A. Dainese*

One of the most promising channels for open charm detection is the $D^0 \to K^-\pi^+$ decay (and charge conjugate) that has a BR of 3.8%. The expected yields (BR$\times$d$N$/d$y$ at $y = 0$), in pp collisions at $\sqrt{s} = 14$ TeV and in central Pb–Pb (0–5% $\sigma^{\rm tot}$) at $\sqrt{s_{\rm NN}} = 5.5$ TeV are $7.5 \times 10^{-4}$ and $5.3 \times 10^{-1}$ per event, respectively [14].

The main feature of this decay topology is the presence of two tracks with impact parameters $d_0 \sim 100$ $\mu$m. The detection strategy [15] to cope with the large combinatorial background from the underlying event is based on the selection of displaced-vertex topologies, i.e. two tracks with large impact parameters and good alignment between the $D^0$ momentum and flight-line, and on invariant-mass analysis to extract the signal yield. This strategy was optimized separately for pp and Pb–Pb collisions, as a function of the $D^0$ transverse momentum, and statistical and systematic errors were estimated [16, 17]. The results, in terms of $p_T$ coverage and statistical precision, are found to be similar for the two colliding systems [16, 17].

Figure 3 (left) shows the expected sensitivity of ALICE for the measurement of the $D^0$ $p_T$-differential cross section in pp collisions, along with NLO pQCD [18] calculation results corresponding to different choices of the charm quark mass and of renormalization and factorization scales. In the right-hand panel of the figure we present the ratio 'data/theory' ('default parameters/theory parameters') which better allows to compare the different $p_T$-shapes obtained by changing the input 'theory parameters' and to illustrate the expected sensitivity of the ALICE measurement. The estimated experimental errors are much smaller than the theoretical uncertainty band. We note that the data cover the region at low transverse momentum where the accuracy of the pQCD calculation becomes poorer and where novel effects, determined by the high partonic density of the initial state at LHC energies, may play an important role (see "Small-$x$ effects in heavy quark production" section of this report).

---

[2]Given that ALICE is dedicated to the study of nucleus–nucleus collisions at the LHC, some of the presented results are relative to Pb–Pb collisions at $\sqrt{s} = 5.5$ TeV per nucleon–nucleon collisions. These results can be taken as lower limits for the performance in pp collisions, where the background contributions are much lower.

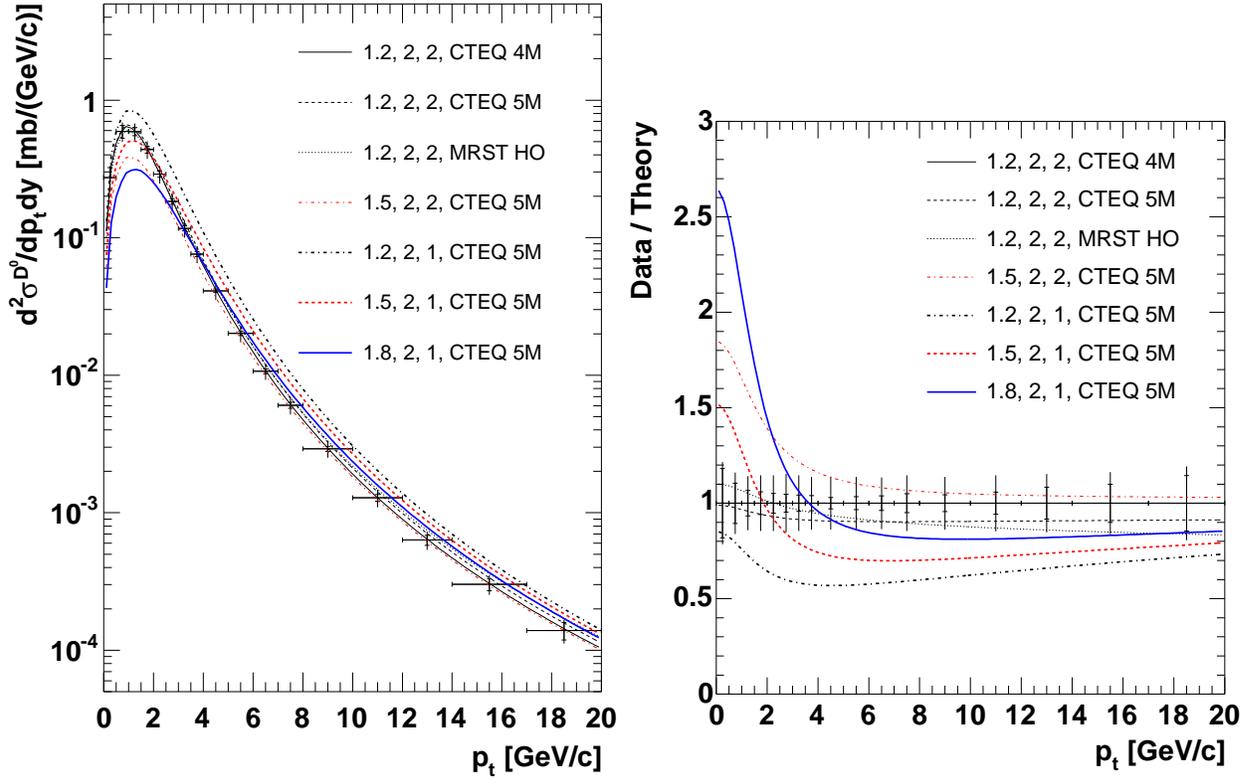

Fig. 3: ALICE sensitivity on $\mathrm{d}^2\sigma^{D^0}/\mathrm{d}p_T\mathrm{d}y$, in pp at 14 TeV, compared to the pQCD predictions obtained with different sets of the input parameters $m_c$ [GeV], $\mu_F/\mu_0$, $\mu_R/\mu_0$ and PDF set ($\mu_0$ is defined in the text). The inner bars represent the statistical error, the outer bars the quadratic sum of statistical and $p_T$-dependent systematic errors. A normalization error of 5% is not shown. The panel on the right shows the corresponding 'data/theory' plot.

### 4.2 Beauty production measurements in ALICE

*author: A. Dainese*

The expected yields (BR $\times \mathrm{d}N/\mathrm{d}y$ at $y = 0$) for $B \to e^\pm + X$ plus $B \to D(\to e^\pm + X) + X'$ in pp collisions at $\sqrt{s} = 14$ TeV and in central Pb–Pb (0–5% $\sigma^{\mathrm{tot}}$) at $\sqrt{s_{\mathrm{NN}}} = 5.5$ TeV are $2.8 \times 10^{-4}$ and $1.8 \times 10^{-1}$ per event, respectively [14].

The main sources of background electrons are: (a) decays of D mesons; (b) neutral pion Dalitz decays $\pi^0 \to \gamma e^+ e^-$ and decays of light mesons (e.g. $\rho$ and $\omega$); (c) conversions of photons in the beam pipe or in the inner detector layers and (d) pions misidentified as electrons. Given that electrons from beauty have average impact parameter $d_0 \simeq 500$ $\mu$m and a hard momentum spectrum, it is possible to obtain a high-purity sample with a strategy that relies on: electron identification with a combined $\mathrm{d}E/\mathrm{d}x$ and transition radiation selection, which allows to reduce the pion contamination by a factor $10^4$; impact parameter cut to reject misidentified pions and electrons from sources (b) and (c); transverse momentum cut to reject electrons from charm decays.

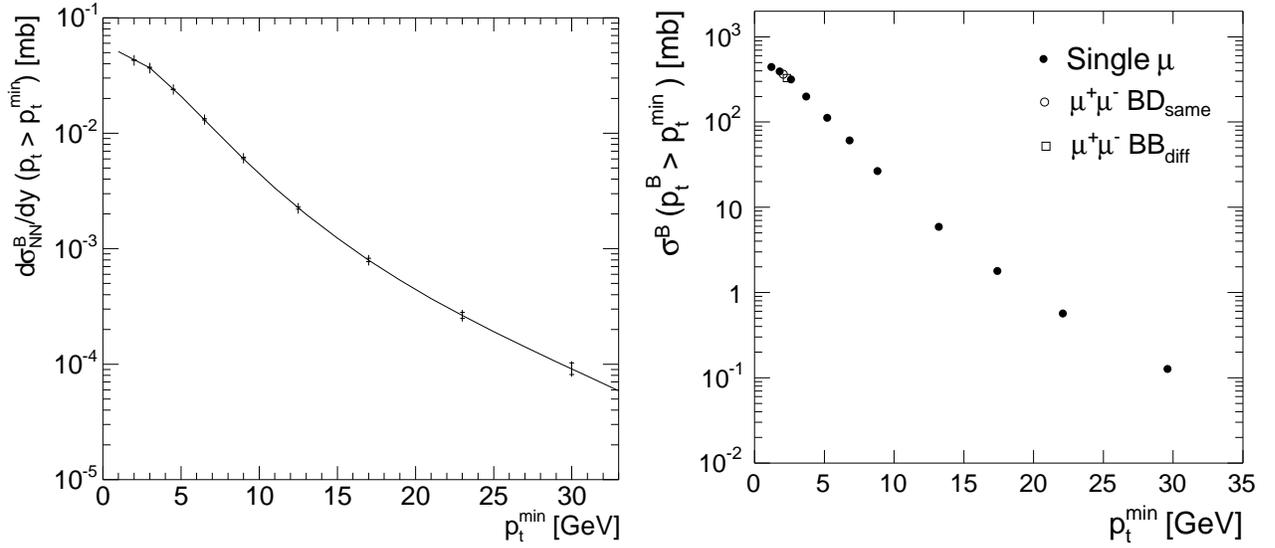

Fig. 4: B production cross section vs. $p_T^{\min}$ reconstructed by ALICE in $10^7$ central Pb–Pb events. Left: $\mathrm{d}\sigma_{\mathrm{NN}}^{\mathrm{B}}/\mathrm{d}y$ at $y=0$ (normalized to one nucleon–nucleon collision) from single electrons in $|\eta|<0.9$; statistical (inner bars) and quadratic sum of statistical and $p_T$-dependent systematic errors (outer bars) are shown; a 9% normalization error is not shown. Right: $\sigma^{\mathrm{B}}$ integrated in $2.5<y^{\mathrm{B}}<4$ (not normalized to one nucleon–nucleon collision) from single muons and di-muons in $2.5<\eta<4$; only (very small) statistical errors shown.

As an example, with $d_0 > 200$ $\mu$m and $p_T > 2$ GeV, the expected statistics of electrons from B decays is $8\times 10^4$ for $10^7$ central Pb–Pb events, allowing the measurement of electron-level $p_T$-differential cross section in the range $2 < p_T < 18$ GeV. The residual contamination of about 10%, acculated in the low-$p_T$ region, of electrons from prompt charm decays and from misidentified charged pions can be evaluated and subtracted using a Monte Carlo simulation tuned to reproduce the measured cross sections for pions and $D^0$ mesons. A Monte-Carlo-based procedure can then be used to compute, from the electron-level cross section, the B-level cross section $\mathrm{d}\sigma^{\mathrm{B}}(p_T > p_T^{\min})/\mathrm{d}y$ [17]. In the left-hand panel of Fig. 4 we show this cross section for central Pb–Pb collisions with the estimated uncertainties. The covered range is $2 < p_T^{\min} < 30$ GeV.

B production can be measured also in the ALICE forward muon spectrometer, $2.5 < \eta < 4$, analyzing the single-muon $p_T$ distribution and the opposite-sign di-muons invariant mass distribution [17].

The main backgrounds to the 'beauty muon' signal are $\pi^\pm$, $K^\pm$ and charm decays. The cut $p_T > 1.5$ GeV is applied to all reconstructed muons in order to increase the signal-to-background ratio. For the opposite-sign di-muons, the residual combinatorial background is subtracted using the technique of event-mixing and the resulting distribution is subdivided into two samples: the low-mass region, $M_{\mu^+\mu^-} < 5$ GeV, dominated by muons originating from a single b quark decay through $\mathrm{b} \to \mathrm{c}(\to \mu^+)\mu^-$ (BD$_{\mathrm{same}}$), and the high-mass region, $5 < M_{\mu^+\mu^-} < 20$ GeV, dominated by $\mathrm{b}\bar{\mathrm{b}} \to \mu^+\mu^-$, with each muon coming from a different quark in the pair (BB$_{\mathrm{diff}}$). Both samples have a background from $\mathrm{c}\bar{\mathrm{c}} \to \mu^+\mu^-$ and a fit is done to extract the charm- and beauty-component yields. The single-muon $p_T$ distribution has three components with different slopes: K and $\pi$, charm, and beauty decays. Also in this case a fit technique allows to extract a $p_T$ distribution of muons from B decays. A Monte Carlo procedure, similar to that used for semi-electronic decays, allows to extract B-level cross sections for the data sets (low-mass $\mu^+\mu^-$, high-mass $\mu^+\mu^-$, and $p_T$-binned single-muon distribution), each set covering a specific B-meson $p_T > p_T^{\min}$ region, as preliminarily shown in Fig. 4 (right). Since only minimal cuts are applied, the reported statistical errors are very small and high-$p_T$ reach is excellent. Systematic errors are currently under study.

### 4.3 Study of $\mathrm{b}\bar{\mathrm{b}}$ correlations in ATLAS

*author: Th. Lagouri*

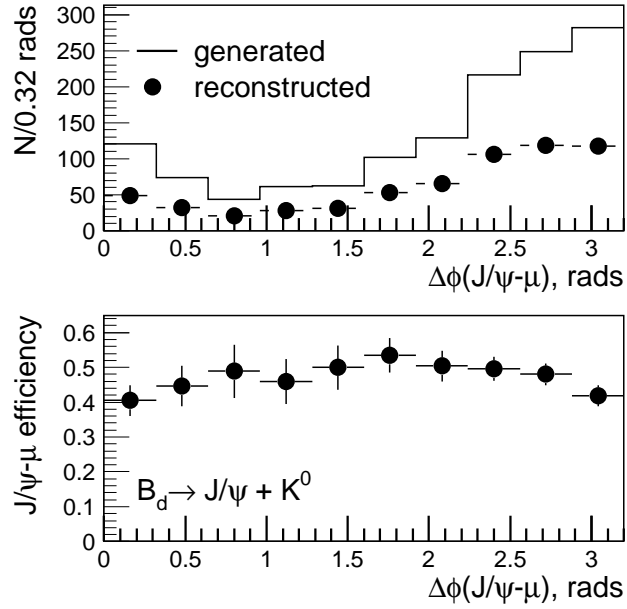

Fig. 5: Distribution of the opening angle $\Delta\phi(J/\psi\text{-}\mu)$ between the $J/\psi$ from the decay $B_d \to J/\psi K^0$ and the muon coming from the associated B hadron decay to muon, both direct $b \to \mu$ and indirect $b \to c \to \mu$.

The ATLAS detector [19] is well engineered for studies of b-production, and together with the huge rate of b-quark production that will be seen at LHC, offers great potential for the making of novel precise b production measurements. Correlations between b and $\overline{b}$ quarks and events with more than one heavy-quark pair, $b\overline{b}b\overline{b}$, $b\overline{b}c\overline{c}$, $b\overline{b}s\overline{s}$, that were difficult to access in previous experiments due to limited statistics, will be investigated in detail. A new technique has been developed in ATLAS for measuring correlations, and this will yield results that will shed new light on our understanding of the QCD cross-section for $b\overline{b}$-production.

A detailed study investigated a possibility of $b\overline{b}$ correlations measurement using the $\Delta\phi(J/\psi\text{-}\mu)$ distribution, the azimuthal separation of a $J/\psi$ and a muon [20, 21, 22]. This technique is expected to be superior to earlier methods used at the Tevatron Run-1 based on muon–muon or muon–b-jet correlations. The new method does not require separation cuts between the two objects. Such cuts were necessary to control the background, but they required a model-dependent extrapolation of the results to full azimuthal space [23]. Using a full simulation of the Inner Detector and the Muon Spectrometer of the ATLAS detector [19] it is shown that such a distribution can be extracted from heavy flavour events at LHC.

ATLAS studies were done for two channels selected to measure the azimuthal angle difference $\Delta\phi(b\overline{b})$ between b and $\overline{b}$ quarks:

$\overline{b} \to B_d \to J/\psi(\to \mu\mu)K^0$, $b \to \mu + X$
$\overline{b} \to B_s \to J/\psi(\to \mu\mu)\phi$, $b \to \mu + X$

The numbers of events expected for 30 fb$^{-1}$ as might be achieved after 3 years of running at a luminosity of $10^{33}$ cm$^2$s$^{-1}$ are $4.8 \times 10^4$ and $3.2 \times 10^4$ respectively for these channels. No isolation cuts are needed to separate exclusively reconstructed B-decays from the muon produced in the semi-leptonic decay of the other B-particle in the event. The reconstruction efficiency remains high in topologies where the azimuthal angle difference $\Delta\phi(J/\psi\text{-}\mu)$ between $J/\psi$ and the muon is small.

Special attention was devoted to background events in which the muon is produced from the decays $K^\pm$, $\pi^\pm \to \mu^\pm + X$ instead of $b \to \mu + X$. The study showed that this background is not problematic in $B_d$ decays, however it is important in the case of $B_s^0$ meson.

In summary, the results of the analysis suggest that backgrounds from $K/\pi$ decays are small, and that backgrounds from events containing 4 b quarks are relatively flat in $\Delta\phi(J/\psi\text{-}\mu)$. The efficiency of the recon-struction of muons with this technique is also relatively flat in $\Delta\phi(J/\psi\text{-}\mu)$ and so we conclude that corrections to the measured $\Delta\phi(J/\psi\text{-}\mu)$ distribution are likely to be small.

## 4.4 Quarkonia measurements in ALICE
*author: A. Morsch*

Heavy quarkonia states are hard, penetrating probes which provide an essential tool to study the earliest and hottest stages of heavy-ion collisions [24]. They can probe the strongly interacting matter created in these reactions on short distance scales and are expected to be sensitive to the nature of the medium, i.e. confined or de-confined [25, 26]. The suppression (dissociation) of the heavy-quark resonances is considered as one of the most important observables for the study of the QGP at the LHC (see Ref. [2] for a recent review).

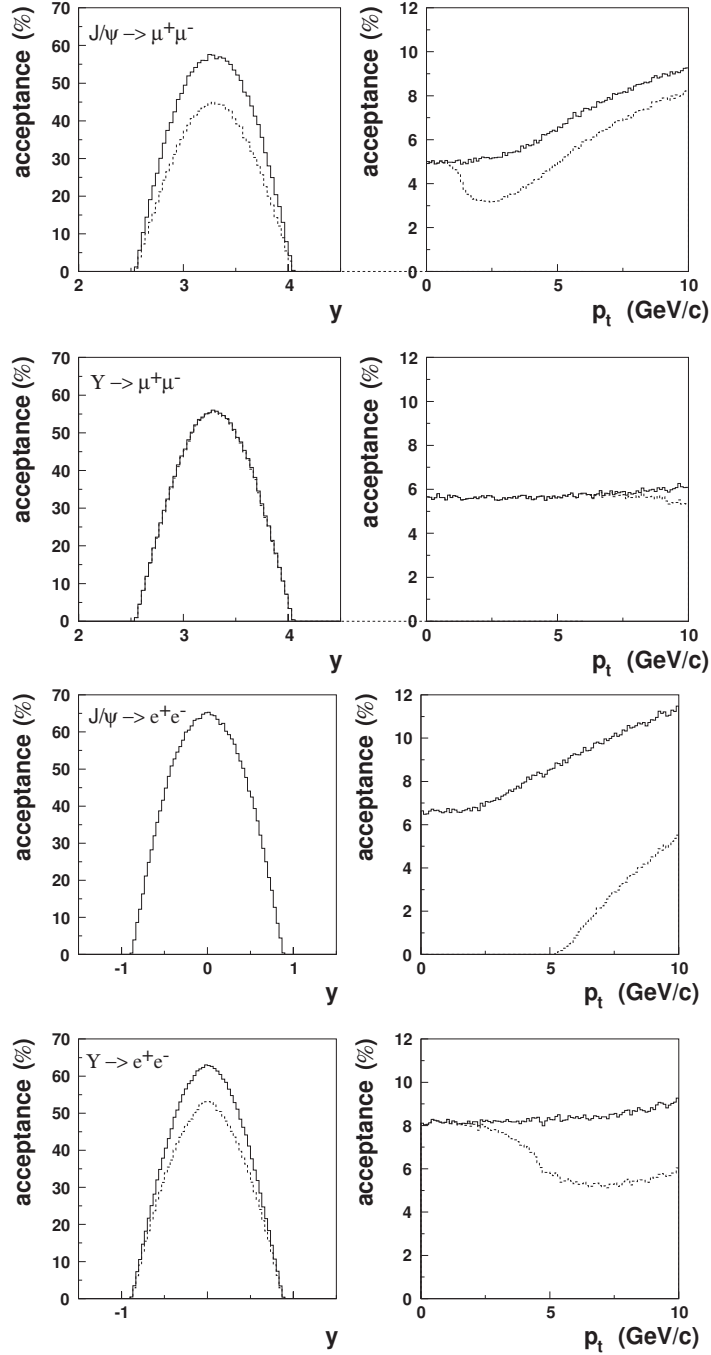

Fig. 6: Acceptance for J/$\psi$ and $\Upsilon$ as a function of $y$ and $p_T$ for measurements in the di-muon channel and di-electron channels. To give an idea of the effect of the trigger, the acceptances are shown without (solid) and with (dashed) a sharp cut on the transverse momentum of single muons of 1 GeV/$c$ (2 GeV/$c$) for J/$\psi$ ($\Upsilon$) and for single electrons of 3 GeV.

In ALICE, quarkonia will be measured in the di-electron channel using a barrel ($|\eta| < 0.9$) Transition Radiation Detector (TRD) [2] and in the di-muon channel using a forward Muon Spectrometer ($2.5 < \eta < 4$) [2]. The complete spectrum of heavy-quark vector mesons (J/$\psi$, $\psi'$, $\Upsilon$, $\Upsilon'$, $\Upsilon''$) can be measured down to zero $p_T$. In particular the good mass resolution allows to resolve the Upsilon family.

The Muon Spectrometer uses a low-$p_T$ trigger threshold, $p_T > 1$ GeV, on single muons for charmonia and a high-$p_T$ trigger, $p_T > 2$ GeV, for bottonia detection. The TRD can trigger on single electrons with $p_T > 3$ GeV, which results in a minimum transverse momentum of triggered charmonia of 5.2 GeV. Electron identification combined with the excellent vertexing capabilities of the inner tracking system allows ALICE to distinguish direct charmonium production from secondary charmonium production through B decays.

The energy density dependence will be studied by varying the impact parameters and by studying in addition to the heaviest collision system (Pb–Pb) also intermediate mass and low mass A–A systems. To determine the primary production cross-section of the resonances and the amount of pre-resonance absorption, corresponding measurement have to be performed for pA and pp collisions.

Table 1: Main characteristics of quarkonia detection with the TRD and the Muon Spectrometer in ALICE.

|  | Muon Spectrometer | TRD |
|---|---|---|
| Acceptance | $2.5 < \eta < 4$ | $|\eta| < 0.9$ |
| Mass Resolution J/$\psi$ | 72 MeV | 34 MeV |
| Mass Resolution $\Upsilon$ | 99 MeV | 93 MeV |

Table 1 shows the main quarkonia detection characteristics of the TRD and the Muon Spectrometer and the acceptances for J/$\psi$ and $\Upsilon$ as a function of $y$ and $p_T$ are shown in Fig. 4.4. In one year of pp running at $\langle L \rangle = 3 \times 10^{30}$ cm$^{-2}$s$^{-1}$ ALICE will detect several $10^6$ J/$\psi$s and several $10^4$ $\Upsilon$s in the di-muon channel. A $\Upsilon$ statistics of $10^2$–$10^3$ can be obtained in the di-electron channel. For the J/$\psi$ we expect $\approx 10^4$ untriggered low-$p_T$ and $\approx 10^4$ high-$p_T$ triggered events.

# Small-$x$ effects in heavy quark production

*A. Dainese[a], K. J. Eskola[b], H. Jung[c], V. J. Kolhinen[b], K. Kutak[d,e], A. D. Martin[f], L. Motyka[c], K. Peters[c,g], M. G. Ryskin[h], and R. Vogt[i,j]*

[a] University of Padova and INFN, Padova, Italy
[b] Department of Physics, University of Jyväskylä, Jyväskylä, Finland
[c] Deutsches Elektronen-Synchroton DESY, Hamburg, FRG
[d] University of Hamburg, Hamburg, FRG
[e] Henryk Niewodniczanski Institute of Nuclear Physics, Polish Academy of Sciences, Poland
[f] Institute for Particle Physics Phenomenology, University of Durham, Durham, UK
[g] Department of Physics and Astronomy, University of Manchester, UK
[h] Petersburg Nuclear Physics Institute, Gatchina, St. Petersburg, Russia
[i] Lawrence Berkeley National Laboratory, Berkeley, CA, USA
[j] Physics Department, University of California, Davis, CA, USA

**Abstract**

We study small-$x$ effects on heavy flavor production at the LHC in two approaches including nonlinear, saturation-motivated, terms in the parton evolution. One approach is based on collinear factorization, the other on $k_T$ factorization. The prospects for direct experimental study of these effects in $pp$ collisions at the LHC are discussed.

*coordinators: A. Dainese, H. Jung, and R. Vogt*

## 1 Introduction

HERA data are used to constrain the small $x$, moderate $Q^2$ parton densities in two approaches. In the first, HERA $F_2$ data are refit using DGLAP evolution with the first nonlinear recombination terms. Recombination slows the evolution so that, after refitting the data, the gluon distribution is enhanced relative to that obtained by DGLAP alone. The resulting set of parton densities produces charm enhancement in $pp$ collisions at the LHC. On the other hand, assuming $k_T$ factorization, the unintegrated gluon distribution is determined from the HERA $F_2^c$ data, the only inclusive HERA measurements which directly accesses the gluon density. Saturation effects are then included, reducing the small $x$ gluon densities with little distinguishable effect on $F_2$. This approach leads instead to heavy flavor suppression at the LHC. After a short general introduction, both approaches and their predicted effects on heavy quark production are discussed in detail. Direct experimental study of these effects in $pp$ collisions at the LHC may be able to differentiate between the two approaches.

## 2 Small-$x$ partons, absorption and the LHC

*authors: A.D. Martin and M.G. Ryskin*

### 2.1 Partons densities at low $x$?

Almost nothing is known about the behaviour of partons at low $x$. There are essentially no data available for $x < 10^{-4}$ with $Q^2$ in the perturbative region and there is no reliable theory to extrapolate down in $x$.

In the Dokshitzer-Gribov-Lipatov-Altarelli-Parisi (DGLAP)-based [1, 2, 3, 4] global analyses, the small-$x$ behaviour is driven by input distributions at a starting scale $Q = Q_0$. Usually these 'input' distributions are written in the form $x^{-\lambda}(1-x)^\eta$ where $\lambda$ and $\eta$ are free parameters fit to the data. So one can say nothing without data in the $x$ region of interest. Moreover, there may be large low-$x$ contributions to the gluon of the form $(\alpha_s \ln(1/x))^n$ – the so-called Balitsky-Fadin-Kuraev-Lipatov (BFKL) effects [5, 6, 7, 8], beyond the DGLAP approximation.

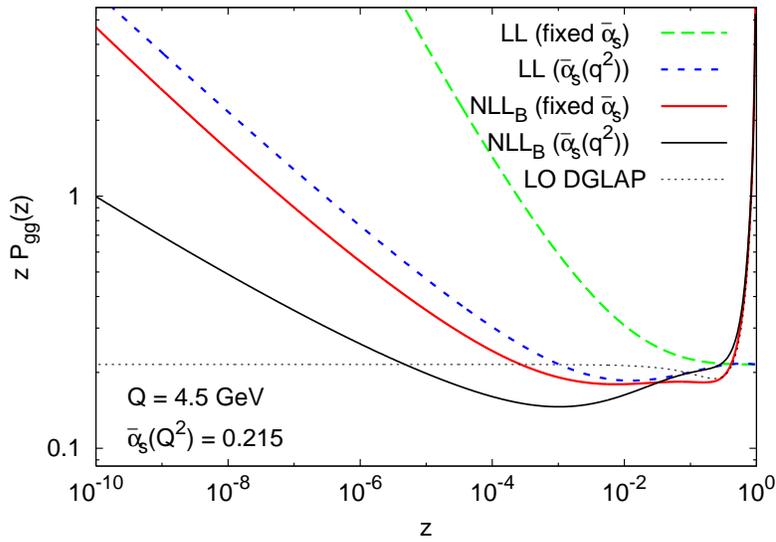

Fig. 1: The gluon-gluon splitting function, $P_{gg}$, with fixed and running coupling in the LL and resummed NLL BFKL approximations, compared with the LO DGLAP behaviour. The figure is taken from Ciafaloni et al. [9, 10, 11, 12, 13, 14]. The subscript B refers to scheme B which ensures energy-momentum conservation in the splitting.

Thus it seems better to discuss low-$x$ behaviour in terms of BFKL-evolution. However there are also problems here. The next-to-leading logarithm (NLL) corrections to the leading order (LO) BFKL (CCFM) amplitude are known to be very large and one needs to resum such corrections to obtain a relatively stable result. We cannot justify the perturbative QCD approach at low $Q^2$ so that the solution of the BFKL equation need to be matched to some non-perturbative amplitude at $Q = Q_0$. This non-perturbative distribution (analogous to the 'input' in the DGLAP case) is not known theoretically. Either it has to be fit to low $x$ data or some phenomenological model such as a Regge-based parametrization has to be used.

After a reasonable resummation of the NLL corrections in the region where the starting virtuality $Q_0$ is not close to the final value of $Q$, $Q > Q_0$, the resummed BFKL amplitude turns out to be similar to that resulting from DGLAP evolution [9, 10, 11, 12, 13, 14]. For example, the preasymptotic effects on the resummation of the gluon-gluon splitting function are so large that the NLL BFKL power growth only sets in for $z < 10^{-5}$, as can be seen from Fig. 1. Moreover, the behaviour of the convolution $P_{gg} \otimes g/g$, normalized to $g$, in the NNLO DGLAP and NLL approximations is exactly the same down to $z \sim 10^{-4}$ [15].

Thus, in practice, the DGLAP and BFKL based approaches are rather close to each other in the HERA kinematic regime. In both cases, the main problem is the low-$x$ behaviour of the amplitude at $Q = Q_0$ where we need to phenomenologically determine possible non-perturbative contributions, power corrections and so on.

## 2.2 The puzzle of the $x^{-\lambda}$ behaviour

Since the BFKL amplitude grows as a power of $x$, $A \propto x^{-\lambda}$, it will violate unitarity as $x \to 0$. Indeed, even after the NLL resummation, the expected power, $\lambda \simeq 0.3$, is rather large. Thus, we first discuss absorption effects which tame the violation of unitarity. The upper limit of the small $x$ behaviour of the parton distributions $a = g, q$ is given by the extrapolation

$$xa(x, q^2) = \left(\frac{x_0}{x}\right)^{0.3} x_0 a(x_0, q^2) \qquad (1)$$

below $x_0 = 0.001$. The distributions are reliably determined from global parton analyses at $x > x_0$.

On the other hand, it is reasonable to expect that at $Q \lesssim Q_0 \sim 1$ GeV the behaviour will reflect that of hadron-hadron interactions: $\lambda = 0.08$ [16]. Most likely the lower value of $\lambda$ is explained by absorptive/screening effects in hadron-hadron collisions. So, for extrapolation down to $x \sim 10^{-7} - 10^{-6}$ we may regard $\lambda = 0.08$ as a lower limit since, in DIS, we expect smaller absorptive effects than those in hadron-hadron interactions.

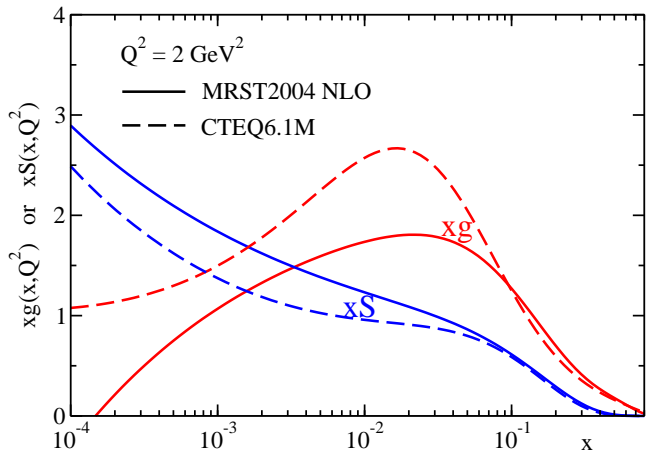

Fig. 2: The behaviour of the gluon and sea quark distributions at $Q^2 = 2$ GeV$^2$ found in the CTEQ6.1M [17] and MRST2004 NLO [18] global analyses. The valence-like behaviour of the gluon is evident.

However, present global analyses, which do not allow for absorption effects, reveal that at $Q \sim 1 - 1.5$ GeV and low $x$, the sea quarks have a Pomeron-like behaviour, $xq \sim x^{-0.2}$, whereas the gluon distribution has a valence-like behaviour, $xg \sim \sqrt{x}$. This different behaviour is evident from Fig. 2, which shows the behaviour of the gluon and sea quark distributions, $xS = 2x(\bar{u} + \bar{d} + \bar{s})$ for $Q^2 = 2$ GeV$^2$. Such a result looks strange from the Regge viewpoint where the same vacuum singularity (Pomeron) should drive both the sea quarks and the gluons since the same power is expected for sea quarks and gluons, $\lambda_g = \lambda_q$.

This difference demonstrates that the actual situation is even more complicated. It is worth noting that a simultaneous analysis of inclusive and diffractive DIS data indicates that, after accounting for screening effects and allowing for some power corrections, it may be possible to describe the HERA data with $\lambda_g = \lambda_q = 0$ [19]. The absorptive effects, estimated from fitting the diffractive DIS data, enlarge the input gluon distribution at low $x$.

It may initially seem strange that accounting for absorptive effects gives a *larger* gluon density[1] at low $x$ and $Q^2$. The point is that the only way to describe the data, *which are sensitive to absorptive effects*, within the framework of DGLAP evolution without absorption, is to choose a very low 'input' gluon density in order to *mimic* the screening corrections 'hidden' in the data. Indeed, there is a tendency for the gluon distribution to even become negative at low $x$ and $Q^2$. On the other hand, allowing for absorption during DGLAP evolution (with the help of the Gribov-Levin-Ryskin (GLR) and Mueller-Qiu (MQ), GLRMQ, equations [22, 23]) the same data may be described with a larger and definitely positive input gluon density at $Q = Q_0$.

### 2.3 Estimates of absorptive effects: GLRMQ to BK

The saturation of parton densities ($\lambda = 0$) may be obtained using the Balitski-Kovchegov (BK) [24, 25] equation, based on the BFKL equation, as well as the aforementioned GLRMQ equations. The latter equation is based on DGLAP evolution. These equations sum the set of fan diagrams which describe the rescattering of intermediate partons on the target nucleon. The screening caused by these rescatterings prohibits the power growth of the parton densities.

The GLR equation for the gluon may be written symbolically as

$$\frac{\partial xg}{\partial \ln Q^2} = P_{gg} \otimes g + P_{gq} \otimes q - \frac{81\alpha_s^2}{16R^2 Q^2} \int \frac{dy}{y} [y\, g(y, Q^2)]^2 . \tag{2}$$

---

[1]The same result was obtained in Ref. [20, 21] – note, however, it was based on LO evolution and the large NLO correction to $P_{qg}$ will change the $q, g$ relationship.

The nonlinear shadowing term, $\propto -[g]^2$, arises from perturbative QCD diagrams which couple four gluons to two gluons so that two gluon ladders recombine into a single gluon ladder. The minus sign occurs because the scattering amplitude corresponding to a gluon ladder is predominantly imaginary. The parameter $R$ is a measure of the transverse area $\pi R^2$ where the gluon density is sufficiently dense for recombination to occur.

The BK equation is an improved version of the GLR equation. It accounts for the more precise triple-pomeron vertex (first calculated in Ref. [26, 27, 28]) and can be used for the non-forward amplitude. The GLR equation was in momentum space, whereas the BK equation is written in coordinate space in terms of the dipole scattering amplitude $N(\mathbf{x}, \mathbf{y}, Y) \equiv N_{\mathbf{xy}}(Y)$. Here $\mathbf{x}$ and $\mathbf{y}$ are the transverse coordinates of the two $t$-channel gluons which form the colour-singlet dipole and $Y = \ln(1/x)$ is the rapidity. The BK equation reads

$$\frac{\partial N_{\mathbf{xy}}}{\partial Y} = \frac{\alpha_s N_c}{\pi} \int \frac{d^2\mathbf{z}}{2\pi} \frac{(\mathbf{x}-\mathbf{y})^2}{(\mathbf{x}-\mathbf{z})^2(\mathbf{y}-\mathbf{z})^2} \{N_{\mathbf{xz}} + N_{\mathbf{yz}} - N_{\mathbf{xy}} - N_{\mathbf{xz}} N_{\mathbf{yz}}\} \ . \tag{3}$$

For small dipole densities, $N$, the quadratic term in the brackets may be neglected and Eq. (3) reproduces the conventional BFKL equation. However for large $N$, that is $N \to 1$, the right-hand side of Eq. (3) vanishes and we reach saturation when $N = 1$. The equation sums up the set of fan diagrams where at small $Y$ the target emits any number of pomerons (i.e. linear BFKL amplitudes) while at large $Y$ we have only one BFKL dipole.

Starting from the same initial conditions, the solution of the BK equation gives *fewer* small-$x$ partons than that predicted by its *parent* linear BFKL/CCFM equation[2].

In principle, it would appear more appropriate to use the BFKL-based BK equation to describe the parton densities at low $x$. Unfortunately, however, the BK equation is only a model. It cannot be used for numerical predictions. We discuss the reasons below.

### 2.4 Status of the BK equation

The Balitski-Kovchegov (BK) equation [24, 25] is an attempt to describe saturation phenomena. However it is just a 'toy model' and cannot, at present, be used to reliably estimate absorptive effects at small $x$. The reasons are as follows:

- The BK equation is based on the LO BFKL/CCFM equation. We know that the NLL corrections are large. We need to know the NLL corrections not only for the linear part of the evolution, but also for the nonlinear term.
- Even neglecting the NLL corrections, we need to match the solution to a boundary condition at rather low $Q^2$. This boundary condition is not theoretically known.
- It sums a limited set of diagrams. The selection of diagrams (the fan graphs) was justified in the region where absorptive effects are relatively small. When these corrections become important, as in the saturation region, one has to allow for many other graphs whose contributions become comparable to the fan diagram contributions[3].
- To solve the BK equation we need an initial condition at fixed $x$ and all $Q^2$. These conditions are not well enough known. In particular, the maximum (saturation) value of the gluon density depends on the radius: $xg(x, q^2) \propto R^2 q^2$. At the moment, the radius $R$ is a free parameter. It may be small — the so-called 'hot spot' scenario. Moreover, we should account for the possibility of dissociation of the target proton[4]. The contribution coming from the dissociation is expected to have a smaller $R$.

### 2.5 Relevance to, and of, the LHC

How do the uncertainties at low $x$ affect the predictions for the LHC? Fortunately for inclusive production of possible massive new particles with $M \gtrsim 100$ GeV, the partons are sampled at $x$ values and scales $M$ reliably determined from NLO and NNLO global analyses. For illustration, we discuss $W$ production which has been studied in detail [29, 30, 31]. Central $W$ production ($y_W = 0$) at the LHC samples partons at $x = M_W/\sqrt{s} = 0.006$. However to predict the total cross section, $\sigma_W$, we need to integrate over rapidity, important for $|y_W| \lesssim 4$ so that $\sigma_W$ has some sensitivity to partons as low as $x \sim 10^{-4}$. The total uncertainty on the NNLO prediction of $\sigma_W$ has been estimated to be $\pm 4\%$ [29]. Therefore $W$ production at the LHC can serve as a good luminosity monitor. To reduce the uncertainty in the prediction of $\sigma_W$ will require a better theoretical understanding of low $x$ partons.

---

[2]Analogously, starting from the same input (and not fitting the input to the data) the GLR equation gives fewer small-$x$ partons than that predicted by the parent linear DGLAP equation.

[3]Unfortunately the problem of summing all relevant diagrams has not been solved, even in the simpler case of Reggeon field theory.

[4]We know that these channels provide more than $30 - 40\%$ of $F_2^D$ measured at HERA.

Of course, if the new particles are sufficiently massive, $M \gtrsim 1$ TeV, and produced by gluon fusion, then the uncertainties due to the PDFs will be larger. However, there are situations where the scale is considerably lower such as exclusive double-diffractive Higgs production which depends on the unintegrated gluon at $Q^2 \approx 5$ GeV$^2$ with $x \sim M_H/\sqrt{s} \sim 0.01$. The absorptive effects are also expected to be small here.

Turning the discussion around, is it possible for the LHC experiments to determine the behaviour of partons in the $x$ region below $10^{-4}$ at low scales? One possibility is $\mu^+\mu^-$ Drell-Yan production in which events are observed with the $\mu^+\mu^-$ invariant mass as low as possible and the rapidity as large as possible. For example, for $M_{\mu\mu} = 4$ GeV and $y_{\mu\mu} = 3$, we sample quarks at $x = 1.4 \times 10^{-5}$. This process predominantly samples the sea quark distributions. To study the small $x$ behaviour of the gluon at low scales we may consider $\chi_c$ production, or prompt photon production driven by the subprocess $gq \to \gamma q$.

In practice, rather than $\chi_c$, it may be better to study $pp \to J/\psi X$ as a function of $y_{J/\psi}$. This process is also sensitive to the gluon distribution through the subprocesses $gg \to J/\psi g$, $gg \to \chi \to J/\psi \gamma$. There are also contributions from the subprocesses $gg \to b\bar{b}$ with $b \to J/\psi$, and $q\bar{q} \to J/\psi$. The analysis of such data will be considerably helped by the detailed observations of prompt $J/\psi$ and $J/\psi$ from $b$ in central production at the Tevatron [32]. In fact, the first ever NLO global parton analysis [33] used $J/\psi$ data as a function of rapidity to constrain the gluon distribution.

The LHCb detector covers the rapidity region of $2 < \eta < 5$ [34], and may be able to perform some of the above measurements. There is another possibility. Since LHCb will operate at a luminosity of $2 \times 10^{32}$ cm$^{-2}$s$^{-1}$, there will usually be a single collision per bunch crossing and thus practically no 'pile-up' problems. Installing a forward detector at LHCb would offer the possibility of observing asymmetric events with one very large rapidity gap to probe the region of $x_{I\!P} \leq 10^{-5}$.

## 3  Including nonlinear terms in gluon evolution: the GLRMQ and BK approaches

### 3.1  GLRMQ approach

*authors: K.J. Eskola and V.J. Kolhinen*

The DGLAP [1, 2, 3, 4] evolution equations describe the scale evolution of the parton distribution functions (PDFs) well in the region of large interaction scale, $Q^2 \gtrsim 4$ GeV$^2$ [35, 17, 36]. However, toward small values of $x$ and $Q^2$, the gluon recombination terms start to play an increasingly important role. The inclusion of correction terms which arise from fusion of two gluon ladders leads to nonlinear power corrections to the DGLAP evolution equations. The first of these nonlinear corrections are the GLRMQ terms.

With the GLRMQ corrections, the gluon evolution equation becomes

$$\frac{\partial xg(x,Q^2)}{\partial \ln Q^2} = \frac{\partial xg(x,Q^2)}{\partial \ln Q^2}\bigg|_{\text{DGLAP}} - \frac{9\pi}{2}\frac{\alpha_s^2}{Q^2}\int_x^1 \frac{dy}{y} y^2 G^{(2)}(y,Q^2). \tag{4}$$

We model the two-gluon density in the latter term on the right-hand side as

$$x^2 G^{(2)}(x,Q^2) = \frac{1}{\pi R^2}[xg(x,Q^2)]^2, \tag{5}$$

where $R = 1$ fm is the radius of the proton (we comment further on this later). The corrections to the sea quark distributions are

$$\frac{\partial xq(x,Q^2)}{\partial \ln Q^2} \approx \frac{\partial xq(x,Q^2)}{\partial \ln Q^2}\bigg|_{\text{DGLAP}} - \frac{3\pi}{20}\frac{\alpha_s^2}{Q^2}x^2 G^{(2)}(x,Q^2).$$

We have assumed that the higher-twist gluon term, $G_{HT}$ [23], is negligible.

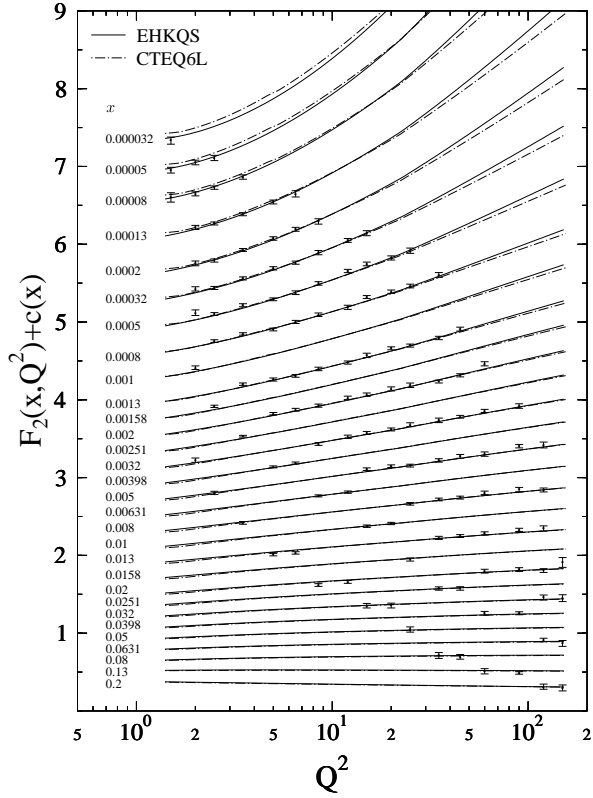

Fig. 3: Calculated $F_2(x,Q^2)$ values compared with the H1 data.

Since these correction terms are negative, they slow down the evolution of the PDFs. Due to the $1/Q^2$ dependence, they also die out in the evolution so that at large scales Eqs. (4) and (6) relax into the linear DGLAP equations.

In order to study the interplay between the nonlinear corrections and the initial PDFs and observe the nonlinear effects in fits to the DIS data, in Ref. [37] we compared the structure function $F_2(x,Q^2) = \sum_q e_q^2[xq(x,Q^2) + x\bar{q}(x,Q^2)]$, calculated with the nonlinearly-evolved PDFs, to the HERA H1 data [38]. As reference distributions we used the CTEQ5L and CTEQ6L PDF sets at large scales. We chose these sets because the CTEQ collaboration uses only the large scale, $Q^2 > 4$ GeV$^2$, data in their fits, thus avoiding some of the possible nonlinear effects appearing in the small $x$, $Q^2 < 4$ GeV$^2$ region in their initial distributions.

At small $x$, sea quarks dominate $F_2$ and the gluon distribution dictates its scale evolution. At leading order (LO), the DGLAP contribution can be approximated as $\partial F_2(x,Q^2)/\partial \ln Q^2 \approx (10\alpha_s/27\pi)xg(2x,Q^2)$ [39]. Larger $xg(x,Q^2)$ values correspond to faster $F_2(x,Q^2)$ evolution. The scale evolution of $F_2(x,Q^2)$ at small $x$ computed with CTEQ5L is too fast due to a large small-$x$ small-$Q^2$ gluon distribution. The newer CTEQ6L set has much smaller gluon distribution in this region (see Fig. 4 (left)), giving a slower evolution and hence a good fit to the H1 data.

Our goal in Ref. [37] was to determine whether the good fit to the data could be maintained using the GLRMQ-corrected DGLAP scale evolution together with initial scale PDFs differing from CTEQ6L. We constructed a new set of PDFs using the CTEQ5L and CTEQ6L distributions piecewise as baselines at scales $Q^2 \sim 3-10$ GeV$^2$ where the linear terms dominate the evolution and evolved them nonlinearly to lower $Q^2$. We then interpolated between the sets in $x$ and assumed a power-like dependence at small-$x$ for gluons and sea quarks. These initial PDF candidates were then evolved to higher scales and compared to the data. This iterative procedure was repeated until a sufficiently good fit to the data was found.

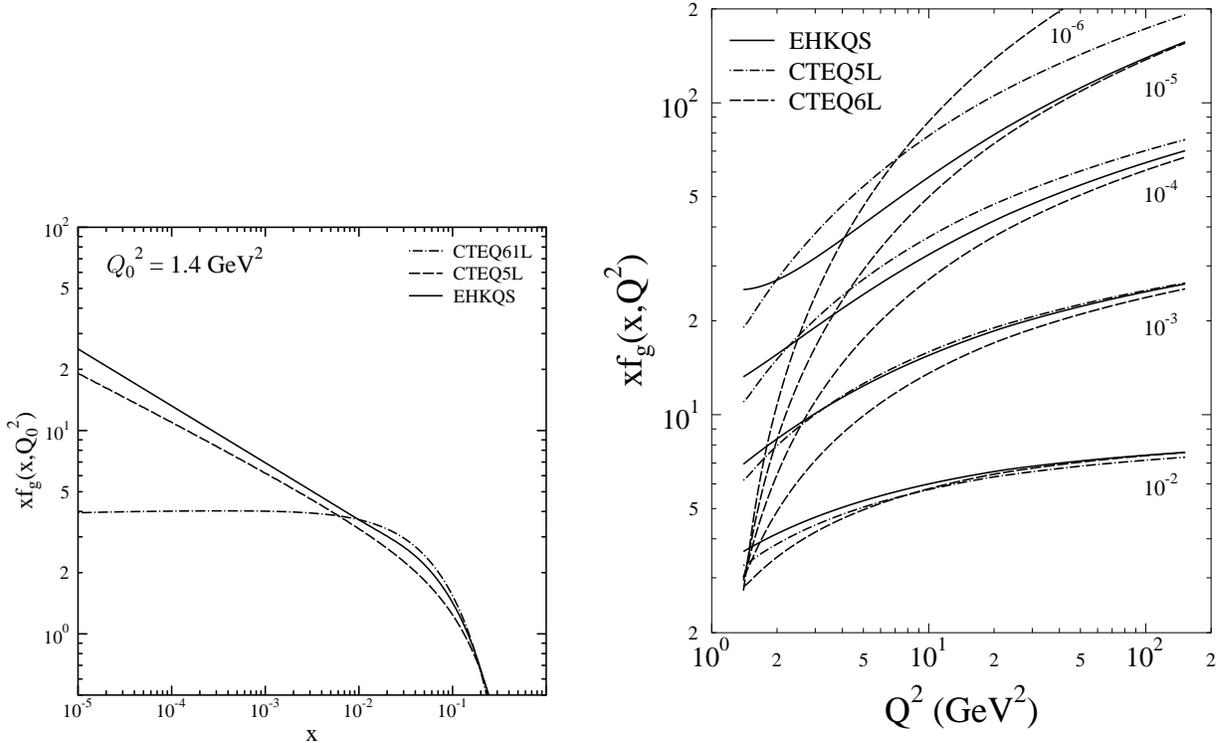

Fig. 4: Left: initial gluon distributions at $Q_0^2 = 1.4$ GeV$^2$. Right: evolution of gluon distributions for several fixed values of $x$ shows that the effect of the nonlinear terms vanishes as $Q^2$ increases.

As a result, we obtained a new set of initial PDFs[5], called EHKQS, which, when evolved using the nonlinear DGLAP+GLRMQ evolution equations, produced an equally good or even better fit to the H1 data relative to CTEQ6L, shown in Fig. 3. At $Q^2 \sim 1.4$ GeV$^2$ and $x \sim 10^{-5}$, a good fit to the HERA data requires the nonlinear evolution to be compensated by a larger gluon distribution than obtained with DGLAP alone. The enhancement is a factor of $\sim 6$ relative to CTEQ6L, as shown in Fig. 4 (left). The $Q^2$ dependence of EHKQS is compared to CTEQ6L and CTEQ5L in Fig. 4 (right) for several different values of $x$.

We used $R = 1$ fm as the free proton radius in the two-gluon density term. We did not repeat the calculations with different $R$ but, depending on the transverse matter density of the free proton, some $\sim 20\%$ uncertainty in $R$ can be expected. Since the nonlinear contributions decrease as $R$ increases, a larger $R$ would lead to reduced enhancement of the small $x$ gluons below $Q^2 \sim 10$ GeV$^2$. Thus, minimizing the $\chi^2$ of the fit with respect to $R$ is a future task.

### 3.2 BK approach

*authors: H. Jung, K. Kutak, K. Peters, L. Motyka*

A theoretical framework capable of describing the QCD evolution of parton densities taking gluon rescattering (corresponding to nonlinear effects) into account is the Balitsky-Kovchegov (BK) equation [24, 40, 41, 42, 25]. The equation, based on the BFKL approach [6, 43, 7], may be used to determine the unintegrated gluon density. The BK equation resums the BFKL pomeron fan diagrams with the triple pomeron vertex derived in the high energy limit of QCD. In the doubly logarithmic limit, the BK equation reduces [25] to the collinear

---

[5]In fact, we produced three new sets of initial distributions, differing by the charm quark mass and parton distribution at the initial scale, see Ref. [37] for more details. All sets produced equally good fits to the HERA data.

Gribov-Levin-Ryskin (GLR) equation [22]. It is the non-collinear limit, however, which gives the dominant contribution to the triple pomeron vertex [44, 45]. We conclude that GLR approach misses an essential part of the nonlinear gluon dynamics.

The solution to the BK equation, constrained by the low-$x$ HERA data will be used to extrapolate the parton densities to the LHC kinematical domain. Extrapolation is necessary as the LHC may probe very low values of $x$, down to $10^{-7}$ for $M = 10$ GeV and $\eta \sim 9$, where unitarity corrections may be important even at relatively large scales of a few GeV$^2$. Last but not least, unitarity corrections may also break $k_T$ factorization. We will also discuss which processes may be affected.

This section is organized as follows. First we give a brief description of the formalism used to determine the gluon evolution. Within this formalism, we fit the HERA charm structure function, $F_2^c$, data, the most relevant inclusive measurement directly sensitive to the gluon density. Using further assumptions about the sea quarks, $F_2$ can also be described well. The resulting gluon density is then used to compute heavy quark production and to investigate the nonlinear effects. First we estimate $b\bar{b}$ production at CDF and D0. Then, cross sections for heavy quark production at various LHC experiments are estimated, tracing the impact of the unitarity corrections. Finally, conclusions are given.

The standard framework to determine parton evolution is the collinear DGLAP formalism. It works rather well for inclusive quantities but, for more exclusive processes, the $k_T$-factorization scheme is more appropriate because both the longitudinal and transverse components of the gluon momenta are considered. In this framework, the process-independent quantity is the unintegrated gluon distribution, connected to the process-dependent hard matrix element via the $k_T$-factorization theorem. Linear evolution of the unintegrated gluon distribution may be described by one of the small $x$ evolution equations using the $k_T$-factorization scheme, the BFKL and CCFM [46, 47, 48, 49] equations. These equations are based on resummation of large logarithmic pQCD corrections, $\alpha_s^n \ln^m(1/x)$, and are equivalent at the leading logarithmic level.

The very small $x$ kinematic region is also the regime where the growth of the gluon density must be tamed in order to preserve unitarity. Recently, a successful description of unitarity corrections to DIS was derived within the color dipole formulation of QCD. This is the Balitsky-Kovchegov (BK) equation which describes the BFKL evolution of the gluon in a large target, including a nonlinear term corresponding to gluon recombination at high density.

In our analysis, we determine the unintegrated gluon distribution from the BK equation unified with the DGLAP equation following KMS (Kwieciński, Martin and Staśto) [50, 51, 52, 53]. We use the abbreviation KKMS (Kutak, Kwieciński, Martin and Staśto) [52, 53] for the unified nonlinear equation. The linear part of this equation is given by the BFKL kernel with subleading $\ln(1/x)$ corrections, supplemented by the non-singular parts of the DGLAP splitting functions. Thus resummation of both the leading $\ln Q^2$ and $\ln(1/x)$ terms are achieved. The subleading terms in $\ln(1/x)$ are approximated by the so-called consistency constraint and the running coupling constant. The nonlinear part is taken directly from the BK equation, ensuring that the unitarity constraints are preserved. One expects that this framework provides a more reliable description of the gluon evolution at extremely small $x$, where $\ln(1/x) \gg 1$ and the unitarity corrections are important, than does DGLAP.

We give a short review of the KKMS equation, starting from the impact parameter dependent BK equation. The equation for the unintegrated gluon density, $h(x, k^2, b)$, at impact parameter $b$ from the center of the target, becomes

$$\frac{\partial h(x, k^2, b)}{\partial \ln 1/x} = \frac{\alpha_s N_c}{\pi} k^2 \int_{k_0^2} \frac{dk'^2}{k'^2} \left\{ \frac{h\left(x, k'^2, b\right) - h\left(x, k^2, b\right)}{|k'^2 - k^2|} + \frac{h\left(x, k^2, b\right)}{[4k'^4 + k^4]^{\frac{1}{2}}} \right\}$$
$$- \pi \alpha_s \left(1 - k^2 \frac{d}{dk^2}\right)^2 k^2 \left[\int_{k^2}^{\infty} \frac{dk'^2}{k'^4} \ln\left(\frac{k'^2}{k^2}\right) h(x, k'^2, b)\right]^2, \quad (6)$$

the BFKL equation at LL$x$ accuracy, extended by the negative recombination term. The (dimensionless) unintegrated gluon distribution is obtained from $h(x, k^2, b)$ by integration over $b$,

$$f(x, k^2) = \int d^2b \, h(x, k^2, b). \quad (7)$$

A comment about the impact parameter treatment is in order. In Eq. (7), we assume that the evolution is local in $b$. However, the complete BK equation results in some diffusion in the impact parameter plane. This diffusion effect may be neglected if the target is much larger than the inverse of the saturation scale. In this scheme, the impact parameter dependence enters through the initial condition at large $x_0$, $h(x_0, k^2, b) = f(x_0, k^2) S(b)$ where $f(x_0, k^2)$ is the unintegrated gluon distribution. Note that, due to nonlinearities, the $b$ dependence of $h(x, k^2, b)$ does not factorize from $x$ and $k$ at low $x$.

The input profile function is assumed to be Gaussian, $S(b) = \exp(-b^2/R^2)/\pi R^2$, where $R^2$ corresponds to the square of the average nucleon radius. Since the size of the target, $R$, sets the magnitude of the initial parton density in the impact parameter plane, $h(x_0, k^2, b)$, the unitarity corrections depend on $R$. At smaller $R$, gluons are more densely packed in the target and the nonlinear effects are stronger.

References [52, 53] proposed to combine Eq. (6) with the unified BFKL-DGLAP framework developed in Ref. [50]. In this scheme, the (linear) BFKL part is modified by the consistency constraint [54, 55], resulting in the resummation of most of the subleading corrections in $\ln(1/x)$ which arise from imposing energy-momentum conservation on the BFKL kernel [56, 57, 58, 59]. In addition, we assume that the strong coupling constant runs with scale $k^2$, another source of important NLL$x$ corrections. Finally, the non-singular part of the leading order DGLAP splitting function and quark singlet distribution were included in the $x$ evolution. The final improved nonlinear equation for the unintegrated gluon density is

$$
\begin{aligned}
h(x, k^2, b) = &\ \tilde{h}^{(0)}(x, k^2, b) + \\
&+ \frac{\alpha_s(k^2) N_c}{\pi} k^2 \int_x^1 \frac{dz}{z} \int_{k_0^2} \frac{dk'^2}{k'^2} \left\{ \frac{h(\frac{x}{z}, k'^2, b) \Theta(\frac{k^2}{z} - k'^2) - h(\frac{x}{z}, k^2, b)}{|k'^2 - k^2|} + \frac{h(\frac{x}{z}, k^2, b)}{|4k'^4 + k^4|^{\frac{1}{2}}} \right\} + \\
&+ \frac{\alpha_s(k^2)}{2\pi} \int_x^1 dz \left[ \left( P_{gg}(z) - \frac{2N_c}{z} \right) \int_{k_0^2}^{k^2} \frac{dk'^2}{k'^2} h(\frac{x}{z}, k'^2, b) + P_{gq}(z) \Sigma\left(\frac{x}{z}, k'^2, b\right) \right] + \\
&- \pi \left(1 - k^2 \frac{d}{dk^2}\right)^2 k^2 \int_x^1 \frac{dz}{z} \left[ \int_{k^2}^\infty \frac{dk'^2}{k'^4} \alpha_s(k'^2) \ln\left(\frac{k'^2}{k^2}\right) h(z, k'^2, b) \right]^2 .
\end{aligned}
\tag{8}
$$

The second line of the equation corresponds to the BFKL evolution. The theta function, $\Theta(\frac{k^2}{z} - k'^2)$, reflects the consistency constraint that generates the dominant part of the subleading BFKL corrections. The third line corresponds to the DGLAP effects generated by the part of the splitting function, $P_{gg}(z)$, that is not singular in the limit $z \to 0$ and also by the quarks where $\Sigma(x, k^2, b^2)$ corresponds to the impact-parameter dependent singlet quark distribution. The nonlinear screening contribution following from the BK equation is given in the last term. The inhomogeneous contribution, defined in terms of the integrated gluon distribution, carries information about the transverse profile of the proton,

$$
\tilde{h}^{(0)}(x, k^2, b) = \frac{\alpha_s(k^2)}{2\pi} S(b) \int_x^1 dz P_{gg}(z) \frac{x}{z} g\left(\frac{x}{z}, k_0^2\right) ,
\tag{9}
$$

at $k_0^2 = 1$ GeV$^2$. The initial integrated density at $k_0^2$ is parameterized as

$$
xg(x, k_0^2) = N(1-x)^\rho
\tag{10}
$$

where $\rho = 2.5$. The size of the dense gluon system inside the proton is assumed to be $R = 2.8$ GeV$^{-1}$, in accord with the diffractive slope, $B_d \simeq 4$ GeV$^{-2}$, of the elastic $J/\psi$ photoproduction cross section at HERA. In this process, the impact parameter profile of the proton defines the $t$ dependence of the elastic cross section, $B_d \simeq R^2/2$, by Fourier transform. In the 'hot-spot' scenario, the radius can be smaller, $R = 1.5$ GeV$^{-1}$. We also use the hot spot value to compare with measurements and make predictions for the LHC.

Equation (8) was solved numerically both in the linear approximation and in full. The method for solving Eq. (8) was developed in Refs. [50, 52]. In Fig. 5, the effects of linear and nonlinear evolution on the unintegrated gluon distribution are given as a function of $x$ for $k^2 = 5$ and 50 GeV$^2$. Nonlinear evolution leads to sizeable suppression at the smallest $x$ values. While the nonlinear effects are small in the HERA $x$ range, they may be important at the LHC. In the following sections, we address the importance of these nonlinear effects.

The initial distribution in Eq. (10) was obtained by fitting the HERA $F_2^c$ measurements [60, 61] using the Monte Carlo CASCADE [62, 63] for evolution and convolution with the off-shell matrix elements. We find $\chi^2$ per degree of freedom of 0.46 (1.17) for H1 (ZEUS). The fits were repeated both with the standard KMS evolution without the nonlinear contribution and with extended KMS evolution including the nonlinear part. The predicted $F_2^c$ is equivalent for both linear and nonlinear evolution, independent of $R$. Thus nonlinear evolution is only a small effect at HERA, even in the hot-spot scenario with $R = 1.5$ GeV$^{-1}$.

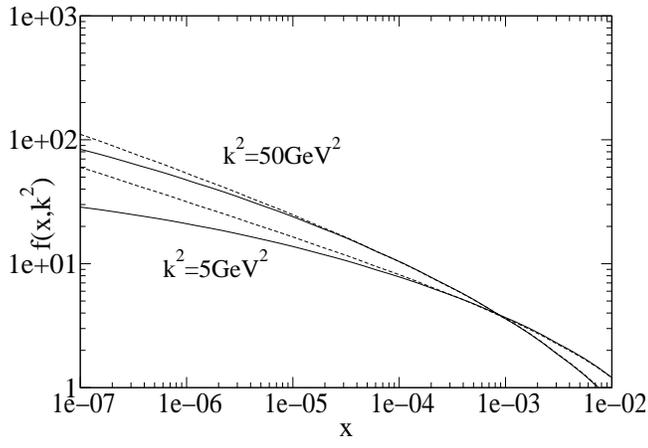

Fig. 5: The unintegrated gluon distribution obtained from Eq. (8) as a function of $x$ for different values of $k_T^2$. The solid lines correspond to the solution of the nonlinear equation with $R = 2.8$ GeV$^{-1}$ while the dashed lines correspond to the linear part.

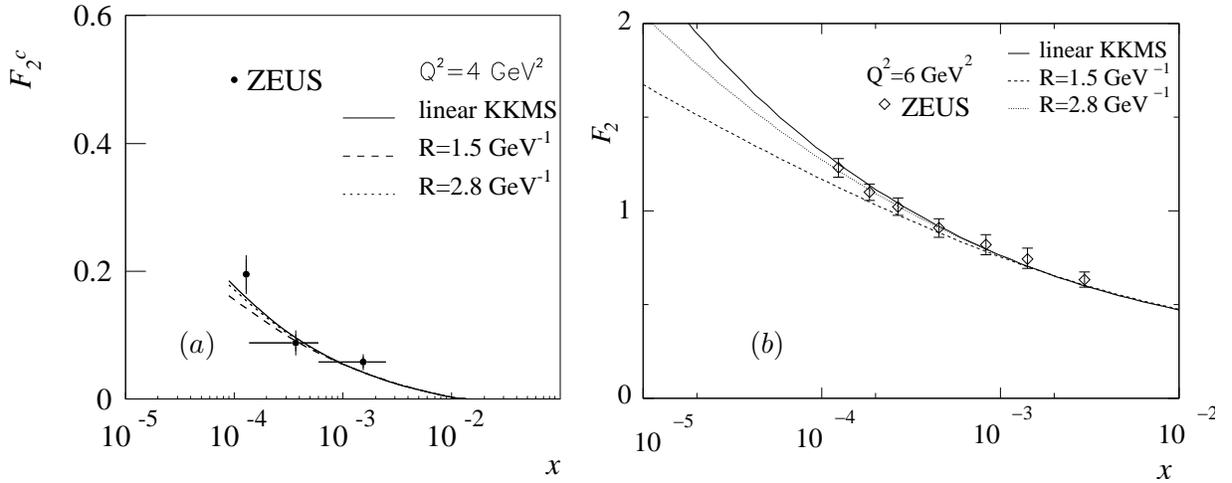

Fig. 6: The charm structure function, $F_2^c$, [61] at $Q^2 = 4$ GeV$^2$ (a) and $F_2$ [64] at $Q^2 = 6$ GeV$^2$ (b) obtained for KKMS evolution with different values of $R$.

In Fig. 6(a) we compare the measured $F_2^c$ [61] to our prediction at $Q^2 = 4$ GeV$^2$. We have determined our initial distribution from $F_2^c$ since it is the only inclusive measurement at HERA directly sensitive to the gluon distribution. However, we can also describe $F_2$ [64] by making further assumptions about the sea quark distribution, following the KMS approach. The agreement with the data, shown in Fig. 6(b), is also quite good. There is only a small effect for $Q^2 > 5$ GeV$^2$, even in the hot-spot scenario with $R = 1.5$ GeV$^{-1}$.

Next, this constrained gluon density was used to calculate $gg \to b\bar{b}$ production at the Tevatron as a cross check of the fit and the evolution formalism. We use $m_b = 4.75$ GeV and a renormalization scale in $\alpha_s$ of $Q^2 = 4m_b^2 + p_T^2$. The predicted cross section was then compared to both CDF [65, 66] and D0 [67] measurements. The predictions agree well with the data.

In Fig. 7(a) the cross section for $B$ decays to $J/\psi$ is shown as a function the $J/\psi$ $p_T$ [65, 66]. The KKMS gluon density fits the data well in all three scenarios with deviations only appearing for $p_T > 12$ GeV. It is interesting to note that the approach described here gives even better agreement than the NLO collinear approach [68].

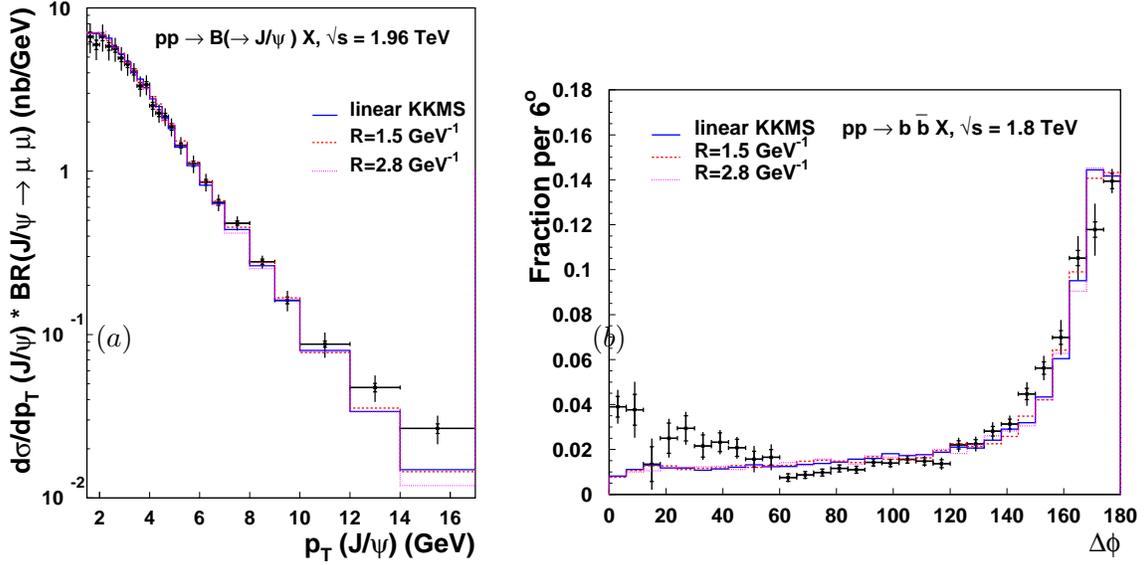

Fig. 7: Bottom production, measured by CDF, is compared to predictions using CASCADE with linear and nonlinear KKMS evolution, including two values of $R$. (a) The $p_T$ distribution of $B$ meson decays to $J/\psi$. (b) The azimuthal angle, $\Delta\phi$, distribution of $b\bar{b}$ pair production smeared by the experimental resolution.

In Fig. 7(b), the azimuthal angle distribution between the $b$ and $\bar{b}$ quarks, $\Delta\phi$, is given. The $\Delta\phi$ and $b\bar{b}$ $p_T$ distributions are correlated since $\Delta\phi < 180°$ corresponds to higher pair $p_T$. Since the $k_T$-factorization formula allows the incoming gluons to have sizable transverse momenta, the calculated $\Delta\phi$ distribution agrees very well with the data for $\Delta\phi > 60°$ with only smearing due to the experimental resolution. The enhancement of the data relative to the calculations at low $\Delta\phi$ requires further study.

Both plots compare linear (solid histograms) and nonlinear KKMS evolution (dotted and dashed histograms) for $R = 1.5$ GeV$^{-1}$ and 2.8 GeV$^{-1}$ respectively. The nonlinear part of the evolution also has no impact in this kinematic region.

## 4 Phenomenological applications: heavy quark production at the LHC

### 4.1 GLRMQ approach

*author: R. Vogt*

Since the HERA $F_2$ data can be described by both linear DGLAP and nonlinear DGLAP+GLRMQ evolution, as shown in Fig. 3 of Section 3.1, additional independent probes are needed. Here, we discuss how charm quark production in $pp$ collisions could probe the gluon enhancement predicted in Section 3.1 and described in detail in Ref. [20, 21]. Charm production is an ideal choice since the charm mass is low and its production is dominated by gluons. Assuming factorization, the inclusive differential charm cross section is

$$d\sigma_{pp \to c\bar{c}X}(Q^2, \sqrt{s}) = \sum_{i,j,k=q,\bar{q},g} f_i(x_1, Q^2) \otimes f_j(x_2, Q^2) \otimes d\hat{\sigma}_{ij \to c\bar{c}\{k\}}(Q^2, x_1, x_2) \quad (11)$$

where $\hat{\sigma}_{ij \to c\bar{c}\{k\}}(Q^2, x_1, x_2)$ are the perturbatively calculable partonic cross sections for charm production at scales $Q^2 \sim m_T^2 \gg \Lambda_{\rm QCD}^2$, $x_1$ and $x_2$ are the parton momentum fractions and $f_i(x, Q^2)$ are the proton parton densities. We assume that the renormalization and factorization scales are equal. Only the leading order $gg$ and $q\bar{q}$ channels are considered here.

The values of the charm quark mass and scale used in the calculations are chosen to give good agreement with the total cross section data at NLO: $m = 1.2$ GeV and $Q^2 = 4m^2$ for standard DGLAP-evolved NLO PDFs such as CTEQ6M [69] and MRST [70]. Nearly equivalent agreement may be obtained with $m = 1.3$ GeV and $Q^2 = m^2$ [71, 72]. Both choices assure that the PDFs are evaluated above the minimum scales. While scales proportional to $m$ are used in the total cross section, inclusive calculations of distributions also depend on the transverse momentum scale, $p_T$, so that $m_T = \sqrt{m^2 + p_T^2}$ is used instead [73].

To illustrate the effects of the nonlinear EHKQS distributions [37] of Section 3.1 on charm production at the LHC, we show

$$R(y) \equiv \frac{d\sigma(\text{EHKQS})/dy}{d\sigma(\text{CTEQ61L})/dy} \quad \text{and} \quad R(p_T) \equiv \frac{d\sigma(\text{EHKQS})/dp_T}{d\sigma(\text{CTEQ61L})/dp_T} \qquad (12)$$

in Fig. 8 where $y$ is the charm quark rapidity. The results are calculated for the maximum LHC $pp$, $p$Pb and Pb+Pb energies, $\sqrt{S} = 14$ (solid), 8.8 (dashed) and 5.5 (dot-dashed) TeV respectively. The results for $m = 1.2$ GeV and $Q^2 = 4m_T^2$ are on the left-hand side while those with $m = 1.3$ GeV and $Q^2 = m_T^2$ are on the right-hand side.

The change in the slope of $R(y)$ occurs when one $x$ drops below the minimum value of the EHKQS set where further nonlinearities become important, $x_{\min}^{\text{EHKQS}} = 10^{-5}$, and enters the unconstrained $x$ region. The minimum $x$ of CTEQ61L is lower, $x_{\min}^{\text{CTEQ61L}} = 10^{-6}$. While the EHKQS gluon distribution is fixed at its minimum for $x < x_{\min}^{\text{EHKQS}}$, the CTEQ61L distribution continues to change until $x_{\min}^{\text{CTEQ61L}}$. In inclusive kinematics with an identified charm quark and fixed $x_T = 2m_T/\sqrt{S}$, the unconstrained $x$-region contributes to charm production in the region

$$y_l \equiv \ln\left(1/x_T - \sqrt{1/x_T^2 - 1/x_{\min}}\right) \leq |y| \leq \ln\left(1/x_T + \sqrt{1/x_T^2 - 1/x_{\min}}\right) . \qquad (13)$$

The upper limit is close to the phase space boundary. Expanding the lower limit, $y_l$, in powers of $x_T^2/x_{\min} \ll 1$, $y_l \approx \ln[m_T/(x_{\min}\sqrt{S})] \geq \ln[m/(x_{\min}\sqrt{S})]$. If $m = 1.2$ GeV, the small $x$ region contributes to charm production at $|y| \geq y_l = 2.2$, 2.6 and 3.1 for $\sqrt{S} = 14$, 8.8 and 5.5 TeV, respectively. The rather sharp turnover in $R(y)$ indicates where the $x < 10^{-5}$ region begins to contribute. For $|y| > y_l$ and $Q^2 > 4$ GeV$^2$, as $x$ decreases, the CTEQ61L gluon distribution increases considerably above that of the EHKQS distribution. Thus $R(y) < 1$ at large rapidities when $Q^2 = 4m_T^2$. At midrapidity $R(y)$ is insensitive to the EHKQS extrapolation region, $x < x_{\min}^{\text{EHKQS}}$. Since $R(y)$ is integrated over $p_T$, it not only reflects the enhancement at $m_T = m$ because $Q^2 \propto m_T^2$ and the $p_T$ distribution peaks around $p_T \approx 1$ GeV. When $Q^2 = m_T^2$, the ratios are broad because the CTEQ61L gluon distribution is relatively flat as a function of $x$ for $Q^2 \sim 2 - 3$ GeV$^2$. The enhancement decreases and broadens with decreasing energy.

Since the rapidity distributions are rather flat, there are still important contributions to the $p_T$ distributions from the extrapolation region, up to $\sim 30\%$ at $\sqrt{S} = 14$ TeV for $m = 1.2$ GeV and $Q^2 = 4m^2$. Thus the sensitivity of $R(p_T)$ to the unconstrained region should be kept in mind. At the largest $\sqrt{S}$, the contribution from the $x < 10^{-5}$ region is greatest and if $Q^2 \geq 4m^2$, $xg^{\text{CTEQ61L}}(x, Q^2) > xg^{\text{EHKQS}}(x, Q^2)$. Because the contribution from the region $x < 10^{-5}$ decreases with $\sqrt{S}$, at low $p_T$ $R(p_T)$ decreases with energy. In contrast, for $Q^2 = m_T^2$, $xg^{\text{EHKQS}}(x, Q^2) > xg^{\text{CTEQ61L}}(x, Q^2)$ and the enhancement decreases with energy.

Because the DGLAP gluon distributions are already well constrained by HERA data, they cannot absorb additional large effects. Therefore we conclude that, if a low-$p_T$ enhancement in the charm cross section relative to the DGLAP-based result is observed in future experiments, it is a signal of nonlinear effects on the PDF evolution.

### 4.2 BK approach

*authors: H. Jung, K. Kutak, K. Peters, L. Motyka*

Since the Tevatron measurements are well described using the unintegrated parton densities constrained by HERA and convoluted with the off-shell matrix elements, the same approach may be used for heavy quark production at the LHC at *e.g.* $\sqrt{s} = 14$ TeV. As discussed previously, see also Fig. 5, heavy quark production at this energy is already in the region where saturation effects may be relevant. In the kinematic regions, such as at the LHC, where nonlinear evolution may become important, the cross section will be suppressed due to the negative sign of the nonlinear term in Eq. (8).

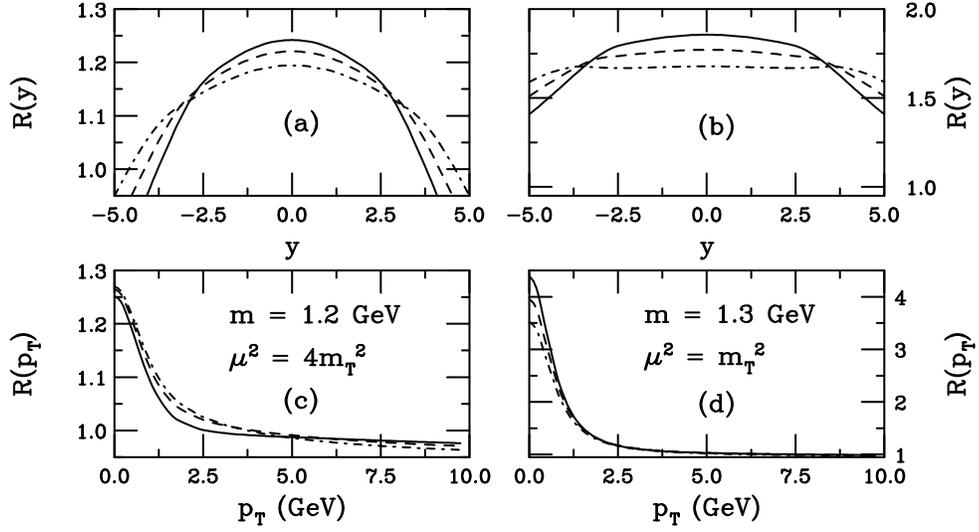

Fig. 8: We present $R(y)$, (a) and (c), and $R(p_T)$, (b) and (d), in $pp$ collisions at $\sqrt{S} = 14$ (solid), 8.8 (dashed) and 5.5 (dot-dashed) TeV. The left-hand side shows $m = 1.2$ GeV and $Q^2 = 4m_T^2$, the right-hand side $m = 1.3$ GeV and $Q^2 = m_T^2$.

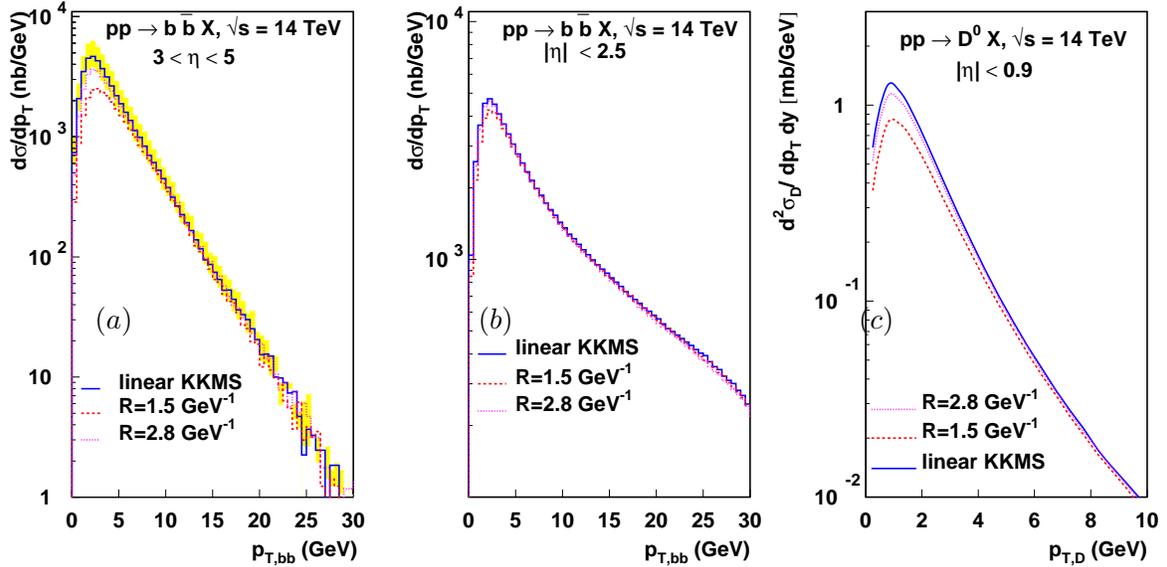

Fig. 9: (a) and (b) show $b\bar{b}$ production as a function of pair $p_T$ without cuts in $3 < |\eta| < 5$ (a) and in the ATLAS/CMS acceptance (b). The $D^0$ meson $p_T$ distribution in the ALICE acceptance is shown in (c).

First, we compute the $b\bar{b}$ production cross section at 14 TeV without any experimental cuts. In Fig. 9(a) the $b\bar{b}$ differential cross section is shown as a function of pair $p_T$ in the forward region, $3 < |\eta| < 5$. We compare linear evolution (solid histogram), nonlinear evolution with $R = 1.5$ GeV$^{-1}$ (dashed histogram) and $R = 2.8$ GeV$^{-1}$ (dotted histogram). The grey band shows the uncertainty in the linear result due to the $b$ quark mass. We take a central value of 4.75 GeV (the solid histogram) and vary $m_b$ from 4.5 to 5 GeV. The $b\bar{b}$ pair results are shown since the pair $p_T$ is most sensitive to the gluon $k_T$ and thus to the saturation effects. In the hot-spot scenario, saturation effects are visible for $p_{T_{b\bar{b}}} < 5$ GeV. These saturation effects grow with rapidity, increasing the suppression to a factor of $3 - 4$ in the higher rapidity regions. For larger $R$, the saturation effects are not very significant.

In Fig. 9(b), the $b\bar{b}$ production cross section is computed within the ATLAS and CMS acceptance ($p_T > 10$ GeV and $|\eta| < 2.5$ for both the $b$ and $\bar{b}$ quarks, see Ref. [74]). With these cuts, the observed suppression due to nonlinear effects nearly vanishes. This result suggests that $k_T$ factorization can safely be applied in the central $\eta$ region. Thus saturation effects due to nonlinear gluon evolution are seen only for $p_{T_{b\bar{b}}} < 10$ GeV and at high $\eta$. This regime is accessible with upgraded ATLAS/CMS detectors or in LHCb where the $b$ quark $p_T$ can be measured to 2 GeV for $1.9 < \eta < 4.9$. In this kinematic regime, the hot-spot scenario predicts a factor of two suppression of the $b\bar{b}$ cross section.

Similarly, we investigated $c\bar{c}$ production at ALICE. In ALICE, it will be possible to measure the $D^0$ down to $p_T \sim 0.5$ GeV in $|\eta| < 0.9$. The result is shown in Fig. 9(c) with $m_c = 1.5$ GeV. In the hot-spot scenario (dashed curve), a factor of two suppression occurs at $p_T \sim 1$ GeV.

## 5 Perspectives for experimental observation at LHC

*authors: A. Dainese and R. Vogt*

### 5.1 Introduction

In Section 4.1, charm production in $pp$ collisions at the LHC was suggested as a promising way to study the effects of nonlinear evolution on the parton densities. Due to gluon dominance of charm production and the small values of $x$ and $Q^2$ probed, $x \approx 2 \times 10^{-4}$ and $Q^2 \approx 1.69 - 6$ GeV$^2$ at midrapidity and transverse momentum[6] $p_T \approx 0$, charm production at the LHC is sensitive to the gluon enhancement arising from nonlinear evolution. The resulting charm enhancement was quantified in Ref. [20, 21] by the LO ratios of the differential cross sections computed with the nonlinearly-evolved EHKQS PDFs [37], obtained from DGLAP+GLRMQ evolution, relative to the DGLAP-evolved CTEQ61L PDFs.

The enhancement of the nonlinearly-evolved gluons increases as $x$ and $Q^2$ decrease. Consequently, the charm enhancement increases with center of mass energy, $\sqrt{S}$. Thus the maximum enhancement at the LHC will be at $\sqrt{S} = 14$ TeV and small charm quark transverse momentum. The sensitivity of the charm enhancement to the value of the charm quark mass, $m$, as well as to the choice of the factorization, $Q_F^2$, and renormalization, $Q_R^2$, scales was studied in Ref. [20, 21] assuming $Q^2 = Q_F^2 = Q_R^2 \propto m_T^2$ where $m_T^2 = p_T^2 + m^2$. The most significant charm enhancement occurs when $m$ and $Q^2/m_T^2$ are both small. A comparison of the NLO total cross sections with low energy data shows that the data prefer such small $m$ and $Q^2$ combinations [71, 72]. The largest enhancement is obtained with $m = 1.3$ GeV and $Q^2 = m_T^2$, see Fig. 8 in Section 4.1.

In Section 4.1, only charm enhancement was described. Neither its subsequent hadronization to $D$ mesons nor its decay and detection were considered. In this section, we address these issues to determine whether the charm enhancement survives hadronization and decay to be measured in the ALICE detector [75].

---
[6]Here we use $p_T$ for the transverse momentum of the charm quark and $p_T^D$ for the transverse momentum of the $D$ meson.

The calculation described in Section 4.1 was to leading order since the EHKQS sets are evolved according to the LO DGLAP+GLRMQ equations using a one-loop evaluation of $\alpha_s$. Thus these LO distributions should generally not be mixed with NLO matrix elements and the two-loop $\alpha_s$. However, the charm quark total cross section is increased and the $p_T$ distribution is broadened at NLO relative to LO [76]. Thus, to determine whether or not the enhancement is experimentally measurable, we assume that the enhancement is the same at NLO and LO and employ a NLO cross section closest to the calculation of the enhancement in Ref. [20, 21].

As described in Ref. [76], the theoretical $K$ factor may be defined in more than one way, depending on how the LO contribution to the cross section is calculated. In all cases, the $\mathcal{O}(\alpha_s^3)$ contribution to cross section is calculated using NLO PDFs and the two-loop evaluation of $\alpha_s$. If the LO contribution is also calculated using NLO PDFs and a two-loop $\alpha_s$, this is the "standard NLO" cross section. It is used in most NLO codes, both in the global analyses of the NLO PDFs and in evaluations of cross sections and rates [76]. The $K$ factor formed when taking the ratio of the "standard NLO" cross section to the LO cross section with the NLO PDFs [76], $K_0^{(1)}$, indicates the convergence of terms in a fixed-order calculation [77]. On the other hand, if the LO contribution to the total NLO cross section employs LO PDFs and the one-loop $\alpha_s$, we have a cross section which we refer to here as "alternative NLO". The $K$ factor calculated taking the ratio of the "alternative NLO" cross section to the LO cross section with LO PDFs [76], $K_2^{(1)}$, indicates the convergence of the hadronic cross section toward a result. If $K_0^{(1)} > K_2^{(1)}$, convergence of the hadronic cross section is more likely [77]. This is indeed the case for charm production [76]. We also note that $K_2^{(1)}$ is a much weaker function of energy than $K_0^{(1)}$. Since, in the absence of nonlinear NLO PDFs, the "alternative NLO" cross section is more consistent with the calculated enhancement, we use this cross section to calculate the NLO $D$ meson rates and $p_T$ spectra. We note also that, in both cases, the $p_T$ distributions have the same slope even though $K_2^{(1)}$, for the alternative NLO cross section, is somewhat smaller. Thus, using a non-standard NLO calculation will not change the slope of the $p_T$ distributions, distorting the result.

The LO and NLO calculations used to obtain the full NLO result in both cases can be defined by modification of Eq. (11) in Section 4.1. We define the full LO charm production cross section as

$$d\sigma_{\rm LO}^{\rm 1L} = \sum_{i,j=q,\overline{q},g} f_i^{\rm LO}(x_1, Q_F^2) \otimes f_j^{\rm LO}(x_2, Q_F^2) \otimes d\hat{\sigma}_{ij \to c\overline{c}}^{\rm LO}(\alpha_s^{\rm 1L}(Q_R^2), x_1, x_2) \quad (14)$$

where the superscript "LO" on $d\hat{\sigma}_{ij \to c\overline{c}}$ indicates the use of the LO matrix elements while the superscript "1L" indicates that the one-loop expression of $\alpha_s$ is used. The LO cross section typically used in NLO codes employs the NLO PDFs and the two-loop (2L) $\alpha_s$ so that

$$d\sigma_{\rm LO}^{\rm 2L} = \sum_{i,j=q,\overline{q},g} f_i^{\rm NLO}(x_1, Q_F^2) \otimes f_j^{\rm NLO}(x_2, Q_F^2) \otimes d\hat{\sigma}_{ij \to c\overline{c}}^{\rm LO}(\alpha_s^{\rm 2L}(Q_R^2), x_1, x_2) \ . \quad (15)$$

In either case, the NLO contribution, $\mathcal{O}(\alpha_s^3)$ for heavy quark production, is

$$d\sigma_{\mathcal{O}(\alpha_s^3)} = \sum_{i,j=q,\overline{q},g} f_i^{\rm NLO}(x_1, Q_F^2) \otimes f_j^{\rm NLO}(x_2, Q_F^2) \otimes \sum_{k=0,q,\overline{q},g} d\hat{\sigma}_{ij \to c\overline{c}k}^{\rm NLO}(\alpha_s^{\rm 2L}(Q_R^2), Q_F^2, x_1, x_2) \quad (16)$$

where the superscript "NLO" on $d\hat{\sigma}_{ij \to c\overline{c}k}$ indicates the use of the NLO matrix elements. The additional sum over $k$ in Eq. (16) includes the virtual ($k = 0$) and real ($k = q$, $\overline{q}$ or $g$ depending on $i$ and $j$) NLO corrections. In the calculations of $d\sigma_{\rm LO}^{\rm 2L}$ and $d\sigma_{\mathcal{O}(\alpha_s^3)}$, we use the value of $\Lambda_{\rm QCD}^{(4)}$ given for the NLO PDFs and work in the $\overline{\rm MS}$ scheme. The standard NLO cross section is then

$$d\sigma_{\rm NLO}^{\rm std} = d\sigma_{\rm LO}^{\rm 2L} + d\sigma_{\mathcal{O}(\alpha_s^3)} \quad (17)$$

while our "alternative NLO" cross section is defined as

$$d\sigma_{\rm NLO}^{\rm alt} = d\sigma_{\rm LO}^{\rm 1L} + d\sigma_{\mathcal{O}(\alpha_s^3)} \ . \quad (18)$$

Since the enhancement in Ref. [20, 21] was defined using $d\sigma_{\rm LO}^{\rm 1L}$ only, the best we can do is to use the alternative NLO cross section in our analysis, as described below.

We now discuss how the enhancement is taken into account in the context of the NLO computation. We calculate the LO inclusive charm $p_T$ distribution, $d^2\sigma/dp_T dy$, with the detected charm (anticharm) quark in the rapidity interval $\Delta y$ with $|y| < 1$, motivated by the pseudorapidity acceptance of the ALICE tracking barrel, $|\eta| < 0.9$. The rapidity, $y_2$, of the undetected anticharm (charm) quark is integrated over. The charm enhancement factor $R(p_T, \Delta y)$ is then

$$R(p_T, \Delta y) = \frac{\int_{\Delta y} dy \int dy_2 \frac{d^3\sigma(\text{EHKQS})}{dp_T dy dy_2}}{\int_{\Delta y} dy \int dy_2 \frac{d^3\sigma(\text{CTEQ61L})}{dp_T dy dy_2}} \ . \tag{19}$$

Next, we assume that the enhancement calculated at LO is the same when calculated at NLO. This is the only reasonable assumption we can make to test whether the enhancement can be detected with ALICE which will measure the physical $p_T^D$ distribution. The alternative NLO cross section is therefore the closest in spirit to the LO computation in Ref. [20, 21]. Thus, the enhanced NLO charm $p_T$ distribution is

$$R(p_T, \Delta y) \ d\sigma_{\text{NLO}}^{\text{alt}}(\Delta y)/dp_T \ . \tag{20}$$

Our results are obtained with the same parameters used in Section 4.1, $m = 1.2$ GeV and $Q^2 = 4m_T^2$ as well as $m = 1.3$ GeV and $Q^2 = m_T^2$. These two choices are the baseline results against which other parameter choices will be compared to see if the enhancement is detectable.

## 5.2 From charm to $D$ enhancement

To make a more realistic $D$ meson distribution, we have modified the charm $p_T$ distribution by the heavy quark string fragmentation in PYTHIA [78]. Charm events in $pp$ collisions at $\sqrt{S} = 14$ TeV are generated using PYTHIA (default settings) with the requirement that one of the quarks is in the interval $|y| < 1$. The charm quarks are hadronized using the default string model. Since $c$ and $\overline{c}$ quarks fragment to $D$ and $\overline{D}$ mesons[7], respectively, in each event related $(c, D)$ and $(\overline{c}, \overline{D})$ pairs can easily be identified[8]. These pairs are reweighted to match an arbitrary NLO charm quark $p_T$ distribution, $dN_{\text{NLO}}^c/dp_T$. If $dN_{\text{PYTHIA}}^c/dp_T$ is the charm $p_T$ distribution given by PYTHIA, each $(c, D)$ pair is assigned the weight

$$\mathcal{W}(p_T) = \frac{dN_{\text{NLO}}^c/dp_T}{dN_{\text{PYTHIA}}^c/dp_T} \tag{21}$$

where $p_T$ is the transverse momentum of the charm quark of the pair. Therefore, the reweighted final-state $D$ distribution corresponds to the one that would be obtained by applying string fragmentation to the NLO $c$-quark distribution. The resulting $D$ distribution is significantly harder than that obtained using the Peterson fragmentation function [79]. The enhancement survives after fragmentation although the $D$ enhancement is somewhat lower than that of the charm because for a given $p_T^D$, the $D$ spectrum receives contributions from charm quarks with $p_T \gtrsim p_T^D$, where the charm enhancement is smaller.

## 5.3 Sensitivity to the enhancement

Figure 10 shows the double-differential $D^0$ cross section, $d^2\sigma_D/dp_T^D dy$, in $|y| < 1$ as a function of the transverse momentum. The points represent the expected "data" measured by ALICE, obtained from the alternative NLO cross section scaled by the enhancement factor $R(p_T, \Delta y)$ defined in Eq. (19), and modified by string fragmentation. The solid and dashed curves are obtained by applying string fragmentation to the alternative NLO and standard NLO $c\overline{c}$ cross sections, respectively. Thus, the "data" points include the enhancement while the curves do not. The horizontal error bars indicate the bin width, the vertical error bars represent the statistical error and the shaded band gives the $p_T$-dependent systematic error. The 5% $p_T$-independent systematic error on the normalization is not shown. (See Ref. [80] for a discussion of the error analysis. The standard NLO cross section, Eq. (17), and the $\mathcal{O}(\alpha_s^3)$ contribution to the alternative NLO cross section, Eq. (16), were calculated using the HVQMNR code [81] with CTEQ6M and $\Lambda_{\text{QCD}}^{(4)} = 0.326$ GeV. The LO contribution to the alternative NLO cross section, Eq. (14), was calculated using the CTEQ61L PDFs. Fragmentation was included as described in Section 5.2. The enhancement, the difference between the data and the solid curves for $p_T^D \lesssim 3$ GeV, is more pronounced for the larger mass and lower scale, on the right-hand side of Fig. 10.

---
[7] Here $D \equiv D^+, D^0$.
[8] Events containing charm baryons were rejected.

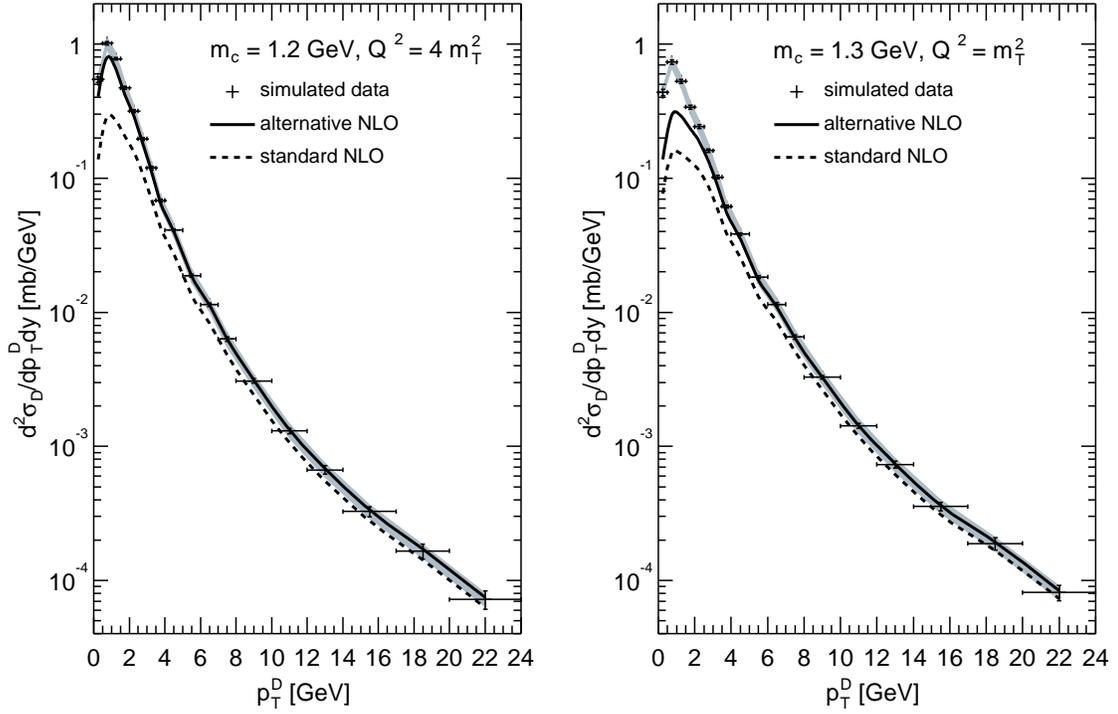

Fig. 10: Comparison of the simulated ALICE data generated from $R(p_T, \Delta y)d\sigma_{\text{NLO}}^{\text{alt}}$ with the alternative (solid) and standard (dashed) NLO calculations. The effect of string fragmentation is included in the "data" points as well as in the curves. The left-hand side shows the result for $m = 1.2$ GeV and $Q^2 = 4m_T^2$ while the right-hand side is the result for $m = 1.3$ GeV and $Q^2 = m_T^2$. The error bars on the data represent the statistical error and the shaded band represents the $p_T$-dependent systematic error. The 5% normalization error is not shown.

There is a significant difference between the alternative and standard NLO distributions. Part of the difference is due to the one- and two-loop evaluations of $\alpha_s$ since $\alpha_s^{2\text{L}} < \alpha_s^{1\text{L}}$. However, the most important contribution is the large differences between the LO and NLO gluon distributions, especially at low scales [80].

In order to address the question of the experimental sensitivity to the effect of nonlinear gluon evolution on low-$p_T$ charm production, we consider, as a function of $p_T^D$, the ratio of the simulated data, including the enhancement, to alternative NLO calculations using a range of $m$ and $Q^2$ with PYTHIA string fragmentation. We denote this ratio as "Data/Theory" and try to reproduce this ratio with NLO calculations employing recent linearly-evolved PDFs and tuning $m$ and $Q^2$.

Since the enhancement has disappeared for $p_T^D \gtrsim 5$ GeV, we refer to this unenhanced region as high $p_T^D$. The $p_T^D$ region below 5 GeV, where the enhancement is important, is referred to as low $p_T^D$. If no set of parameters can describe both the high- and low-$p_T^D$ components of the distribution equally well, and, if the set that best reproduces the high-$p_T^D$ part underestimates the low-$p_T^D$ part, this would be a strong indication of the presence of nonlinear effects.

The Data/Theory plots are shown in Fig. 11. The points with the statistical (vertical bars) and $p_T$-dependent systematic (shaded region) error correspond to the data of Fig. 10, including the enhancement, divided by themselves, depicting the sensitivity to the theory calculations. The black squares on the right-hand sides of the lines $\text{Data}/\text{Theory} = 1$ represent the 5% $p_T$-independent error on the ratio coming from the cross section normalization. This error is negligible relative to present estimates of other systematic uncertainties

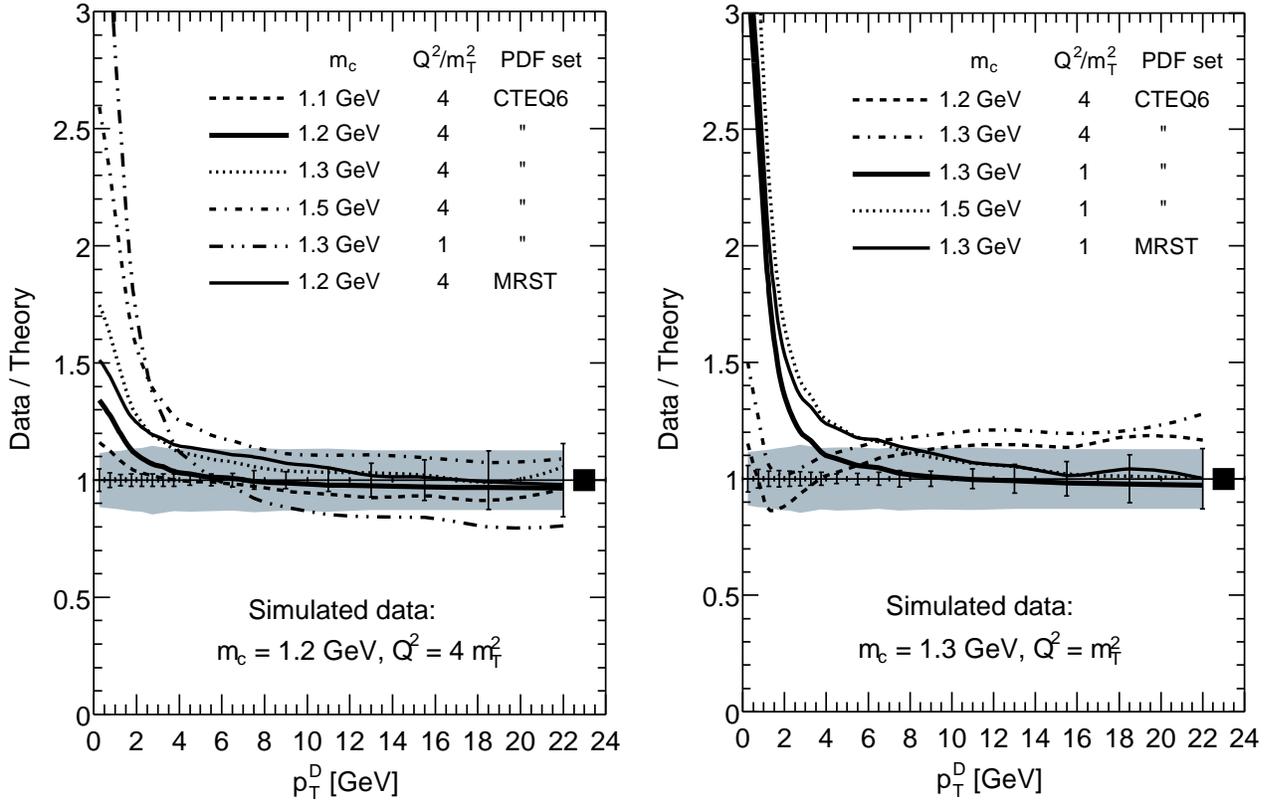

Fig. 11: Ratio of the generated ALICE data relative to calculations of the alternative NLO cross sections with several sets of parameters and PYTHIA string fragmentation. The left-hand side shows the result for $m = 1.2$ GeV and $Q^2 = 4m_T^2$ while the right-hand side is the result for $m = 1.3$ GeV and $Q^2 = m_T^2$.

($\simeq 13\%$).

On the left-hand side, the thick solid curve with $m = 1.2$ GeV and $Q^2 = 4m_T^2$ best agrees with the high-$p_T^D$ ratio by construction since $R \approx 1$ at large $p_T^D$. It also shows the effect of the enhancement well beyond the error band for $p_T^D \lesssim 2$ GeV. Better agreement with the data over the entire $p_T^D$ range can be achieved only by choosing a charm quark mass lower than 1.2 GeV, below the nominal range of charm masses, as illustrated by the dashed curve for $m = 1.1$ GeV. Higher masses with $Q^2 = 4m_T^2$ produce much larger Data/Theory ratios than the input distribution. The ratio with $m = 1.3$ GeV and $Q^2 = m_T^2$ (dot-dot-dashed curve) gives a much larger ratio at low $p_T^D$ and drops below $\approx 1$ for $p_T^D > 8$ GeV.

We also present the ratio using the MRST parton densities (MRST2001 LO [36] in Eq. (14) and MRST2002 NLO [82] in Eq. (16)) with $m = 1.2$ GeV and $Q^2 = 4m_T^2$. We find that this result also agrees reasonably well with the CTEQ6 results for the same $m$ and $Q^2$. Thus, the enhancement seems to be rather independent of the PDF. The CTEQ61L and the MRST2001 LO distributions are similar at low $x$, suggesting that non-linearly evolved PDFs based on MRST2001 LO would produce an enhancement like that of Ref. [20, 21].

On the right-hand side of Fig. 11 the thick solid curve, employing the same parameters as the data, gives the best agreement at high $p_T^D$. We note that even though the results with $Q^2 = 4m_T^2$ and $m \leq 1.3$ GeV lie closer to the data at low $p_T^D$ and within the combined statistical and systematic error at higher $p_T^D$, the curves with these parameters have the wrong slopes for $p_T^D \lesssim 8$ GeV. The statistical sensitivity is expected to be good enough to distinguish the difference in curvature. The results with the MRST PDFs do not alter the conclusions.

### 5.4 Conclusions

We have studied whether the EHKQS gluon distributions [37] could generate an observable $D$ meson enhancement in $pp$ collisions at the LHC. Using the EHKQS LO PDFs and LO matrix elements for charm quark production and PYTHIA string fragmentation for $D$ meson hadronization, the enhancement indeed survives to the $D$ mesons.

The $D$ meson enhancement, however, drops rapidly with transverse momentum. Therefore, $D$ measurement capability at small $p_T^D$ is necessary to verify the effect experimentally. The ALICE detector can directly reconstruct $D^0 \to K^-\pi^+$. We have demonstrated that, in the most optimistic case, the enhancement can be detected above the experimental statistical and systematic errors. When the charm mass is somewhat smaller, $m = 1.2$ GeV, but the scale is larger, $Q^2 = 4m_T^2$, it is more difficult to detect the enhancement over the experimental uncertainties.

**Acknowledgments**

The work of K. Kutak was supported in part by the Graduiertenkolleg Zukünftige Entwicklungen in der Teilchenphysik. M.G. Ryskin would like to thank the IPPP at the University of Durham for hospitality, and A.D. Martin thanks the Leverhulme Trust for an Emeritus Fellowship. Their work was supported by the Royal Society, by grants INTAS 00-00366, RFBR 04-02-16073, and by the Federal Program of the Russian Ministry of Industry, Science and Technology SS-1124.2003.2. The work of RV was supported in part by the Director, Office of Energy Research, Division of Nuclear Physics of the Office of High Energy and Nuclear Physics of the U. S. Department of Energy under Contract Number DE-AC02-05CH11231.

# Heavy quark fragmentation

*J. Bracinník$^1$, M. Cacciari$^2$, M. Corradi$^3$, G. Grindhammer$^1$*

$^1$Max-Planck-Institut für Physik, München, Germany
$^2$LPTHE - Université P. et M. Curie (Paris 6), Paris, France
$^3$INFN Bologna, via Irnerio 46, Bologna, Italy

**Abstract**
The fragmentation of heavy quarks into hadrons is a key non-perturbative ingredient for the heavy quark production calculations. The formalism is reviewed, and the extraction of non-perturbative parameters from $e^+e^-$ and from $ep$ data is discussed.

*coordinator: M. Corradi*

## 1 Introduction
*author: M. Cacciari*

When we try to describe in QCD the production of a hadron we are always faced with the necessity to take into account the non-perturbative hadronization phase, i.e. the processes which transform perturbative objects (quarks and gluons) into real particles. In the case of light hadrons the QCD factorization theorem [1, 2, 3, 4, 5, 6] allows to factorize these non-perturbative effects into universal (but factorization-scheme dependent) *fragmentation functions* (FF):

$$\frac{d\sigma_h}{dp_T}(p_T) = \sum_i \int \frac{dx}{x} \frac{d\sigma_i}{dp_T}\left(\frac{p_T}{x};\mu\right) D_{i\to h}(x;\mu) + \mathcal{O}\left(\frac{\Lambda}{p_T}\right) . \tag{1}$$

In this equation, valid up to higher twist corrections of order $\Lambda/p_T$ ($\Lambda$ being a hadronic scale of the order of a few hundred MeV and $p_T$ for instance a transverse momentum), the partonic cross sections $d\sigma_i/dp_T$ for production of the parton $i$ are calculated in perturbative QCD, while the fragmentation functions $D_{i\to h}(x;\mu)$ are usually extracted from fits to experimental data. Thanks to their universality they can be used for predictions in different processes. The artificial factorization scale $\mu$ is a reminder of the non-physical character of both the partonic cross sections and the fragmentation functions: it is usually taken of the order of the hard scale $p_T$ of the process, and the fragmentation functions are evolved from a low scale up to $\mu$ by means of the DGLAP evolution equations.

This general picture becomes somewhat different when we want to calculate the production of heavy-flavoured mesons. In fact, thanks to the large mass of the charm and the bottom quark, acting as a cutoff for the collinear singularities which appear in higher orders in perturbative calculations, one can calculate the perturbative prediction for heavy *quark* production. Still, of course, the quark $\to$ hadron transition must be described. Mimicking the factorization theorem given above, it has become customary to complement the perturbative calculation for heavy quark production with a non-perturbative fragmentation function accounting for its hadronization into a meson:

$$\frac{d\sigma_H}{dp_T}(p_T) = \int \frac{dx}{x} \frac{d\sigma_Q^{pert}}{dp_T}\left(\frac{p_T}{x}, m\right) D_{Q\to H}^{np}(x) . \tag{2}$$

It is worth noting that at this stage this formula is not given by a rigorous theorem, but rather by some sensible assumptions. Moreover, it will in general fail (or at least be subject to large uncertainties) in the region where the mass $m$ of the heavy quark is not much larger than its transverse momentum $p_T$, since the choice of the scaling variable is not unique any more, and $\mathcal{O}(m/p_T)$ corrections cannot be neglected.

Basic arguments in QCD allow to identify the main characteristics of the non-perturbative fragmentation function $D^{np}_{Q \to H}(x)$. In 1977 J.D. Bjorken [7] and M. Suzuki [8] independently argued that the average fraction of momentum lost by the heavy quark when hadronizing into a heavy-flavoured hadron is given by

$$\langle x \rangle^{np} \simeq 1 - \frac{\Lambda}{m} \ . \tag{3}$$

Since (by definition) the mass of a heavy quark is much larger than a hadronic scale $\Lambda$, this amounts to saying that the non-perturbative FF for a heavy quark is very hard, i.e. the quark loses very little momentum when hadronizing. This can also be seen with a very simplistic argument: a fast-traveling massive quark will lose very little speed (and hence momentum) when picking up from the vacuum a light quark of mass $\Lambda$ to form a heavy meson[1].

This basic behaviour is to be found as a common trait in all the non-perturbative heavy quark FFs derived from various phenomenological models. Among the most commonly used ones we can cite the Kartvelishvili-Likhoded-Petrov [12], Bowler [13], Peterson-Schlatter-Schmitt-Zerwas [14] and Collins-Spiller [15] fragmentation functions. These models all provide some functional form for the $D^{np}_{Q \to H}(x)$ function, and one or more free parameters which control its hardness. Such parameters are usually not predicted by the models (except perhaps on an order-of-magnitude basis), and must be fitted to the experimental data.

During the '80s many such fits were performed, and these and similar functions were also included in many Monte Carlo event generators. Eventually, some 'best' set of parameter values (for instance for the PSSZ form) was determined [16] and subsequently widely used.

These first applications, given the limited accuracy of the available data, tended to overlook two aspects which have become more important in recent years, when the accuracy of the data has vastly improved:

- A non-perturbative FF is designed to describe the heavy quark → hadron transition, dealing with events mainly populated by soft gluons of energies of a few hundred MeV. However, if a heavy quark is produced in a high energy event it will initially be far off shell: perturbative hard gluons will be emitted to bring it on-shell, reducing the heavy quark momentum and yielding in the process large collinear logarithms (for instance of the form $\alpha_s^n \log^n(p_T/m)$ in a transverse momentum differential cross section). Of course, the amount of gluon radiation is related to the distance between the heavy quark mass scale and the hard scale of the interaction, and is therefore process-dependent. One can (and it was indeed done) either fit different free parameters at different centre-of-mass energies (or transverse momenta), or try to evolve directly the non-perturbative FF by means of the DGLAP equations, hence including into it the perturbative collinear logarithms. However, this is not what non-perturbative fragmentation functions are meant for, and doing so spoils the validity of the relation in eq. (3).
- Since only the final heavy hadron is observed, both the non-perturbative FF and the perturbative cross section for producing the heavy quark must be regarded as non-physical objects. The details of the fitted non-perturbative FF (e.g. the precise value(s) of its free parameter(s)) will depend on those of the perturbative cross sections: different perturbative calculations (leading order, next-to-leading order, Monte Carlo, ...) and different perturbative parameters (heavy quark masses, strong coupling, ...) will lead to different non-perturbative FFs. These in turn will have to be used only with a perturbative description similar to the one they have been determined with. Hence the limited accuracy (and hence usefulness) of a 'standard' determination of the parameters [17].

The first point was addressed by Mele and Nason in a paper [18] which deeply changed the field of heavy quark fragmentation, and essentially propelled it into the modern era. Mele and Nason observed that, in the limit where one neglects heavy quark mass terms suppressed by a large energy scale, a heavy quark cross section can be factored into a massless, $\overline{MS}$-subtracted cross section for producing a light parton, and a process-independent[2], *perturbative* heavy quark fragmentation function describing the transition of the massless

---
[1]More modern and more rigorous derivations of this result can be found in [9, 10, 11].
[2]Mele and Nason extracted this function from the $e^+e^-$ cross section, convincingly conjecturing its process independence, which was successively established on more general grounds in [19]

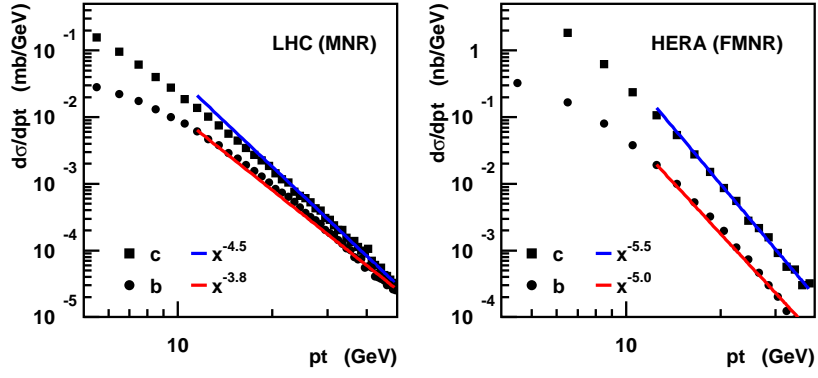

Fig. 1: Power-law fits to the heavy quark $p_T$ distributions at LHC (left) and HERA (right) obtained with the NLO programs MNR and FMNR. The resulting exponents are $N = 4.5/3.8$ for charm/beauty at LHC and $N = 5.5/5.0$ for $c/b$ at HERA.

parton into the heavy quark:

$$\frac{d\sigma_Q^{pert,res}}{dp_T}(p_T, m) = \sum_i \int \frac{dx}{x} \frac{d\sigma_i}{dp_T}\left(\frac{p_T}{x}; \mu\right) D_{i \to Q}(x; \mu, m) + \mathcal{O}\left(\frac{m}{p_T}\right). \quad (4)$$

The key feature of this equation is that it is entirely perturbative: every term can be calculated in perturbative QCD. The perturbative fragmentation functions $D_{i \to Q}(x; \mu, m)$ (not to be confused with the non-perturbative one $D_{Q \to H}^{np}(x)$) can be evolved via DGLAP equations from an initial scale of the order of the heavy quark mass up to the large scale of the order of $p_T$. This resums to all orders in the strong coupling the collinear logarithms generated by the gluon emissions which bring the heavy quark on its mass shell, leading to a more accurate theoretical prediction for $d\sigma_Q/dp_T$.

Once a reliable perturbative cross section for the production of a heavy quark is established, one is simply left with the need to account for its hadronization. For this purpose one of the functional forms listed above can be used for the non-perturbative FF, and implemented as in eq. (2), but using the improved, resummed cross section given by eq. (4). Since most of the the scaling-violation logarithms are accounted for by the evolution of the perturbative FF, the non-perturbative one can now be scale-independent and only contain the physics related to the hadronization of the heavy quark. It will always, however, depend on the details of the perturbative picture used.

## 2 Extraction of heavy quark fragmentation parameters from $e^+e^-$ and their impact on HERA and LHC

*author: M. Corradi*

### 2.1 Importance of $\langle x \rangle^{\mathrm{np}}$

According to the factorization of the fragmentation functions (FF), the differential cross section $d\sigma/dp_T$ for the production of a heavy hadron $H$ can be written as the convolution of the perturbative heavy quark differential cross section $d\sigma^{\mathrm{pert}}/dp_T$ and the non-perturbative fragmentation function $D^{\mathrm{np}}(x)$:

$$\frac{d\sigma}{dp_T}(p_T) = \int \frac{dx}{x} D^{\mathrm{np}}(x) \frac{d\sigma^{\mathrm{pert}}}{dp_T}\left(\frac{p_T}{x}\right). \quad (5)$$

This convolution neglects mass terms $\mathcal{O}(m_Q/p_T)$ and non-perturbative terms $\mathcal{O}(\Lambda_{\mathrm{qcd}}/m_Q)$.

The heavy quark $p_T$ distribution behaves at large $p_T$ like a power law $d\sigma^{\mathrm{pert}}/dp_T = C p_T^{-N}$. Figure 1 shows power-law fits to the $p_T$ distributions of heavy quarks at LHC and in photoproduction at HERA as obtained from the NLO programs MNR [20] and FMNR [21]. For $p_T > 10$ GeV $N$ was found to range from 3.8 ($b$ at LHC) to 5.5 ($c$ at HERA). Combining this power-law behavior with eq. (5), the hadron $p_T$ distribution

| Function | $D(x)$ | parameters | $x$ range |
|---|---|---|---|
| Kartvelishvili | $(1-x)x^\alpha$ | $\alpha = 2/\delta - 3$ | $[0,1]$ |
| Peterson | $\frac{1}{x}\left(1 - \frac{1}{x} - \frac{\epsilon}{(1-x)}\right)^{-2}$ | $\epsilon$ | $[0,1]$ |
| Gauss | $\exp(-(\frac{x-\mu}{2\sigma})^2)$ | $\mu = 1-\delta$ $\sigma = \delta/2$ | $[-\infty, \infty]$ |
| Box | const. | - | $[1-2\delta, 1]$ |
| Triangular: | $x - x_0$ | $x_0 = 1 - 3\delta$ | $[1-3\delta, 1]$ |

Table 1: Test functions used in Fig. 2. The functions assume a value different from zero in the range given by the third column.

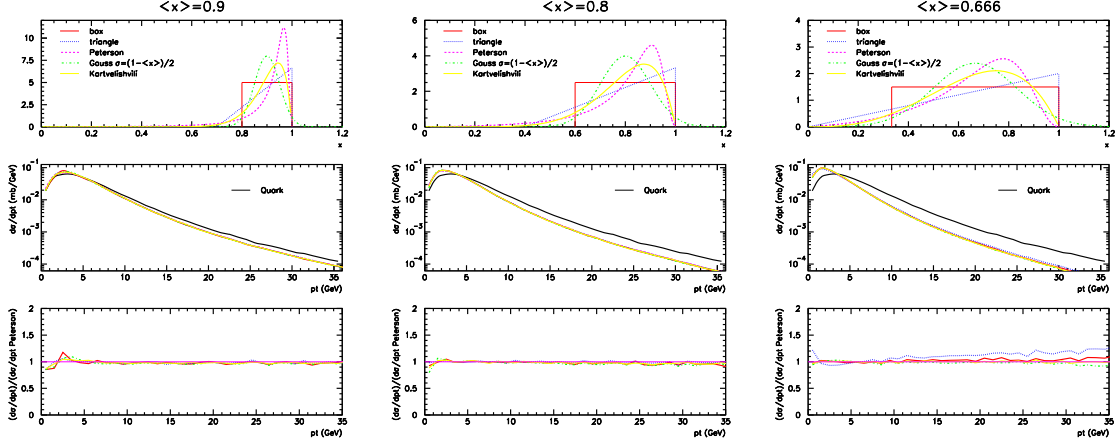

Fig. 2: Effect of the convolution of the heavy quark transverse momentum distribution with different test functions for different values of the non-perturbative FF $\langle x \rangle = 0.9$ (left), $0.8$ (center), $0.666$ (right). For each $\langle x \rangle$, the upper plot shows the test functions, the middle plot shows the perturbative $p_T$ distribution obtained with the MNR program for beauty at LHC and the hadron $p_T$ distributions after the convolution with the test functions. The lower plot shows the ratio of the different hadron $p_T$ distributions to the result obtained with the Peterson one.

is given by

$$\frac{d\sigma}{dp_T}(p_T) = \int dx\, x^{N-1}\, D^{\mathrm{np}}(x)\, C p_T^{-N} = \frac{d\sigma^{\mathrm{pert}}}{dp_T} \hat{D}_N^{\mathrm{np}}, \quad (6)$$

where $\hat{D}_N^{\mathrm{np}} = \int dx\, x^{N-1}\, D^{\mathrm{np}}(x)$ is the $N^{th}$ Mellin moment of the non-perturbative FF.

The hadron distribution is therefore governed by the $4^{th}$ - $5^{th}$ Mellin moments of $D^{\mathrm{np}}(x)$. It is interesting to translate the Mellin moments into more intuitive central moments

$$\mu_n = \int dx\, (x - \langle x \rangle)^n\, D^{\mathrm{np}}(x) \quad \text{for} \quad n \geq 2 \quad (7)$$

where $\langle x \rangle = \int dx\, x D^{\mathrm{np}}(x)$ is the mean value. The first Mellin moments, written in terms of $\langle x \rangle$ and $\mu_n$, are: $\hat{D}_1 = 1$, $\hat{D}_2 = \langle x \rangle$, $\hat{D}_3 = \langle x \rangle^2 + \mu_2$, $\hat{D}_4 = \langle x \rangle^3 + 3\mu_2 \langle x \rangle + \mu_3$.

In heavy quark fragmentation, the mean value of $D^{\mathrm{np}}(x)$ can be written as $\langle x \rangle = 1 - \delta$ where $\delta = \mathcal{O}(\Lambda_{\mathrm{qcd}}/m_Q)$ is small [11]. For any positive function with $\langle x \rangle = 1 - \delta$, defined in the interval $[0,1]$, the central moments are limited by $\delta$, $|\mu_n| \leq \delta$. In practice, reasonable heavy quark fragmentation functions are concentrated in a small region around $1 - \delta$ and therefore the higher central moments are small. To be specific, if the function is different from zero in a region of size $\pm K\delta$ (with $K = \mathcal{O}(1)$) around $1-\delta$ then $|\mu_n| \leq (K\delta)^n$. This means that the Mellin moments of reasonable FFs are given, to a good approximation, by the mean value to the $N-1$ power:

$$\hat{D}_N = \langle x \rangle^{N-1} + \mathcal{O}(\delta^2). \quad (8)$$

The expansion to $\delta^2$ involves the second central moment $\mu_2$: $\hat{D}_N = \langle x \rangle^{N-1} + \frac{(N-1)!}{2(N-3)!}\mu_2 \langle x \rangle^{N-3} + \mathcal{O}(\delta^3)$. For a reasonable FF and a perturbative distribution falling with the power $-N$, Eq. 6 and 8 give

$$\frac{d\sigma}{dp_T}(p_T) = \frac{d\sigma^{\mathrm{pert}}}{dp_T^Q}(p_T)\, (\langle x \rangle^{\mathrm{np}})^{N-1} + \mathcal{O}(\delta^2). \quad (9)$$

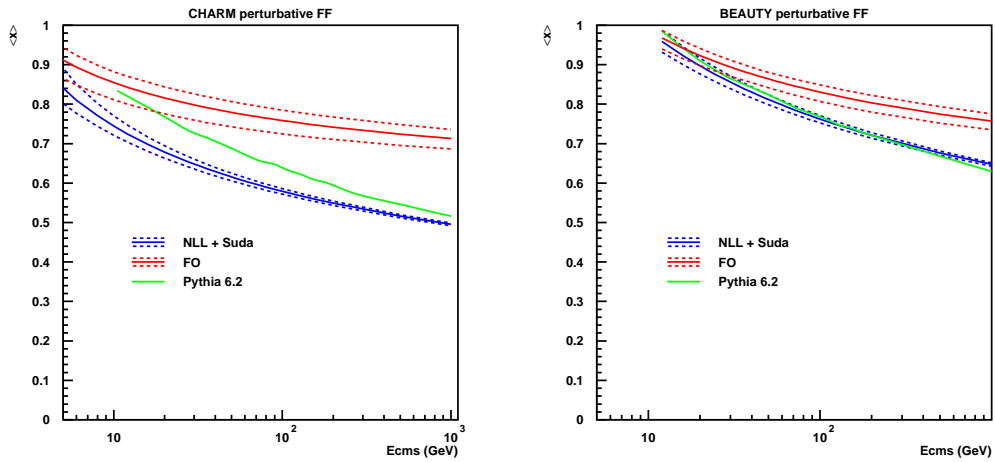

Fig. 3: Average fragmentation function from the perturbative calculations for charm (left) and beauty (right) as a function of the $e^+e^-$ center of mass energy.

Therefore the effect of the non-perturbative FF is to introduce a shift in the normalisation that depends on the average $x$, while the details of the shape of $D(x)$ have negligible effect. To check that this reasoning works with realistic fragmentation functions and realistic perturbative $p_T$ distributions, various functions with the same $\langle x \rangle$ but different shapes have been tested in convolution with the perturbative $p_T$ spectrum for $b$ production at LHC obtained with the NLO program MNR. The test functions considered are the Peterson [14] and Kartvelishvili [12] fragmentation functions, a Gaussian distribution with $\sigma = 1 - \mu$, a flat and a triangular distribution. Table 1 gives more detail about these functions. Three average values were chosen: $\langle x \rangle = 0.9, 0.8, 0.666$. Figure 2 shows the result of this test. For each average value, the convolutions are very similar, even if the test functions are very different. For $\langle x \rangle = 0.9, 0.8$, which are typical beauty or charm values, the hadron spectra agree within few %. For the extreme value of $\langle x \rangle = 0.666$, the results for the Peterson and Kartvelishvili functions give very similar hadron spectra while the less realistic Gaussian and Box shapes differ at most 10% from Peterson at large $p_T$ and the extreme Triangular function shows deviations up to $\sim 20\%$.

In conclusion the relevant fragmentation parameter for the inclusive hadron spectra at $pp$ and $ep$ colliders is the mean value $\langle x \rangle^{\mathrm{np}}$ of the non-perturbative FF. The next part will discuss, on the basis of $e^+e^-$ data, what values of $\langle x \rangle^{\mathrm{np}}$ are relevant for different calculations.

## 2.2 Extraction of $\langle x \rangle^{\mathrm{np}}$ from $e^+e^-$ data

In $e^+e^-$ interactions it is convenient to express the factorization ansatz, given for the heavy-hadron $p_T$ in eq. (5), in terms of the heavy-hadron momentum normalized to the maximum available momentum: $x_p = p^H/p^H_{\max}$, where $p^H_{\max} = \sqrt{(\frac{1}{2}E_{\mathrm{cms}})^2 - m_H^2}$:

$$\frac{d\sigma}{dx_p}(x_p) = \int \frac{dx}{x} D^{\mathrm{np}}(x) \frac{d\sigma^{\mathrm{pert}}}{dx_p}\left(\frac{x_p}{x}\right)$$

which corresponds to the following relation for the mean values: $\langle x_p \rangle = \langle x \rangle^{\mathrm{np}} \langle x \rangle^{\mathrm{pert}}$ where $\langle x_p \rangle$ is the mean hadron $x_p$, $\langle x \rangle^{\mathrm{np}}$ is the mean value of the non-perturbative FF and $\langle x \rangle^{\mathrm{pert}} = \int dx\, x\, \frac{d\sigma^{\mathrm{pert}}}{dx_p}$ is the mean value of the perturbative distribution. Then, taking $\langle x_p \rangle$ from experimental data and $\langle x \rangle^{\mathrm{pert}}$ from a particular perturbative calculation, it is possible to extract the value of $\langle x \rangle^{\mathrm{np}}$ valid for that calculation as

$$\langle x \rangle^{\mathrm{np}} = \langle x_p \rangle / \langle x \rangle^{\mathrm{pert}}. \tag{10}$$

Two perturbative calculations will be considered to extract $\langle x \rangle^{\mathrm{pert}}$: a fixed-order (FO) next-to-leading order (i.e. $\mathcal{O}(\alpha_S)$) calculation and a calculation that includes also the resummation of next-to-leading logarithms (NLL) and Sudakov resummation, both obtained with the HVQF program [19]. From the point of view of fragmentation, the FO calculation only considers the emission of a gluon from one of the two heavy quarks generated in the $e^+e^-$ collision while the NLL calculation includes the evolution of the FF from the hard interaction scale down to the scale given by the heavy quark mass. The parameters used for the FO and NLL models are $m_c = 1.5$ GeV, $m_b = 4.75$ GeV, $\Lambda_{\text{QCD}} = 0.226$ GeV and the renormalisation and factorization scales $\mu_R = \mu_F = E_{\text{cms}}$. The starting scale for FF evolution in the NLL model was chosen to be $m_Q$. The theoretical uncertainty was obtained by varying independently the normalisation and factorization scales by a factor 2 and 1/2 and taking the largest positive and negative variations as the uncertainty.

| Charm ($D^*$) measurement | $E_{\text{cms}}$ (GeV) | Measured variable | Value | ISR corr. (%) | $x_E \to x_p$ (%) | $\langle x_p \rangle^{\text{corr}}$ |
|---|---|---|---|---|---|---|
| OPAL [22] | 92 | $\langle x_E \rangle$ | $0.516^{+0.008}_{-0.005} \pm 0.010$ | $+0.4$ | $-0.4$ | $0.516 \pm 0.012$ |
| ALEPH [23] | 92 | $\langle x_E \rangle$ | $0.4878 \pm 0.0046 \pm 0.0061$ | $+0.4$ | $-0.4$ | $0.488 \pm 0.008$ |
| DELPHI [24] | 92 | $\langle x_E \rangle$ | $0.487 \pm 0.015 \pm 0.005$ | $+0.4$ | $-0.4$ | $0.487 \pm 0.016$ |
| TASSO [25] | 36.2 | $\langle x_E \rangle$ | $0.58 \pm 0.02$ | $+6.7$ | $-1.8$ | $0.61 \pm 0.02$ |
| DELCO [26] | 29 | $\epsilon^*_{\text{Pet.}}$ | $0.31^{+0.10}_{-0.08}$ | $+6.3$ | - | $0.55 \pm 0.03$ |
| ARGUS [27] | 10.5 | $\langle x_p \rangle$ | $0.64 \pm 0.03$ | $+4.2$ | - | $0.67 \pm 0.03$ |
| CLEO [28] | 10.5 | $\langle x_p \rangle$ | $0.611 \pm 0.007 \pm 0.004$ | $+4.2$ | - | $0.637 \pm 0.008$ |
| BELLE [29] | 10.58 | $\langle x_p \rangle$ | $0.61217 \pm 0.00036 \pm 0.00143$ | $+4.2$ | - | $0.6379 \pm 0.0016$ |

| Beauty ($B^{wd}$) measurement | $E_{\text{cms}}$ (GeV) | Measured variable | Value | ISR corr. (%) | $x_E \to x_p$ (%) | $\langle x_p \rangle^{\text{corr}}$ |
|---|---|---|---|---|---|---|
| OPAL [30] | 92 | $\langle x_E \rangle$ | $0.7193 \pm 0.0016 \pm 0.0038$ | $+0.3$ | $-0.9$ | $0.715 \pm 0.004$ |
| SLD [31] | 92 | $\langle x_E \rangle$ | $0.709 \pm 0.003 \pm 0.005$ | $+0.3$ | $-0.9$ | $0.705 \pm 0.006$ |
| ALEPH [32] | 92 | $\langle x_E \rangle$ | $0.716 \pm 0.006 \pm 0.006$ | $+0.3$ | $-0.9$ | $0.712 \pm 0.008$ |
| DELPHI [33] | 92 | $\langle x_E \rangle$ | $0.7153 \pm 0.0007 \pm 0.0050$ | $+0.3$ | $-0.9$ | $0.711 \pm 0.005$ |
| JADE [34] | 36.2 | $\langle x_E \rangle$ | $0.76 \pm 0.03 \pm 0.04$ | $+5.4$ | $-3.5$ | $0.77 \pm 0.06$ |
| DELCO [35] | 29 | $\langle x_E \rangle$ | $0.72 \pm 0.05$ | $+4.8$ | $-4.7$ | $0.72 \pm 0.05$ |
| PEP4-TPC [36] | 29 | $\langle x_E \rangle$ | $0.77 \pm 0.04 \pm 0.03$ | $+4.8$ | $-4.7$ | $0.77 \pm 0.07$ |

Table 2: Experimental results on the average fragmentation function in $e^+e^-$ collisions for $D^*$ mesons and weakly decaying beauty hadrons. The table reports, for each experiment, the published variable and the corrections applied to obtain $\langle x_p \rangle^{\text{corr}}$. All the measurement have been corrected for initial state radiation (ISR). Measurements reported in terms of $\langle x_E \rangle$ have been corrected to $\langle x_p \rangle$. In the case of ARGUS the average has been calculated from the full distribution. In the case of TASSO the error on the average was re-evaluated using the full distribution since the published error seems incompatible with the data. DELCO reports a fit with a Peterson distribution that has been translated into $\langle x_p \rangle^{\text{corr}}$. Systematical and statistical uncertainties, where reported separately, have been added in quadrature. The ALEPH beauty measurement refers to $B^+$ and $B^0$ mesons only (i.e. excluding $B_s$ and $\Lambda_b$), a MC study shows that this correspond to underestimating $\langle x_p \rangle^{\text{corr}}$ by $\sim 0.1\%$ only, which is negligible.

The experimental data are also compared to the PYTHIA 6.2 Monte Carlo program [37] which contains an effective resummation of leading-logarithms based on a parton-shower algorithm and which is interfaced to the Lund fragmentation model. In this case the MC model gives directly $\langle x_p \rangle$, while $\langle x \rangle^{\text{pert}}$ has been obtained taking the heavy quark at the end of the parton shower phase. The quark masses have been set to $m_c = 1.5$ GeV and $m_b = 4.75$ GeV, and all the parameters were set to the default values except for specific fragmentation parameters explained below. Three sets of fragmentation parameters were chosen for charm: the default fragmentation (Lund-Bowler), a longitudinal string fragmentation of the Peterson form with $\epsilon = 0.06$ (MSTJ(11)=3, PARJ(54)=-0.06) and the Lund-Bowler fragmentation with parameters retuned by the CLEO collaboration [28] (PARJ(41)=0.178, PARJ(42)=0.393, PARJ(13)=0.627). The two sets chosen for beauty are the default Lund-Bowler fragmentation and the Peterson fragmentation with $\epsilon = 0.002$ (MSTJ(11)=3, PARJ(55)=-0.002). Figure 3 shows $\langle x \rangle^{\text{pert}}$ from the perturbative calculations as a function of the centre of mass energy $E_{\text{cms}}$ for charm and bottom.

## 2.3 Charm

Charm fragmentation data are available from various $e^+e^-$ experiments. The most precise are those at the $Z^0$ pole at LEP (ALEPH [23], OPAL [22], DELPHI [24]) and near the $\Upsilon(4s)$ (ARGUS [27], CLEO [28], BELLE [29]). Less precise data are available in the intermediate continuum region from DELCO [26] at PEP and TASSO [25] at PETRA. Measurements in which the beauty component was not subtracted have been discarded [38, 39, 40]. The experimental data are reported in Table 2. Only measurements relative to the $D^{*\pm}(2010)$ meson are considered, to avoid the complications due to cascade decays that are present for ground state mesons. Charm quarks originating from gluon splitting rather than from the virtual boson from $e^+e^-$ annihilation may be relevant at LEP energies. This contribution is anyway already subtracted in the published data considered here, and it is consistently not considered in the perturbative calculations. Most of the experiments published the mean value of the $x$ distribution. The only exception is ARGUS, for which the mean value was computed from the published distribution. Some of the experiments give the results directly in terms of $x_p$, others in terms of the energy fraction $x_E = 2E_H/E_{\rm cms}$. The latter has been corrected to $x_p$ using the PYTHIA MC. The difference between $\langle x_p \rangle$ and $\langle x_E \rangle$ can be as large as 12% at $E_{\rm cms} = 10.5$ GeV and reduces to less than 1% at $E_{\rm cms} = 92$ GeV. Since the low-mass measurements are already given in terms of $x_p$, the applied corrections from $x_E$ to $x_p$ was always small. QED corrections are also needed to compare the experimental data to the QCD predictions. The initial state radiation (ISR) from the electrons has the effect of reducing the energy available for the $e^+e^-$ annihilation and therefore to reduce the observed value of $\langle x_p \rangle$. A correction, obtained by comparing the PYTHIA MC with and without ISR, was applied to the data to obtain $\langle x_p \rangle^{\rm corr}$. The correction is $\sim 4\%$ at $E_{\rm cms} = 10.5$ GeV, is largest in the intermediate region and is negligible at $E_{\rm cms} = 92$ GeV.

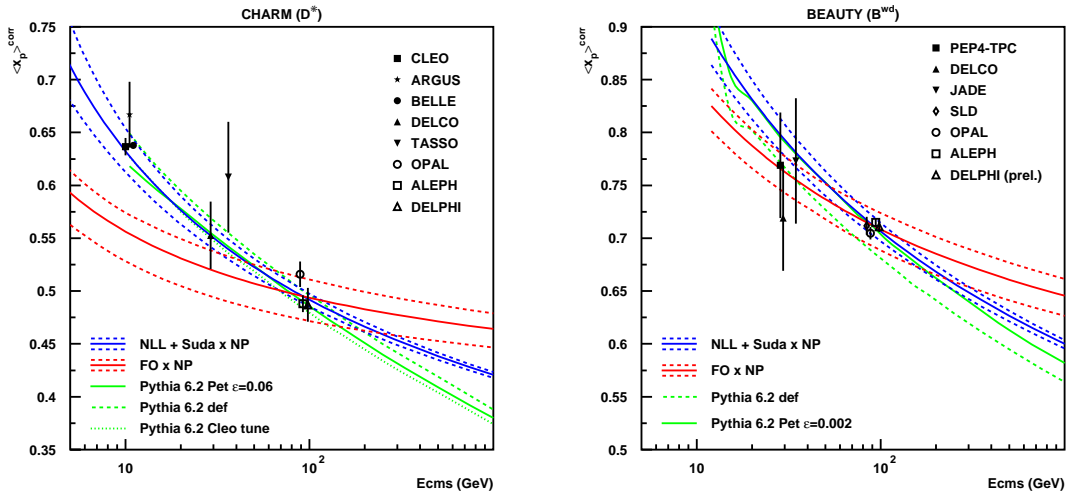

Fig. 4: Average fragmentation function as a function of the center of mass energy for charm (left) and beauty (right). The plots show the experimental results and the curves from NLL and FO theory with a non-perturbative fragmentation obtained using the data at the $Z^0$ energy. The curves from PYTHIA 6.2 with different fragmentation choices are also shown. The experimental points at the $\Upsilon(4s)$ and $Z^0$ resonances are shown slightly displaced in the horizontal axis for better legibility.

Only LEP data at $E_{\rm cms} = 92$ GeV were used to extract $\langle x \rangle^{\rm np}$ since the factorization of the non-perturbative FF could be spoiled by large $\mathcal{O}(m_Q/E_{\rm cms})$ terms at lower energies. Table 3 reports the LEP average $\langle x_p \rangle^{\rm corr}$, the perturbative results at 92 GeV and the resulting $\langle x \rangle^{\rm np}$ for NLL and FO calculations as well as $\langle x \rangle$ and $\langle x \rangle^{\rm pert}$ from PYTHIA. Figure 4 (left) shows $\langle x_p \rangle$ obtained by multiplying the perturbative calculations with the corresponding $\langle x \rangle^{\rm np}$, compared to the experimental data and to the PYTHIA MC with different fragmentation parameters.

With the non-perturbative $\langle x \rangle^{\rm np} = 0.849 \pm 0.018$ obtained at LEP energies, the NLL calculation can reproduce all the data within a quite small theoretical uncertainty. The FO calculation is instead too flat to reproduce the data even considering its large theoretical uncertainty band. The non-perturbative fragmentation $\langle x \rangle^{\rm np}$ obtained at LEP energy for the FO calculation is quite small ($0.65 \pm 0.04$) since it compensates the effect of the FF evolution that is missing in the perturbative part. Therefore FO calculations with $\langle x \rangle^{\rm np}$ extracted at LEP energy undershoot drastically the data at the $\Upsilon(4s)$.

The PYTHIA MC with the Lund-Bowler fragmentation reproduces the data reasonably well. The result with default parameters is slightly above the data while the result with the parameters tuned by the CLEO collaboration is slightly below. Both are compatible within the experimental uncertainty with all the experimental values with the exception of the very precise measurement from Belle from which they differ anyway by less than 2%. PYTHIA with the Peterson fragmentation with $\epsilon = 0.06$ reproduces well the LEP data but is too low at lower energies.

| Charm ($D^*$) @ 92 GeV | $\langle x_p \rangle^{\rm corr}$ | $\langle x_p \rangle^{\rm pert}$ | $\langle x_p \rangle^{\rm np}$ |
| --- | --- | --- | --- |
| Data | $0.495 \pm 0.006$ | - | - |
| NLL | - | $0.583 \pm 0.007$ | $0.849 \pm 0.018$ |
| FO | - | $0.76 \pm 0.03$ | $0.65 \pm 0.04$ |
| PYTHIA Def. | 0.500 | 0.640 | - |
| PYTHIA CLEO | 0.484 | 0.640 | - |
| PYTHIA Pet. $\epsilon = 0.06$ | 0.490 | 0.640 | - |
| | | | |
| Beauty ($B^{wd}$) @ 92 GeV | $\langle x_p \rangle^{\rm corr}$ | $\langle x_p \rangle^{\rm pert}$ | $\langle x_p \rangle^{\rm np}$ |
| Data | $0.7114 \pm 0.0026$ | - | - |
| NLL | - | $0.768 \pm 0.010$ | $0.927 \pm 0.013$ |
| FO | - | $0.83 \pm 0.02$ | $0.85 \pm 0.02$ |
| PYTHIA Def. | 0.686 | 0.773 | - |
| PYTHIA Pet. $\epsilon = 0.002$ | 0.710 | 0.773 | - |

Table 3: Average fragmentation functions at the $Z^0$ resonance for charm (top) and beauty. The table shows the average of the experimental data, the results from the NLL and FO calculations and from the PYTHIA MC with different fragmentation parameters. For the NLL and FO calculations $\langle x_p \rangle^{\rm np}$ is obtained by dividing the average from the experimental data by the perturbative result $\langle x_p \rangle^{\rm np} = \langle x_p \rangle^{\rm corr} / \langle x_p \rangle^{\rm pert}$.

## 2.4 Beauty

In the case of beauty we consider fragmentation measurements for the mix of weakly decaying hadrons $B^{\rm wd}$. Precise measurements are available only at the $Z^0$ peak (SLD [31], ALEPH [32], OPAL [30], DELPHI [33]). Lower energy measurements from PEP (PEP4-TPC [36], DELCO [35]) and PETRA (JADE [34]) have larger uncertainties. As for charm, corrections have been applied for ISR and to convert $\langle x_E \rangle$ to $\langle x_p \rangle$. The data are shown in Table 2 and the results in Table 3 and Figure 4 (right). Since precise data are available only at a single energy, it is impossible to test the energy beaviour of the theoretical predictions. As in the charm case, the energy dependence of PYTHIA and NLL theory are similar, while the FO prediction is much more flat, suggesting that also for beauty the non-perturbative fragmentation obtained for FO at the $Z^0$ could not be applied at lower energy. PYTHIA with Peterson fragmentation with $\epsilon = 0.002$ reproduces the data, while the default Lund-Bowler fragmentation is too soft.

## 2.5 Effect on predictions for heavy quark production at HERA and LHC

Going back to the heavy-hadron production in $ep$ and $pp$ collisions, eq. 9 shows that the uncertainty on the differential heavy-hadron cross section $d\sigma/dp_T$ is related to the uncertainty on the average non-perturbative fragmentation by

$$\Delta(d\sigma/dp_T) = N\Delta(\langle x \rangle^{\rm np}),$$

where $-N$ is the exponent of the differential cross section.

The state of the art calculations for photo- and hadro-production (FONLL [41, 42]) include NLO matrix elements and the resummations of next-to-leading logarithms. The appropriate non-perturbative fragmentation for FONLL is therefore obtained with the NLL theory which has the same kind of perturbative accuracy [43]. Since the NLL calculation gives a good description of $e^+e^-$ data, it seems appropriate to use the value and the uncertainty of $\langle x \rangle^{\rm np}$ as obtained from $e^+e^-$ data at the $Z^0$ peak. The relative error for the $D^*$ fragmentation is $\Delta\langle x \rangle^{\rm np}/\langle x \rangle^{\rm np} = 2\%$ which translates into an uncertainty of 9% on charm production at large $p_T$ at LHC ($N = 4.5$) of 9%. For beauty, the relative uncertainty $\Delta\langle x \rangle^{\rm np}/\langle x \rangle^{\rm np} = 1.4\%$ translates into an uncertainty on large-$p_T$ $B$-hadron production at LHC ($N = 3.8$) of 5.3%. These uncertainty are smaller or of the order of the perturbative uncertainties of the calculation. Nevertheless, it should be noted that this approach is only valid for large transverse momenta. At small transverse momenta the factorization ansatz breaks down and large corrections of order $m_Q/p_T$ may appear. Therefore, for the low-$p_T$ region, the uncertainty on the $p_T$ distribution is large and difficult to evaluate.

|  | $\langle x^{\text{np}} \rangle$ | $\epsilon(\min : \max)$ | $\alpha(\min : \max)$ | $\Delta\langle x^{\text{np}}\rangle/\langle x^{\text{np}}\rangle$ | $\Delta\sigma/\sigma$ (LHC) | $\Delta\sigma/\sigma$ (HERA) |
|---|---|---|---|---|---|---|
| FONLL $D^*$ | $0.849 \pm 0.018$ | $0.0040(0.0027 : 0.0057)$ | $10(9 : 12)$ | 2.1% | 9% | 12% |
| FONLL $B^{wd}$ | $0.927 \pm 0.013$ | $0.00045(0.00026 : 0.00072)$ | $24(20 : 30)$ | 1.4% | 5% | 7% |
| FO-NLO $D^*$ | $0.75 \pm 0.10$ | $0.02(0.004 : 0.08)$ | $5(3 : 10)$ | 13% | 60% | 70% |
| FO-NLO $B^{wd}$ | $0.89 \pm 0.04$ | $0.0015(0.0004 : 0.004)$ | $15(10 : 25)$ | 4.5% | 20% | 22% |

Table 4: Proposed value and uncertainty on $\langle x \rangle^{\text{np}}$ to be used with FO-NLO and FONLL programs for photo- and hadro-production of $D^*$ mesons and weakly decaying $B$ hadrons. The corresponding value and range for the Peterson $\epsilon$ and for the Kartvelishvili $\alpha$ parameters are also reported. The last columns show the corresponding relative uncertainty on $d\sigma/dp_T$ at LHC (assuming a negative power $N = 4.5/3.8$ for charm/beauty) and HERA ($N = 5.5/5.0$ for c/b).

For processes such as DIS and for particular observables FONLL calculations are not available. The best theory available in this case is the fixed order NLO theory. In this case the situation is complex since the equivalent FO calculation for $e^+e^-$ does not reproduce the experimental data. The proposed solution is to vary $\langle x \rangle^{\text{np}}$ from the same value obtained in the NLL case (that would be correct at low $p_T$, where the FF evolution is irrelevant) to the value obtained at the $Z^0$ energy (that would be valid at $p_T \sim 100$ GeV). Therefore we consider for charm $\langle x \rangle^{\text{np}} = 0.075 \pm 0.010$ and for beauty $\langle x \rangle^{\text{np}} = 0.089 \pm 0.004$. When these values are transported to heavy-hadron production at LHC, the corresponding uncertainties on $d\sigma/dp_T$ at large $p_T$ are 60% for charm and 20% for beauty. Therefore the NLO fixed order calculations cannot be used for precise predictions of the charm (and to a lesser extent beauty) production at $pp$ and $ep$ colliders.

In the FO-NLO and FONLL programs the hadron distributions are obtained by reducing the quark momenta according to a given fragmentation functions. Typical fragmentation functions used in these programs are the Peterson and Kartvelishvili forms. Table 4 summarises the proposed values and uncertainties for $\langle x \rangle^{\text{np}}$ to be used with FO-NLO and FONLL calculations and reports the corresponding values and ranges for the Peterson and Kartvelishvili parameters. Similar ranges are used in the calculations presented in the section on "Benchmark cross sections" in these proceedings.

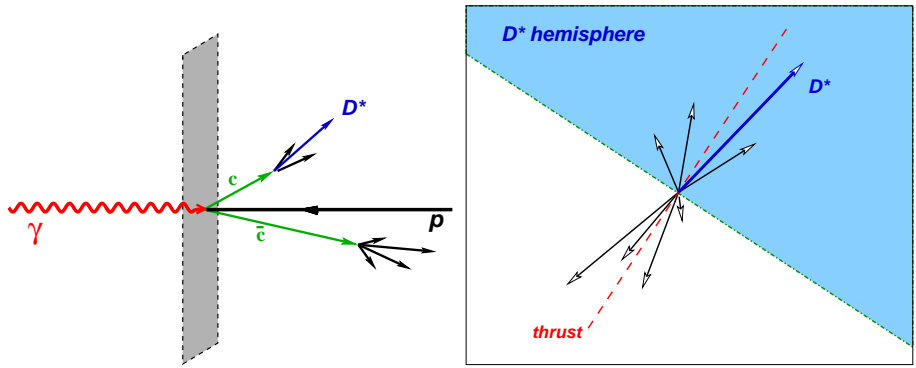

Fig. 5: Hemisphere method

## 3  Measurements of the charm quark fragmentation function at HERA

*authors: J. Braciník and G. Grindhammer*

The differential cross section for the inclusive production of a heavy hadron $H$ from a heavy quark $h$ can be computed in perturbative QCD (pQCD) as a convolution of a short-distance cross section $\hat{\sigma}(p_h)$ with a fragmentation function $D_H^h(z)$:

$$d\sigma(p_H) = \int dz dp_h d\hat{\sigma}(p_h) D_H^h(z) \delta(p - z p_h) \tag{11}$$

The quantity $z$ is the fractional momentum of the heavy quark $h$ which is transferred to the heavy hadron $H$, and the normalized fragmentation function $D_H^h(z)$ gives the probability to observe the hadron $H$ with a momentum fraction $z$.

The precise definition of $D_H^{(h)}(z)$ is in some sense arbitrary. Due to the short and long-distance processes involved, the fragmentation function contains a perturbative and a non-perturbative component. Since the former can be calculated only up to some order in the strong coupling, the non-perturbative component in practice will have to absorb some of the missing higher order corrections. The calculable perturbative part can be absorbed into the definition of $\hat{\sigma}(p_h)$. Since for heavy quarks perturbative gluon emission do not lead to collinear divergencies, the perturbative evolution is well defined, and it is possible to absorb them into $\hat{\sigma}(p_h)$ and to perform perturbative evolution down to a scale of the heavy quark mass $m_h$. In this case the non-perturbative fragmentation function $D_H^h(z)$ accounts for the transition of an almost on-shell quark $h$ into a heavy hadron $H$.

According to the QCD factorization theorem, the non-perturbative fragmentation functions (FF) depend neither on the type of the hard process nor on the scale at which the heavy quark $h$ is originally produced. This implies universality of FF and allows - if valid - to extract fragmentation functions from data for one particular reaction (usually $e^+e^-$ annihilation) and to use them to predict cross sections for other reactions (e.g. in $pp$ and $ep$-collisions). In order to be able to check the reliability of pQCD predictions, it is necessary to check the universality of FF.

In practice, different theoretically motivated functional forms for $D_H^h(z)$ are used, depending on one more free parameters which are fitted to data. Among frequently used expressions are those by Peterson et al. [14] and by Kartvelishvili et al. [12].

From Equation 11 it is clear that $D_H^h(z)$ cannot be measured directly, since all observables are convoluted with the perturbative cross section. In case of $ep$ and $pp$ scattering there are additional convolutions with the parton density functions of one or two interacting hadrons. However, there are some observables which are more sensitive to $D_H^h(z)$ then others.

In $e^+e^-$, a convenient way to study fragmentation is to study the differential cross section of a heavy meson as a function of a scaled momentum or energy z. A customary experimental definition[3] of z is z =

---

[3]Sometime there are slightly different definitions of z [28] in case of heavy meson production close to threshold.

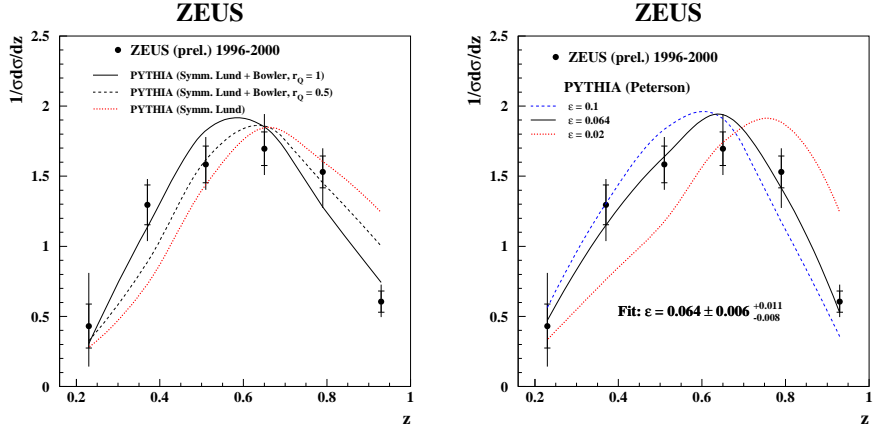

Fig. 6: Normalized differential cross section as a function of $z_{jet}$ as measured by ZEUS in photoproduction for jets with an associated $D^*$-meson with $|\eta_{jet}| < 2.4$ and $E_{T,jet} > 9$ GeV.

$E_H/E_{beam}$, where $E_{beam}$ is the energy of the beams in the center-of-mass system. In leading order, i.e. without gluon emissions, it is also the energy of the charm and anti–charm quark and is equal to $D_H^h(z)$. In contrast to $e^+e^-$ annihilation the choice of a fragmentation observable in $ep$ collisions is more difficult. Two different observables have been used so far, both of them having the feature that in leading order QCD, the z-distributions are equal to $D_H^h(z)$.

In the case of what is called here the jet method, the energy of the charm quark is approximated by the energy of the charm-jet, tagged by a $D^*$-meson, which is considered to be part of the jet. The scaling variable is then defined as $z_{jet} = (E + p_L)_{D^*}/(E + p)_{jet}$.

The idea of the so called hemisphere method (see Figure 5) is to exploit the special kinematics of charm events in $ep$ collisions. The dominant charm production process has been shown to be boson-gluon fusion. If such an event is viewed in the photon-proton center-of-mass frame, the photon puts its full energy into the hard subprocess, while the proton interacts via a gluon, which typically carries only a small fraction of the proton momentum. As the result, both quarks produced, $c$ and $\bar{c}$, move in the direction of the photon. Assuming no initial gluon $k_T$ and no gluon radiation, their transverse momenta are balanced (see Fig. 5, left).

This can be seen best by projecting the quark momenta onto the plane perpendicular to the $\gamma$-p axis. In this plane it is possible to distinguish rather efficiently between the products of the fragmentation of the charm quark and its antiquark. The momenta of all particles are projected onto the plane and the thrust axis in this plane is found (see Fig. 5, right). The plane is then divided into two hemispheres by the line perpendicular to the thrust axis. All particles, lying in the hemisphere containing the $D^*$-meson are marked and their three-momenta and energy are summed-up to give the hemisphere's momentum and energy, which is used to approximate the momentum and energy of the respective charm/anti-charm quark. The scaling variable $z_{hem}$ is then defined as $z_{hem} = (E + p_L)_{D^*}/\sum_{hem}(E + p)$.

The ZEUS collaboration has provided preliminary results [44] on a measurement of normalized differential cross sections of $D^*$-mesons as a function of $z_{jet}$. The measurement was done in photoproduction, in the kinematic range $Q^2 < 1$ GeV$^2$ and $130 < W < 280$ GeV. The $D^*$-mesons were reconstructed using the 'golden channel' $D^* \to D^0\pi_s \to K\pi\pi_s$ and were required to be in the central rapidity region $|\eta| < 1.5$ and to have $p_T > 2$ GeV. Jets were reconstructed using the inclusive $k_\perp$ algorithm. They fulfill the conditions $|\eta_{jet}| < 2.4$ and $E_{T,jet} > 9$ GeV. The jets were reconstructed as massless jets. The beauty contribution to the $D^*$-meson cross section, which amounts to about 9%, was subtracted using the prediction of PYTHIA. The scaling variable was calculated as $z_{jet} = (E + p_L)_{D^*}/(2E_{jet})$. The cross section as a function of $z_{jet}$ is shown in Fig. 6. The uncertainties due to choice of the model used to correct for detector effects, and the subtraction of the beauty component were the largest contributions to the total uncertainty.

The H1 collaboration has recently presented preliminary results [45] on the normalized differential cross section also of $D^*$-mesons as a function of both $z_{\text{hem}}$ and $z_{\text{jet}}$. Their measurement was performed in the kinematic range $2 < Q^2 < 100$ GeV$^2$ and $0.05 < y < 0.7$. The $D^*$-mesons were reconstructed using the 'golden channel' with $|\eta| < 1.5$ in the central rapidity region and $p_T > 1.5$ GeV. The jets were reconstructed using the inclusive $k_\perp$ algorithm in the photon-proton center of mass frame, using the massive recombination scheme. The jets were required to have $E_{\text{T,jet}} > 3$ GeV. The scaling variables were calculated as $z_{\text{jet}} = (E + p_L)_{D^*}/(E + p)_{\text{jet}}$ and $z_{\text{hem}} = (E + p_L)_{D^*}/\sum_{\text{hem}}(E + p)$ and are shown in Fig. 7. The resolved contribution was varied between 10 and 50% and the beauty contribution as predicted by the model was varied by a factor of two. The resulting uncertainties are part of the systematic error of the data points. For these distributions, the contribution of $D^*$-mesons coming from the fragmentation of beauty, as predicted by RAPGAP, was subtracted. It amounts to about $1.3\%$ for the hemisphere method and $1.8\%$ for the jet method. The dominant systematic errors are due to the model uncertainty and the signal extraction procedure.

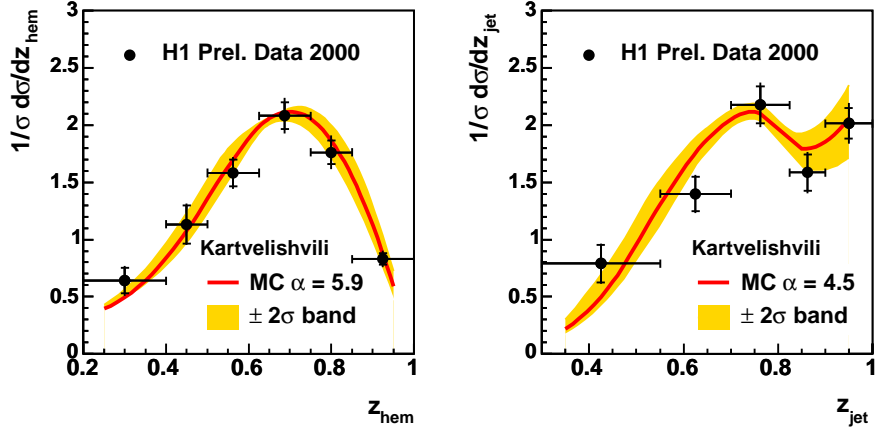

Fig. 7: Normalized differential cross section of $D^*$-meson as a function of $z_{\text{jet}}$ and $z_{\text{hem}}$ in DIS as measured by H1.

| Parametrization | | Hemisphere Method | Jet Method | Suggested range |
|---|---|---|---|---|
| Peterson | $\varepsilon$ | $0.018^{+0.004}_{-0.004}$ | $0.030^{+0.006}_{-0.005}$ | $0.014 < \varepsilon < 0.036$ |
| Kartvelishvili | $\alpha$ | $5.9^{+0.9}_{-0.6}$ | $4.5^{+0.5}_{-0.5}$ | $4 < \alpha < 6.8$ |

Table 5: Extracted fragmentation parameters for $z_{\text{jet}}$ and $z_{\text{hem}}$ from H1.

Both collaborations used the normalized z-distributions to extract the best fragmentation parameters for a given QCD model.

In case of ZEUS, PYTHIA was used together with the Peterson fragmentation function. The MC was fit to the data using a $\chi^2$-minimization procedure to determine the best value of $\epsilon$. The result of the fit is $\epsilon = 0.064 \pm 0.006^{+0.011}_{-0.008}$.

The H1 collaboration used RAPGAP 3.1 interfaced with PYTHIA 6.2. The contribution due to $D^*$-mesons produced in resolved photon processes (in DIS), which amounts to $33\%$ as predicted by the model, has been included in addition to the dominant direct photon contribution. The Peterson and Kartvelishvili parametrizations were both fitted to the data. The results are shown in Table 5.

The parameter of the Peterson fragmentation function as measured by ZEUS and H1 do not agree with each other. This may be due to the different phase-space regions covered by the two measurements (photoproduction versus DIS, $E_{\text{T,jet}} > 9$ GeV versus $E_{\text{T,jet}} > 3$ GeV ) and most importantly, the parameters were extracted for two different models[4]. More detailed investigations are needed to resolve this question.

---

[4]While ZEUS has used the default parameters for PYTHIA, H1 has taken the tuned parameter values of the ALEPH collabora-

The fragmentation function parameters extracted by H1 with the hemisphere and the jet method differ by less than 3 $\sigma$. At the present level of statistical and systematic errors it is not possible to exclude a statistical fluctuation. On the other hand, the potential discrepancy may be a sign of deficiencies in the modelling of the hadronic final state in RAPGAP.

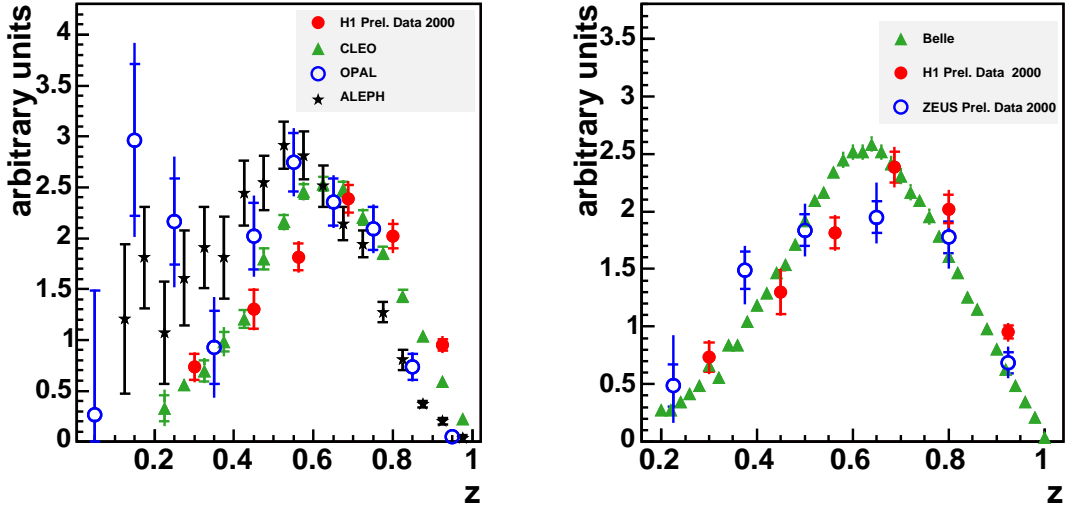

Fig. 8: Comparison of the z-distributions from CLEO, OPAL and ALEPH (left) and Belle and ZEUS (right) with the one from the hemisphere method from H1. All distributions are normalized to unit area from z = 0.4 to z = 1.

The measured $z_{hem}$ distribution of H1 is compared to data from the ALEPH [23], OPAL [22] and CLEO [28] collaborations in Fig. 8 (left) and to ZEUS [44] and Belle [29] in Figure 8 (right)[5]. The results of H1 are in rough agreement with recent data from CLEO and Belle, taken at at 10.5 and 10.6 GeV, corresponding roughly to the average energy of the system at H1. Differences beyond the measurement errors can be observed. However, this may be due to the somewhat different definitions used for the fragmentation observable z, different kinematics, different processes, or it may be a sign of the violation of universality.

| PARAMETRIZATION | | ALEPH | OPAL | BELLE | ZEUS | H1: $z_{hem}$ | H1: $z_{jet}$ |
|---|---|---|---|---|---|---|---|
| Peterson | $\varepsilon$ | $0.034 \pm 0.0037$ | $0.034 \pm 0.009$ | 0.054 | $0.064^{+0.013}_{-0.010}$ | $0.018^{+0.004}_{-0.004}$ | $0.030^{+0.006}_{-0.005}$ |
| Kartvelishvili | $\alpha$ | — | $4.2 \pm 0.6$ | 5.6 | — | $5.9^{+0.9}_{-0.6}$ | $4.5^{+0.5}_{-0.5}$ |

Table 6: Extracted fragmentation parameters from $e^+e^-$ annihilation data by ALEPH [23], OPAL [22] and BELLE [29] and from $ep$ data by ZEUS [44] and H1 [45].

While the z distributions don't need to agree, the fragmentation parameters, which are extracted from them, should agree. This can be expected only, if a model with consistent parameter settings is used which provides an equally good description of the different processes at their respective scales. The values of the Peterson fragmentation parameter, as extracted by different experiments within the PYTHIA/JETSET models, are summarized in Table 6. Contrary to expectations, discrepancies between various experiments can be seen. A consistent phenomenological analysis of these data is therefore needed in order to resolve the reasons for the discrepancies.

The measurement of the charm fragmentation function at HERA provides an important test of our understanding of heavy quark production. We may hope that HERA II data and a phenomenological analysis of existing data will bring new insights in this area.

---

tion [46]

[5]Data points were taken from the figure in [29] and [44].

# Benchmark cross sections for heavy-flavour production

*O. Behnke*[1], *M. Cacciari*[2], *M. Corradi*[3], *A. Dainese*[4], *H. Jung*[5], *E. Laenen*[6], *I. Schienbein*[7], *H. Spiesberger*[8]

[1]Universität Heidelberg, Philosophenweg 12 69120 Heidelberg, FRG;
[2]LPTHE - Université P. et M. Curie (Paris 6), Paris, France;
[3]INFN Bologna, via Irnerio 46, Bologna, Italy;
[4]Università di Padova and INFN, Padova, Italy;
[5]Deutsches Elektronen-Synchroton DESY, Hamburg, FRG;
[6]NIKHEF,Theory Group, Kruislaan 409, 1098 SJ Amsterdam, The Netherlands;
[7]Southern Methodist University Dallas, 3215 Daniel Avenue, Dallas, TX 75275-0175, USA;
[8]Johannes-Gutenberg-Universität Mainz, D-55099 Mainz, FRG

**Abstract**
Reference heavy-flavour cross sections at HERA and LHC have been computed following different theoretical approaches and the results have been compared.

*coordinators: M. Corradi, A. Dainese*

## 1 Introduction

This section presents a comparison of cross sections for HERA and LHC calculated according to different theoretical approaches. Different programs were used to calculate the same reference cross sections, using, as far as possible, the same input parameters and a consistent method to evaluate uncertainties. In this way it is possible to identify processes and kinematical regions in which different approaches give the same answer and regions where they differ. Unified criteria to evaluate the theoretical uncertainty should also allow to understand what approach is expected to be more precise. Moreover these calculations, which incorporate up-to-date parameters and PDF parametrisations, can be used as a reference for experiments and for further theoretical predictions. The cross sections presented here, are available in computer-readable format from the web page `http://www-zeus.desy.de/˜corradi/benchmarks`, where figures in color can also be found.

## 2 Programs

A list of the programs used for the cross section calculations is given below. For further details see the references and the theoretical review on heavy quark production in these proceedings.

- MNR [1] is a fixed-order (FO) NLO program for heavy-flavour hadro-production, it was used for LHC cross sections;
- FMNR [2, 3] is an extension of the previous program to photoproduction, it was used for photoproduction at HERA;
- HVQDIS [4, 5] is a FO-NLO program for heavy-flavour production in deep-inelastic scattering (DIS), it has been used for DIS at HERA;
- FONLL [6, 7] provides matched massive-massless calculations with NLO accuracy and resummation of large $p_T$ logarithms. It is available for hadro- and photo-production and was used for HERA photoproduction and LHC cross sections;
- GM-VFNS [8, 9, 10, 11] is a calculation in the generalised massive variable flavour number scheme. It has been used for charmed hadron $p_T$ spectra at LHC and in photoproduction at HERA;
- CASCADE 1.2009 [12] is a full Monte Carlo program based on unintegrated parton densities and $K_T$ factorisation. It has been used to calculate cross sections for Photoproduction and DIS at HERA and for LHC;

| Parameter | program | central value | lower/upper |
|---|---|---|---|
| $\Lambda_{\mathrm{QCD}}^5$ | all | 0.226 GeV | fix |
| $m_c$ | all | 1.5 GeV | 1.3/1.7 GeV |
| $m_b$ | all | 4.75 GeV | 4.5/5.0 GeV |
| p-PDF | all-CASCADE | CTEQ6.1 [15] | MRST2002 [16]/Alekhin [17] |
|  | CASCADE | CCFM A0 | - |
| $\gamma$-PDF | FMNR, FONLL | AGF [18] | GRV [19] |
| $f(c \to D^*)$ | all | 0.235 | fix |
| $c$ fragmentation: | (F)MNR,HVQDIS | Pet. [20] $\epsilon_c = 0.021$ | 0.002/0.11 |
|  | FONLL | BCFY $r = 0.1$ | 0.06/0.135 |
|  | GM-VFNS | [9] | - |
|  | CASCADE, RAPGAP | Pet. $\epsilon_c = 0.075$ | Def./$\epsilon_c = 0.05$ |
| $b$ fragmentation: | (F)MNR,HVQDIS | Pet. $\epsilon_b = 0.001$ | 0.0002/0.004 |
|  | FONLL | Kart. $\alpha = 29.1$ | 25.6/34.0 |
|  | CASCADE, RAPGAP | Pet. $\epsilon_b = 0.002$ | Def./$\epsilon_b = 0.005$ |

Table 1: The table shows input parameter to the different programs with the corresponding lower and upper values used for the uncertainty: $\Lambda_{\mathrm{QCD}}$, the quark masses, the proton and photon parton densities, the fraction of c quarks decaying into a $D^*$ meson, and the parameters used for fragmentation. The fragmentation form are abbreviated to Pet. for Peterson, Kart. for Kartvelishvili, Def. for the default PYTHIA fragmentation

- RAPGAP 3 [13] is a multi-purpose MC program for $ep$ collisions, it implements heavy-flavour production through the boson-gluon-fusion process $\gamma^* g \to Q\bar{Q}$ at leading order. It has been used for DIS at HERA. Both CASCADE and RAPGAP use PYTHIA [14] routines for fragmentation.

## 3 Parameters and uncertainties

The different calculations were compared using the same input parameters and, where possible, with total uncertainty bands computed in a consistent way. The total uncertainty band includes the effect of the uncertainty on the input parameters and on the missing higher orders in the perturbative expansion.

### 3.1 Perturbative uncertainty

The perturbative uncertainty was obtained by varying the renormalisation and factorisation scales independently in the range $0.5\mu_0 < \mu_F, \mu_R < 2\mu_0$, while keeping $1/2 < \mu_R/\mu_F < 2$, were $\mu_0$ is the nominal value, typically set to the transverse mass $p_T^2 + m_Q^2$ or to $4m^2 + Q^2$ in the DIS case. The largest positive and negative variations were taken as the perturbative uncertainty band.

### 3.2 Input parameters

The uncertainty from the input parameters was obtained by varying each parameter around the central value. An effort was made within the working group to find the best central value and uncertainty for the input parameters. The values used for the perturbative parameters $\Lambda_{\mathrm{QCD}}^5$, $m_c$, $m_b$ as well as the parton distribution functions (PDF) for the proton and for the photon are reported in Table 1.

For practical reasons, rather than using the full treatment of the PDF uncertainty, few different parametrisations were tried and it was checked that the choice of the PDF set always gives a small contribution to the total uncertainty band. In the case of CASCADE, the CCFM A0 parametrisation was used as the central value while the PDF parametrisations A0+ and A0-, obtained from fits to DIS data with different renormalisation scales, were used in conjunction with the variation of the renormalisation scale.

Since the different programs have different perturbative contents, different parameters for the non-perturbative fragmentation function were used. The values were chosen in order to correspond to the same average fragmentation in $e^+e^-$ collisions as explained in the section on heavy quark fragmentation in these proceedings. Table 1 reports the fragmentation form and the corresponding parameters used in the different programs.

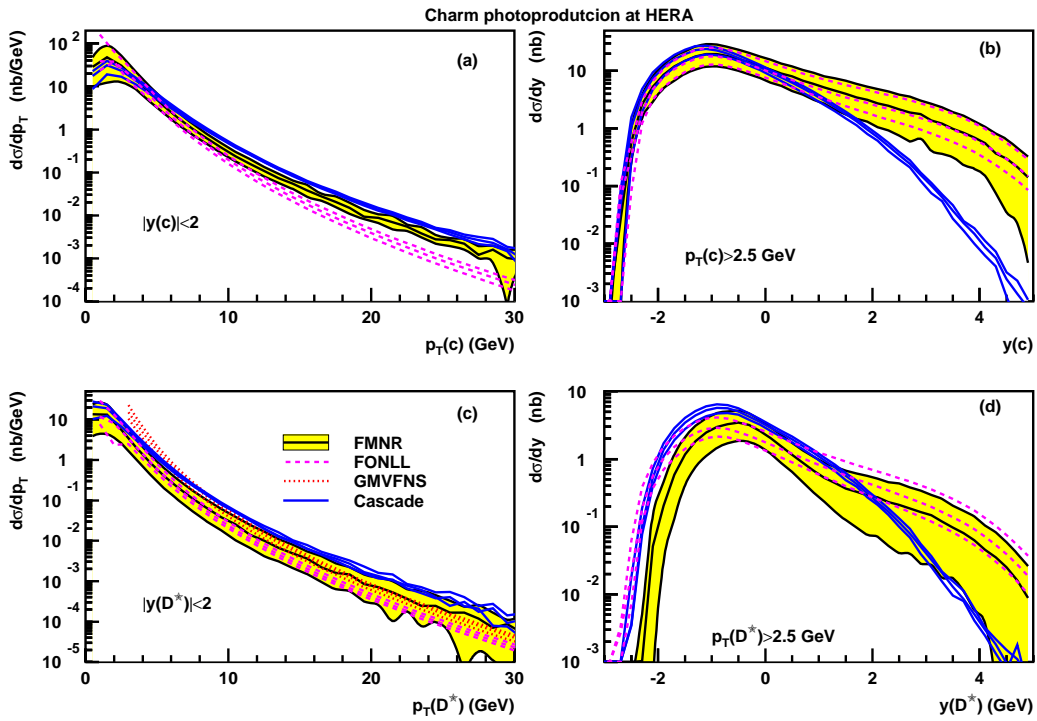

Fig. 1: Cross sections for charm photoproduction at HERA ($Q^2 < 1$ GeV$^2$, $0.2 < y < 0.8$). The differential cross sections as a function of the $p_T$ of the $c$ quark for rapidity $|Y| < 2$ and as a function of the rapidity of the $c$ quark for $p_T > 2.5$ GeV are shown in (a), (b). Plots (c) and (d) show similar cross sections for the production of a $D^*$ meson. The cross sections are shown for FMNR (shaded band), FONLL (empty band with dashed lines), GM-VFNS (empty band with dotted lines) and CASCADE (empty band with full lines).

In the FONLL calculation for charm, the BCFY [21] fragmentation parameter $r$ was varied in conjunction with the variation of the charm mass since different values of $r$ are obtained from $e^+e^-$ data for different $m_c$ [22]. Similarly for beauty, the Kartvelishvili [23] parameter $\alpha$ was varied in conjunction with the variation of the $b$ mass [23]. For GM-VFNS, the fragmentation functions and fractions were taken from [9].

The total uncertainty band was obtained from the sum of the uncertainties added in quadrature coming from the parameter variations and the perturbative uncertainty.

## 4 Results

### 4.1 HERA Photoproduction

The results for HERA Photoproduction are given as $ep$ cross-sections for $0.2 < y < 0.8$ ($y$ is the Bjorken variable while $Y$ is the rapidity in the laboratory frame) and $Q^2 < 1$ GeV$^2$. The beam energies have been set to $E_e = 27.52$ GeV, $E_p = 920$ GeV with the proton beam going in the positive rapidity direction.

Figure 1 shows the differential cross sections as a function of the charm quark transverse momentum (a) and pseudorapidity (b). In (c) and (d) the same cross sections are given for the charmed $D^*$ meson. A meaningful comparison can be performed only for the hadron variables, which are the real physical observables, since the quark level may be defined differently in different approaches. The FO calculation (FMNR) shows a large uncertainty ($\sim 60\%$) at the hadron level due to the related uncertainty on the fragmentation parameters. The resummed programs FONLL and GM-VFNS have much smaller uncertainty and are within the FMNR uncertainty band. The central values from FMNR and FONLL coincide at low transverse momenta. GM-VFNS, instead, tends to grow unphysically at low $p_T(D^*)$. As can be seen in (c), the quark-level disagreement between FO (FMNR) and FONLL calculations is consistently removed at the hadron-level. The unintegrated-PDF Monte Carlo CASCADE tends to be above the other calculations, in particular at large $p_T$. In the case of beauty (Fig. 2) the uncertainty bands are smaller ($\sim 20\%$ for FMNR), CASCADE and FMNR are in good agreement. Due to the large $b$ mass, the resummed calculation FONLL (not shown) is expected to be similar to the fixed-order one (FMNR). For both beauty and charm, FMNR and FONLL show a shoulder at positive rapidities (b, d) due to the "hadron-like" component of the photon that is not present in CASCADE.

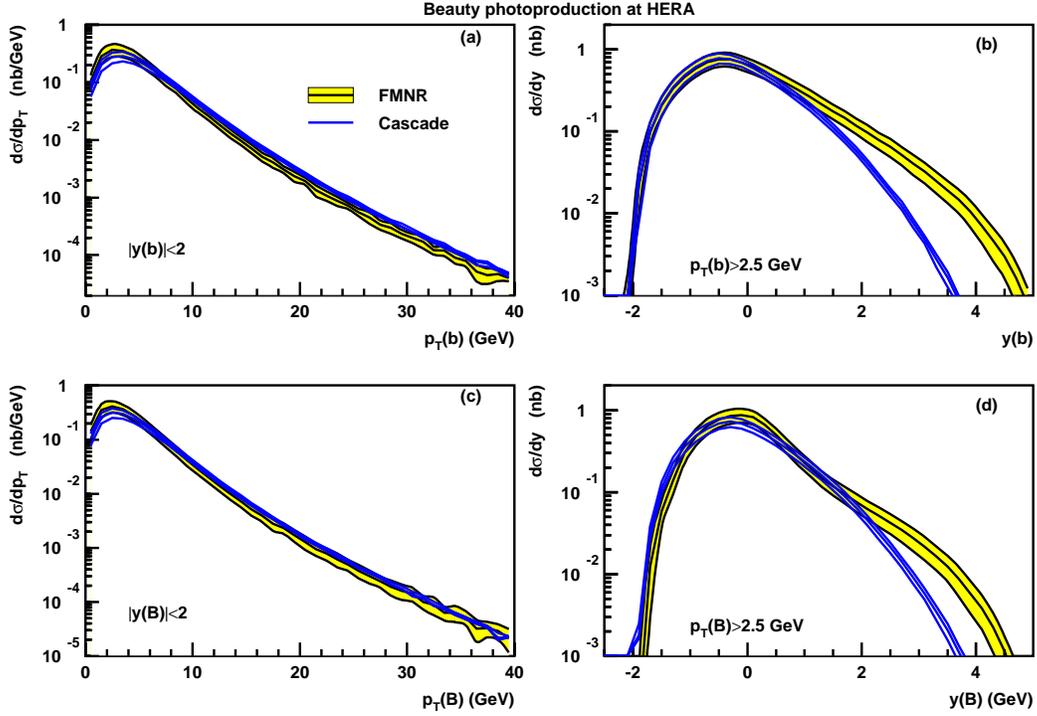

Fig. 2: Cross sections for beauty photoproduction at HERA ($Q^2 < 1$ GeV$^2$, $0.2 < y < 0.8$). The differential cross sections in $p_T$ and rapidity of the $b$ quark are shown in (a), (b). Plots (c) and (d) show the cross sections for the production of a weakly-decaying $B$ hadron as a function of $p_T(B)$ and $Y(B)$. The cross sections are shown for FMNR (shaded band) and CASCADE (empty band with full lines).

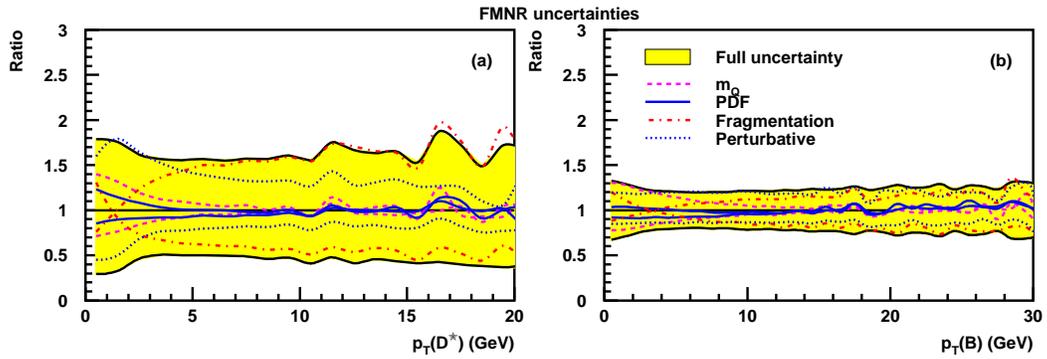

Fig. 3: Breakdown of the different components of the FMNR uncertainty for $d\sigma/dp_T$ for charmed (a) and beauty (b) hadrons in photoproduction at HERA. The plots show the ratio of the upper/lower side of each uncertainty to the nominal value. The following sources of uncertainty are shown: quark mass ($m_Q$), parton density parametrisation (PDF), fragmentation parameter and the perturbative uncertainty from scale variations.

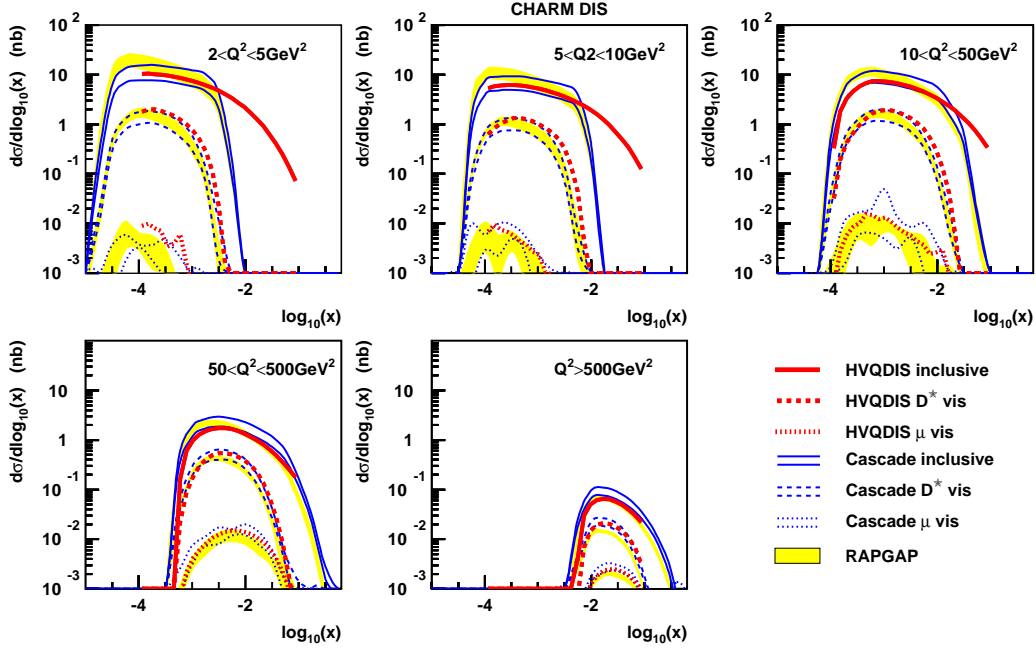

Fig. 4: Charm cross sections in DIS at HERA. Each plot shows the distribution of $\log_{10}(x)$ in a different $Q^2$ range for the inclusive cross-section, the cross-section for a $D^*$ meson in the "visible" range $p_T(D^*) > 1.5$ GeV, $|Y(D^*)| < 1.5$ and the cross-section for a muon from charm decay in the range $p_T(\mu) > 3$ GeV, $|Y(\mu)| < 2$. The thick curves show the central value from HVQDIS, the thin curves represent the uncertainty band from CASCADE and the shaded area shows the uncertainty band from RAPGAP. The fluctuations in the muon cross sections are due to the limited statistics.

Figure 3 shows the different components of the FMNR uncertainty band for charm and beauty. The uncertainties for quark production are typically dominated by the perturbative scale uncertainty with the exception of the low transverse momentum region ($p_T \sim m_Q$) where the uncertainty from the quark-mass can dominate. For hadron production, the fragmentation dominates the FMNR uncertainty at large $p_T$. The PDF uncertainty was found to be small. Resummed calculation have smaller uncertainty bands due to the smaller perturbative and fragmentation contributions at large $p_T$.

### 4.2 HERA DIS

Heavy quark production in DIS is not available in the matched massive-massless approach (except for total cross sections). Therefore the DIS comparison was limited to the FO-NLO program HVQDIS, the unintegrated-PDF MC CASCADE and the RAPGAP Monte Carlo. The DIS cross sections at HERA are reported as $d\sigma/d\log_{10}(x)$ for different bins of $Q^2$ and are intended at the Born level, without electroweak corrections. Figure 4 shows, for each $Q^2$ bin, the inclusive charm cross-section, the cross section for observing a $D^*$ meson in the "visible" range $p_T(D^*) > 1.5$ GeV, $|Y(D^*)| < 1.5$ and for observing a muon in the range $p_T(\mu) > 3$ GeV, $|Y(\mu)| < 2$. The three calculations are compatible at intermediate values of $x$ ($\sim 10^{-3}$). At large $x$ and low $Q^2$, CASCADE and RAPGAP drop to zero much faster than HVQDIS. At low $x$ RAPGAP is significantly larger than HVQDIS while both are within the uncertainty band given by CASCADE. A similar behavior is seen for beauty (Fig. 5). The uncertainty on HVQDIS, not given here, is expected to be small ($\sim 10 - 20\%$ for beauty [24]). The high-$x$ discrepancy between HVQDIS and the other two calculations seems therfore to be beyond the program uncertainties and deserves further investigations.

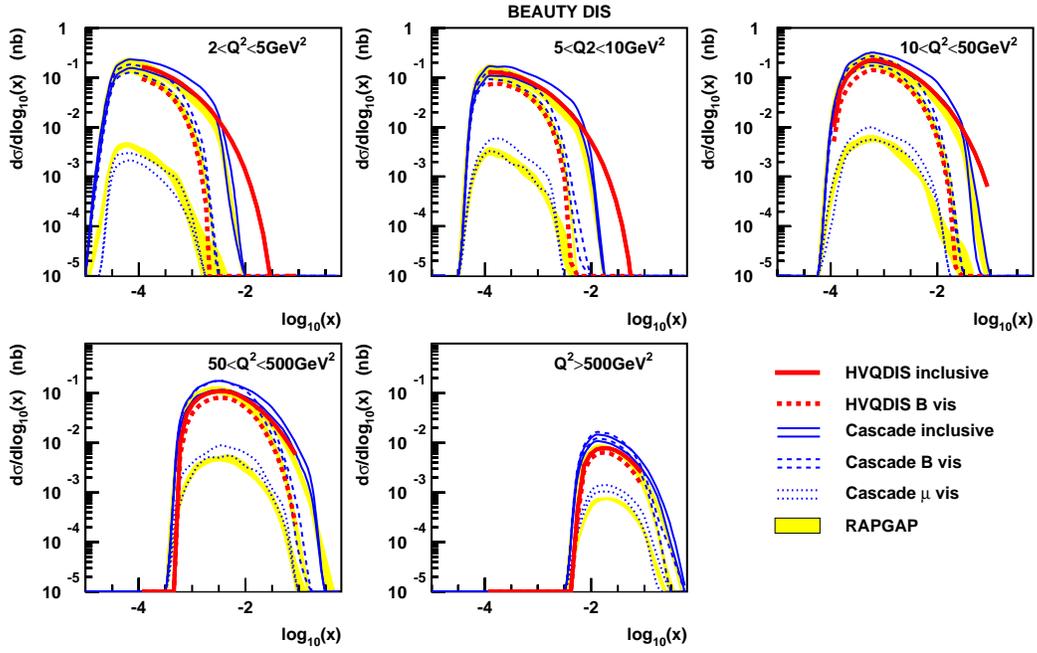

Fig. 5: Beauty cross sections in DIS at HERA. Each plot shows the distribution of $\log_{10}(x)$ in a different $Q^2$ range for the inclusive cross-section, the cross section for a hadron containing a $b$ quark in the "visible" range $p_T(B) > 3$ GeV, $|Y(B)| < 2$ and the cross section for a muon from beauty decay in the range $p_T(\mu) > 3$ GeV, $|Y(\mu)| < 2$. The thick curves show the central value from HVQDIS, the thin curves represent the uncertainty band from CASCADE and the shaded area shows the uncertainty band from RAPGAP. The muon distributions are not given for HVQDIS.

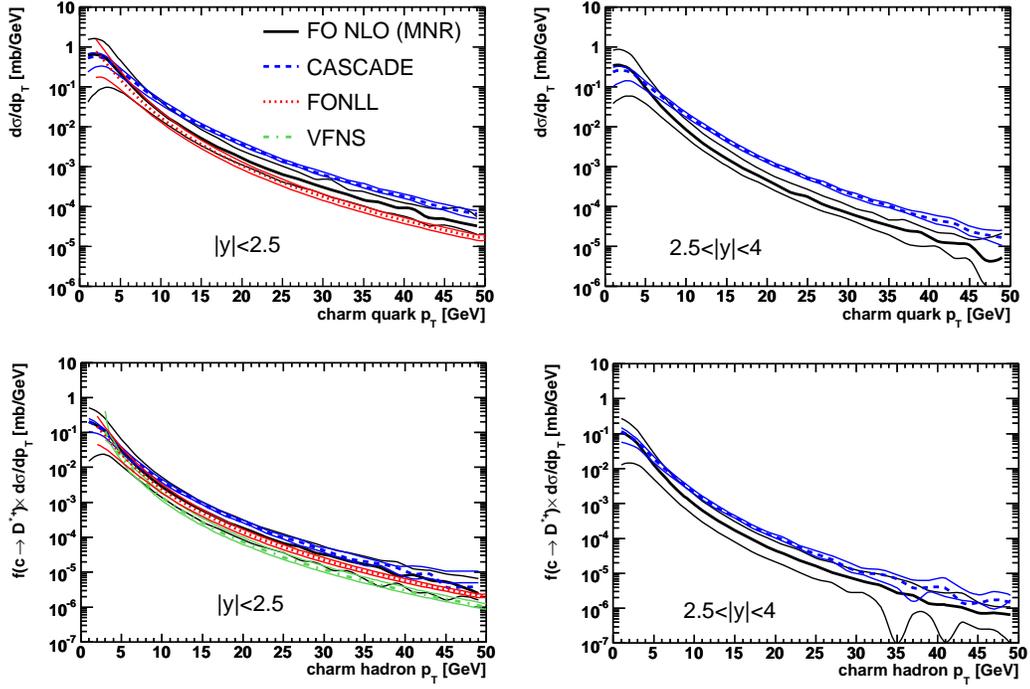

Fig. 6: Cross sections for charm production in $pp$ collisions at the LHC with $\sqrt{s} = 14$ TeV. The differential cross sections in $p_T$ for $c$ quark in the two rapidity ranges $|Y| < 2.5$ and $2.5 < |Y| < 4$ are shown in the upper panels. The lower panels show the cross sections for the production of a $D^*$ meson as a function of $p_T(D^*)$ in the same rapidity ranges.

### 4.3 LHC

For LHC, we computed the cross sections in $pp$ collisions at $\sqrt{s} = 14$ TeV.

Figures 6 and 7 show the single inclusive cross sections as a function of $p_T$, at quark (upper panels) and hadron (lower panels) level, for charm and beauty, respectively. Two rapidity intervals are considered: $|Y| < 2.5$, approximately covering the acceptance of the barrel detectors of ATLAS ($|\eta| < 2.5$), CMS ($|\eta| < 2.5$), and ALICE ($|\eta| < 0.9$); $2.5 < |Y| < 4$, approximately covering the acceptance of LHCb ($2 < \eta < 5$) and of the ALICE muon spectrometer ($2.5 < \eta < 4$).

For charm, we compare the fixed-order NLO results from MNR to the results from the CASCADE event generator, from the GM-VFNS calculation and from the FONLL calculation. The agreement is in general good, in particular in the low-$p_T$ region; at high-$p_T$ CASCADE predicts a larger cross section than the other calculations, especially at forward rapidities. The FONLL central prediction is in agreement with that of the FO NLO calculation at low $p_T$, while deviating from it at high $p_T$, where it gives a smaller cross section.

For beauty, we compare FO NLO (MNR), FONLL and CASCADE. Again, there is agreement at low $p_T$, where, as expected, the FONLL result coincides with the MNR result. At high $p_T$, both CASCADE and FONLL predict a larger cross section than the MNR central values, but all models remain compatible within the theoretical uncertainties. At forward rapidities, for beauty as for charm, CASCADE gives a significantly larger cross section than MNR.

Figure 8 shows the breakdown of the uncertainties for hadron production as obtained with MNR. The perturbative component dominates at LHC. Only the fragmentation component for charm hadron production becomes comparable in size to the perturbative one at large $p_T$.

### 4.4 $Q$-$\bar{Q}$ correlations

The azimuthal separation between the two heavy quarks $\Delta\phi(Q\bar{Q})$ and the transverse momentum of the quark-antiquark system $p_T(Q\bar{Q})$ are particularly sensitive to higher-order effects since at leading order their distri-butions are delta functions peaked at $\Delta\phi(Q\bar{Q}) = \pi$ and $p_T(Q\bar{Q}) = 0$. The distribution of these variables is therefore a direct probe of QCD radiation and is well suited for comparing different calculations.

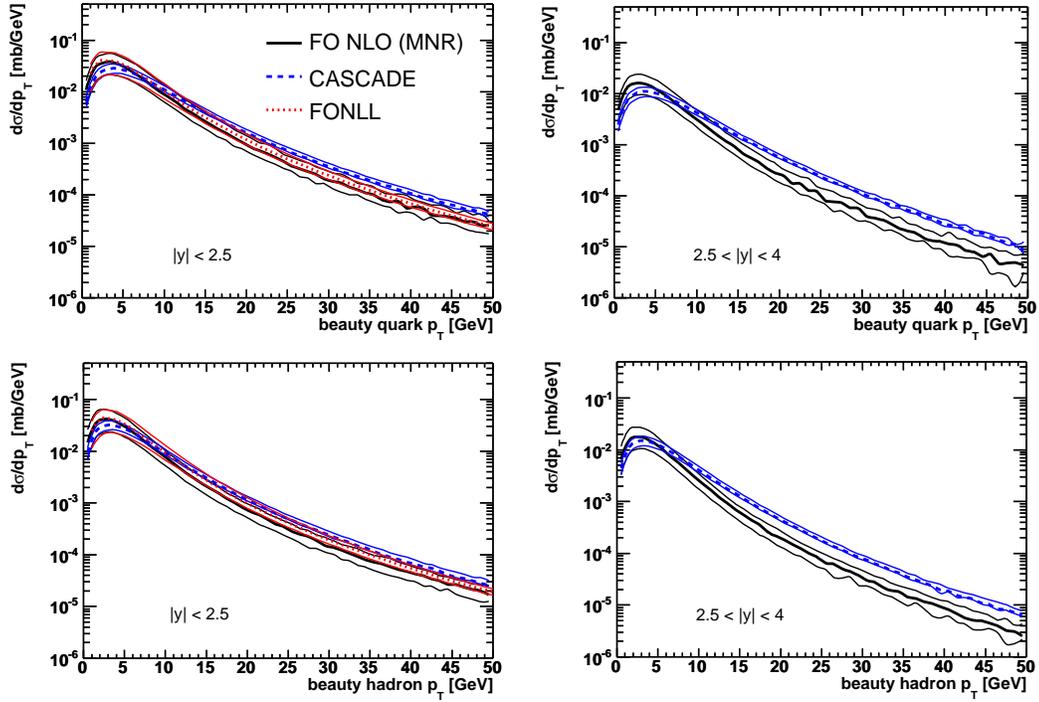

Fig. 7: Cross sections for beauty production in $pp$ collisions at the LHC with $\sqrt{s} = 14$ TeV. The differential cross sections in $p_T$ for $b$ quark in the two rapidity ranges $|Y| < 2.5$ and $2.5 < |Y| < 4$ are shown in the upper panels. The lower panels show the cross sections for the production of a beauty hadron as a function of $p_T$ in the same rapidity ranges.

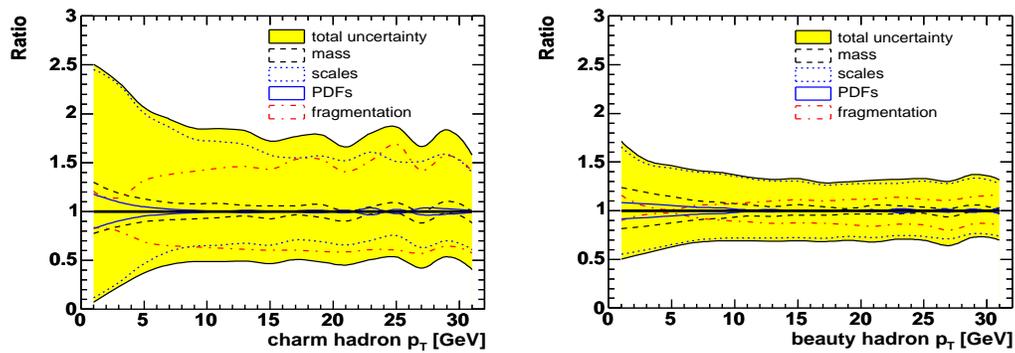

Fig. 8: Breakdown of the different components of the uncertainty on $d\sigma/dp_T$ for charmed (a) and beauty (b) hadrons at LHC as obtained from MNR. The plots show the ratio of the upper/lower side of each uncertainty to the nominal value. The following sources of uncertainty are shown: quark mass ($m_Q$), parton density parametrisation (PDF), fragmentation parameter and the perturbative uncertainty from scale variations.

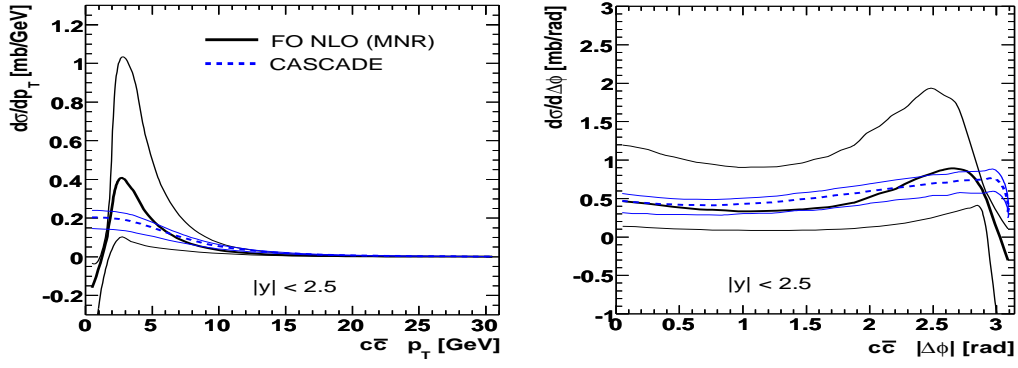

Fig. 9: $Q$-$\bar{Q}$ correlations for charm at LHC: $p_T$ of the $c\bar{c}$ pair (left) and azimuthal angle $\Delta\phi$ between the $c$ and the $\bar{c}$ (right). For both cross sections, the following kinematic cuts are applied: $|Y^c| < 2.5$, $|Y^{\bar{c}}| < 2.5$, $p_T^c > 3$ GeV, $p_T^{\bar{c}} > 6$ GeV.

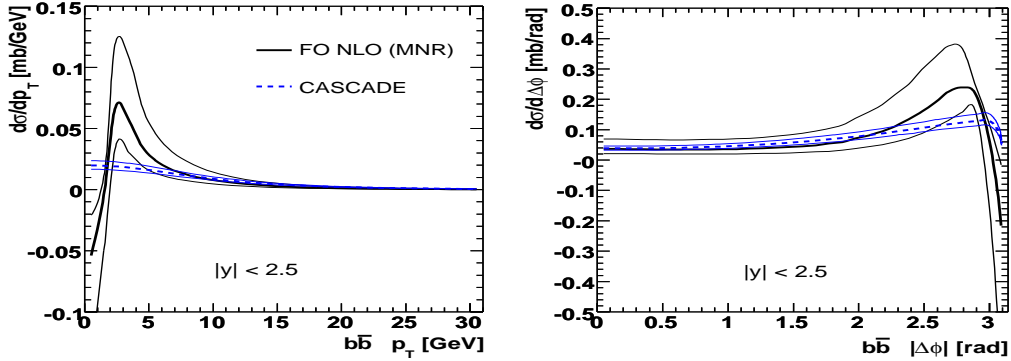

Fig. 10: $Q$-$\bar{Q}$ correlations for charm at LHC: $p_T$ of the $b\bar{b}$ pair (left) and azimuthal angle $\Delta\phi$ between the $b$ and the $\bar{b}$ (right). For both cross sections, the following kinematic cuts are applied: $|Y^b| < 2.5$, $|Y^{\bar{b}}| < 2.5$, $p_T^b > 3$ GeV, $p_T^{\bar{b}} > 6$ GeV.

Figures 9 and 10 show the heavy-quark pair $p_T$ distribution and the quark-antiquark relative azimuthal angle distribution for charm and beauty at LHC, respectively. For both distributions, the two quarks of the pair are required to have $|Y| < 2.5$; also minimum $p_T$ selections are applied to mimic the effect of realistic experimental cuts ($p_T^Q > 3$ GeV and $p_T^{\bar{Q}} > 6$ GeV). In the region near $\Delta\phi(Q\bar{Q}) = \pi$ and $p_T(Q\bar{Q}) = 0$, where the cancellation of soft and collinear divergencies occur, the fixed-order NLO calculation gives an unphysical negative cross section with next to a large positive peak. A larger binning would be needed to average this behavior and produce a more physical results. The CASCADE MC, has a more realistic behavior. Both calculations have a non-zero value at $\Delta\phi(Q\bar{Q}) = 0$ related to "gluon-splitting" events. A similar result was found for HERA as shown in Figure 11. This kind of distribution is expected to be well described by programs that merge NLO matrix elements to the parton-shower MC approach such as MC@NLO [25].

## 5 Conclusions

Heavy-flavour cross sections for HERA and LHC, obtained with fixed-order NLO programs, with matched massive/massless calculations and within the $K_T$-factorisation approach have been compared. Similar results are found for photoproduction at HERA and for the LHC. As expected the resummed calculations were found to be compatible with the fixed-order results but have smaller uncertainties at large $p_T$. Resummed calculations for charm in two different schemes (GM-VFNS and FONLL) are anyway somewhat incompatible both at HERA and LHC, suggesting that their uncertainty may be underestimated. The $K_T$-factorisation program CASCADE predicts larger cross sections than the other approaches at large $p_T$ at LHC and for charm at HERA. The comparison for DIS was limited to FO-NLO and a MC program with leading order matrix elements. Large discrepancies, which deserve further investigations, were found in this case. A comparison with experimental data would be needed for further understanding of the quality of the available calculations.

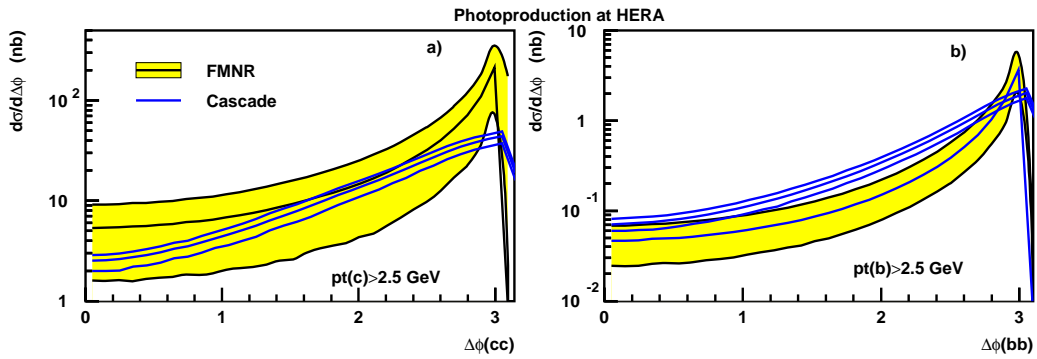

Fig. 11: Azimuthal $Q$-$\bar{Q}$ correlations in photoproduction at HERA for charm (a) and beauty (b). One of the two quarks was required to be in the "visible" region $p_T(Q) > 2.5$ GeV and $|Y(Q)| < 2$.